\documentclass[%
 reprint,
superscriptaddress,
 amsmath,amssymb,
 aps,
prx
]{revtex4-2}

\usepackage{graphicx}
\usepackage{dcolumn}
\usepackage{xcolor}
\usepackage{bm}
\usepackage{makecell}

\newcommand{\abs}[1]{{\vert #1\vert}}

\newcommand{\avg}[1]{{\left\langle #1\right\rangle}}
\newcommand{\avgG}[1]{{\left\langle #1\right\rangle_G}}

\newcommand{\zgc}[0]{{\mathcal{Z}_\text{gc}}}
\newcommand{\at}[0]{\bigg|}
\newcommand{\zg}[0]{{\mathcal{Z}_G}}

\begin{document}

\preprint{APS/123-QED}

\title{Classical theory of electron-ion correlations at electrochemical interfaces: \\ Closing the circuit from double-layer charging to ion adsorption}

\author{Nils Bruch}
\affiliation{Theory and Computation of Energy Materials (IET-3), Institute of Energy Technologies, Forschungszentrum J\"ulich GmbH, 52425 J\"ulich, Germany}
\affiliation{Chair of Theory and Computation of Energy Materials, Faculty of Georesources and Materials Engineering, RWTH Aachen University, 52062 Aachen, Germany}

\author{Michael Eikerling}

\affiliation{Theory and Computation of Energy Materials (IET-3), Institute of Energy Technologies, Forschungszentrum J\"ulich GmbH, 52425 J\"ulich, Germany}
\affiliation{Chair of Theory and Computation of Energy Materials, Faculty of Georesources and Materials Engineering, RWTH Aachen University, 52062 Aachen, Germany}

\author{Tobias Binninger}
\email{t.binninger@fz-juelich.de}
\affiliation{Theory and Computation of Energy Materials (IET-3), Institute of Energy Technologies, Forschungszentrum J\"ulich GmbH, 52425 J\"ulich, Germany}
\date{\today}

\begin{abstract}
The electric double layer (EDL) that forms at the interface between metals and ionic solutions is at the heart of various energy technologies. Recent experimental data have challenged our traditional understanding of the EDL charging behavior, which is based on mean-field Gouy-Chapman-Stern-type (GCS) models. In this article, we present a classical theory for the EDL, derived from first-principles statistical mechanics, that accounts for electron--ion correlation effects using the method of image charges and systematically extends beyond the mean-field level. Such electron--ion correlations introduce an additional interaction between the metal surface and electrolyte ions, significantly altering the EDL structure. Our theory, valid in the limit of dilute electrolyte solutions and weakly charged metal surfaces, achieves quantitative agreement with experimental capacitance data across a wide range of electrode materials and electrolyte solvents, and thus resolves long-standing questions on the origin of discrepancies to GCS predictions. Thereby, the framework conceptually unifies the processes of double-layer charging and ion adsorption (electrosorption), which are typically considered as distinct phenomena, but are shown to be manifestations of the same fundamental electrostatic principles. 
\end{abstract}

\maketitle

\section{Introduction}
Understanding the structure and behavior of the electric double layer (EDL) at metal-electrolyte interfaces is of great importance for emerging electrochemical energy technologies, ranging from batteries to electrolyzers and fuel cells. The classical Gouy-Chapman-Stern (GCS) theory recently celebrated its 100th anniversary and remains a landmark description of the EDL \cite{sternZurTheorieElektrolytischen1924}. In the GCS model the electrolyte region is divided into two parts: an inner region called Helmholtz layer that is considered inaccessible to ions, and an outer diffuse layer, where the electrostatic potential distribution is described by the mean-field (MF) Poisson-Boltzmann (PB) equation. 

In Parsons-Zobel (PZ) plots, the inverse of the (measured) differential capacitance at the potential of zero charge (PZC) is plotted against the inverse of the diffuse layer Gouy-Chapman (GC) capacitance. Often-observed discrepancies between experimental PZ plots and the linear relationship with unity slope predicted by GCS theory are commonly attributed to surface roughness of the electrode \cite{shenEffectSurfaceRoughness2022} or pseudocapacitive contributions due to spurious ion adsorption. Recent electrochemical capacitance data for  Platinum (111) interfaces displayed striking discrepancies with GCS theory \cite{ojhaDoubleLayerPt111Aqueous2020}, whereby surface roughness can be largely excluded for single-crystalline electrodes, and no specific adsorption of electrolyte species is assumed to occur within the ``double-layer window'' of the applied electrode potential \cite{mostanyDeterminationGibbsExcess2003}. These latest results triggered a vivid and controversial discussion on our fundamental understanding of electrochemical charging phenomena. Among various proposed explanations, one hypothesis attributes the discrepancies to weak anion adsorption at the interface \cite{schmicklerEffectWeakAdsorption2021}, while another suggests an additional attractive interaction between ions and the metal surface, the origin of which however remains unclear \cite{doblhoff-dierModelingGouyChapman2021}. A third hypothesis points to specific adsorption of hydroxide ions \cite{huangZoomingInnerHelmholtz2023}. Despite different interpretations of the underlying electrochemical phenomena, all explanatory schemes agree in the postulation of a weak ion–surface interaction that is not captured by traditional GCS theory \cite{doblhoff-dierElectricDoubleLayer2023}. 

These observations suggest that GCS theory is missing important physical phenomena and needs to be refined. Fundamentally, GCS theory is a mean-field (MF) approach, which means that particles interact only with the mean electric field generated by all other charge carriers, thereby neglecting direct correlations due to instantaneous interactions. Recently, we have developed a systematic approach to include ion-ion and ion-solvent correlations in beyond-MF electrolyte models to account for the effects of ionic screening and solvation, respectively, on the properties of the EDL. Thereby, electrolyte correlations were shown to enhance the characteristic features of capacitance curves around the PZC, improving agreement with experimental data \cite{bruchIncorporatingElectrolyteCorrelation2024}. An aspect that is critically missing in state-of-the-art EDL models are electron-ion correlation effects, i.e., interactions due to instantaneous inhomogeneities in the distribution of the electronic surface charge induced by localized ionic charges in direct vicinity of the electrode surface. These are difficult to model, because the quantum mechanical nature of electrons generally requires the use of a fully quantum mechanical framework to describe the specific interactions between electronic orbitals of ions and the metallic band structure of the electrode \cite{gidopoulosKohnShamEquationsMulticomponent1998,binningerFirstprinciplesTheoryElectrochemical2023}. Focusing on \emph{unspecific} electrostatic interactions, in this article we present an approach based on the method of image charges for incorporating electron-ion correlation effects into a classical theory for the EDL.

The method of image charges is a tool to describe metallic and dielectric interfaces where imaginary mirror charges are introduced to ensure that the electrostatic boundary conditions at the interface are satisfied \cite{jacksonClassicalElectrodynamics1999}. In a perfect conductor model of a metal electrode, image charges can be interpreted as \emph{the limit of maximally correlated ion-electron pairs}, whereby the electronic charge of the metal electrode instantaneously adjusts to the (localized) ionic charges in the electrolyte---a ``classical variant'' of the Born-Oppenheimer approximation. The image charge effect causes an attractive interaction between the electrode surface and electrolyte ions that adds to the mean-field interactions mediated by the electrostatic potential. Image-charge models that account for finite Thomas-Fermi (TF) screening of the metal electrode yield an effective image-charge potential profile with a minimum located between an attractive and repulsive component \cite{kornyshevImagePotentialDielectric1977}, whose exact form depends on the TF length of the electrode material and the polarization response of the solvent \cite{hedleyWhatDoesIon2025}. Due to ionic screening in the electrolyte solution, the effects of image charges are strongest at weakly charged interfaces and low electrolyte ion concentrations \cite{attardElectrolytesElectricDouble1996}. Monte Carlo (MC) simulations have shown that accounting for image-charge effects, ion densities at the electrode--electrolyte interface are significantly elevated compared to traditional MF predictions \cite{torrieElectricalDoubleLayers1982,torrieElectricalDoubleLayers1982,henderson*MonteCarloSimulation2005}. Similar behavior was found in recent Molecular Dynamics (MD) simulations \cite{geadaInsightInducedCharges2018,ntimMolecularDynamicsSimulations2023,sonImagechargeEffectsIon2021}, as well as in integral-equation approaches \cite{levineTheoryModifiedPoissonBoltzmann1960,carnieStatisticalMechanicsElectrical1984,outhwaiteModifiedPoissonboltzmannEquation1970,bellLinearizedPotentialEquation1972,hendersonApplicationHypernettedChain1979,kjellanderCorrelationImageCharge1984,attardPoissonBoltzmannImages1988}. To the best of our knowledge, the link between image-charges, electron-ion correlations, and deviations of experimental capacitance plots from GCS theory has not yet been established.

In this article, we explore the role of electron-ion correlations at the interface between a perfect metallic electrode and an electrolyte solution in the maximally correlated limit. To this end, the method of image charges is incorporated into a statistical field theory of charged solid--liquid interfaces, building upon earlier approaches by Netz, Orland, and Lau \cite{netzPoissonBoltzmannFluctuationEffects2000,lauFluctuationCorrelationEffects2008}. The theory is valid in the weak coupling limit, i.e., for dilute electrolyte solutions and weakly charged interfaces around the PZC. It is thus well suited for describing recent experimental results on the electrochemical charging behavior of Pt(111)  \cite{ojhaDoubleLayerPt111Aqueous2020} that have raised questions on the validity of GCS theory in this limit. While the theory is not intended for treating concentrated electrolytes or highly charged interfaces, it systematically bridges the gap between traditional GCS theory and beyond-MF approaches. 

In particular, we employ a Hubbard-Stratonovich (HS) transformation \cite{hubbardCalculationPartitionFunctions1959, stratonovichMethodCalculatingQuantum1957} to convert the statistical partition function of the charged interfacial system into a functional integral involving the electrostatic potential \cite{netzPoissonBoltzmannFluctuationEffects2000}. Such field theoretic approaches have been used to study dielectric interfaces for biological membrane systems \cite{markovichIonicProfilesClose2016,buyukdagliVariationalApproachElectrolyte2010,lueVariationalPerturbationTheory2017, wangEffectsImageCharges2013,wangTheoreticalDescriptionWeakly2015} and recently also for metallic electrodes \cite{zhouImageChargeEffects2024}. Here, for the first time, we demonstrate that electron-ion correlation effects consistently explain long-standing discrepancies between experimental capacitance data and GCS predictions across a wide range of electrode materials in both aqueous and non-aqueous electrolyte solutions, in a limit where classical GCS theory is typically assumed to be applicable. In our model, the strength of the electron-ion correlation effect is controlled by a single parameter that determines the effective minimum distance between an ion and its image charge. Tuning to zero corresponds to the limit of a maximally localized electron-ion pair, establishing a seamless transition from EDL charging to unspecific ion adsorption.

The article is organized as follows. In the theory section (Sec.~\ref{Sec:Theory}), we systematically derive analytical expressions for electrostatic potential, ion densities, and the differential capacitance, as function of the electrode potential including the effect of electron-ion correlations. In the results section (Sec.~\ref{SubSec:Summary1}), predictions of our theoretical capacitance model are compared with experimental capacitance data for aqueous as well as non-aqueous electrolytes, demonstrating that the model quantitatively explains experimental data trends at the PZC, which the GCS model was found incapable of. The additional attractive interaction, previously postulated as the cause of the strongly suppressed Parsons-Zobel slope of Pt(111) capacitance data \cite{doblhoff-dierModelingGouyChapman2021}, is thus suggested to result from interfacial correlations between the metal electronic surface charge and electrolyte ions. In the discussion section (Sec.~\ref{Sec:Discussion}), we analyze the profound consequences of the developed theory for overcoming the traditional conceptual distinction between EDL charging and ion adsorption mechanisms, thus unifying the theoretical description of double-layer capacitance and pseudocapacitance.

\section{Theory}\label{Sec:Theory}

\begin{figure}[t]
	\centering
	\includegraphics[width=\columnwidth]{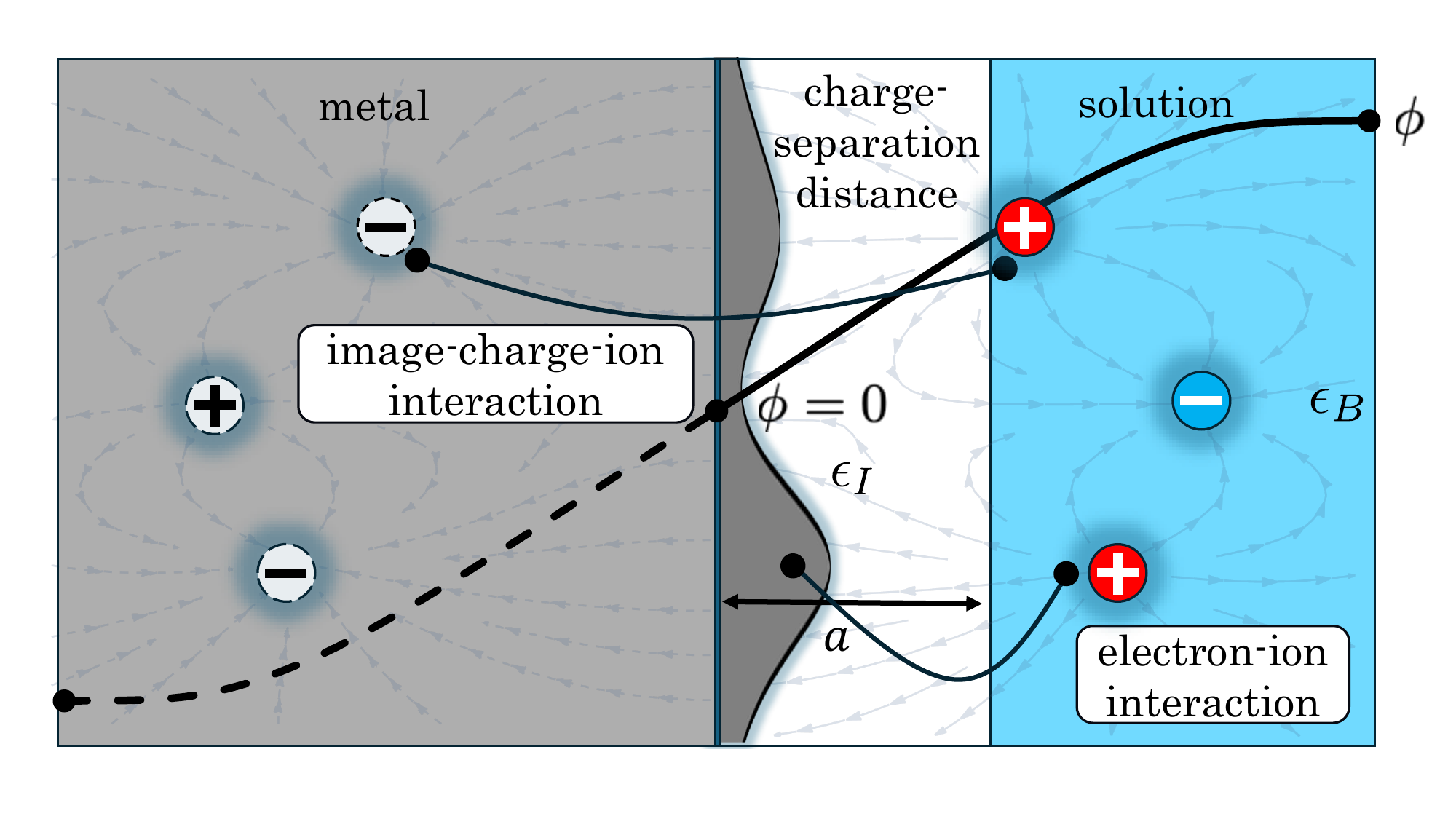}
	\caption{Metal (left) in contact with electrolyte solution (right). Solution and metal region are seperated by a charge-separation distance $a$ with a residual dielectric permittivity $\epsilon_I$. The solution is characterized by a solvent dielectric permittivity $\epsilon_B$. Arrows are the physical electrostatic field vectors for the ionic configuration shown. Image-charges in the metal (charges with dashed boundary) enforce the boundary condition at the metal surface and represent the physical electron charge density (dark gray). Due to the symmetry of the image charge method, the electrostatic potential $\phi$ is antisymmetric under reflection at the image charge plane (black line in the solution, black-dashed in the metal).} 
 \label{fig:SchemcaticModel}
\end{figure}

This section presents the derivation of a classical theory of a perfect metallic conductor in contact with an electrolyte solution, accounting for electron-ion correlations using the method of image-charges. A schematic of the model is illustrated in Fig.~\ref{fig:SchemcaticModel}. On the left-hand side is a metal electrode and on the right-hand side an electrolyte solution. In the image-charge method \cite{jacksonClassicalElectrodynamics1999}, for each charge in the electrolyte (e.g., the anion and two cations in the solution phase in Fig.~\ref{fig:SchemcaticModel}), an image charge of opposite sign is placed on the opposite side but same distance from the metal surface (indicated as dashed charges in Fig.~\ref{fig:SchemcaticModel}). The image-charge method provides the correct \emph{inhomogeneous} electronic surface charge distribution $\sigma_M$ at the metal electrode for any given ion configuration (gray-shaded curve in Fig.\ref{fig:SchemcaticModel}). On the contrary, in classical theories, the electrode charge is typically described by a \emph{homogeneous} surface charge density at the metal surface that is constant in in-plane directions \cite{netzPoissonBoltzmannFluctuationEffects2000,lauFluctuationCorrelationEffects2008,markovichIonicProfilesClose2016}. It will be shown that the present image-charge theory reduces to the constant-surface-charge model at the mean-field level, where ions interact only with mean charge distributions and the electronic surface charge gets smeared out to a constant in-plane averaged value. This means that, conceptually, the image charges already represent the entire electronic surface charge of the metal electrode, and no additional surface charge must be applied. Electrostatic correlations beyond the MF level captured by our framework are thus interpreted as a classical representation of fundamental electron-ion correlations at electrochemical interfaces \cite{binningerFirstprinciplesTheoryElectrochemical2023}. 

To avoid the divergence of the electrostatic energy as ion and image charges get arbitrarily close to each other, a minimum distance $a$ must be introduced to which ions can approach the image plane of the metal surface. While this appears as an effective ``contact layer'' akin to the Helmholtz layer in GCS theory, it is important to note that in the present theory, the parameter $a$ must \emph{not} be interpreted as the geometric minimum distance between electrode surface and the atomic centers of ions. Instead, the parameter $a$ controls the minimum distance between electronic and ionic \emph{charge centers}, which is not restricted in value to typical interatomic distances. In particular, $a$ can assume values of sub-atomic scale due to electron spillover from the electrode surface into the electrolyte region \cite{schmicklerElectronicEffectsElectric1996}. Thereby, $a$ is related to the effective charge screening length at the electrode--electrolyte interface, as described by the Thomas-Fermi length in another recent model \cite{hedleyWhatDoesIon2025}. However, the present theory even allows $a$ to approach values close to zero to describe the limit of charge transfer during ion adsorption, corresponding to (partial) ion discharge as the electronic (image) charge gets fully contracted to the site of the adsorbing ion.

As this work is focused on electron-ion correlations, the effect of the polarizable solvent medium is implicitly described by a dielectric background of permittivity $\epsilon_B$ in the electrolyte region. This approach allows studying the essential influence of electron-ion correlations without interference from other effects such as interfacial solvent structuring \cite{bruchVariationalFunctionalTheory2024}. An explicit description of the dipolar solvent is intended to be included in future extensions of the theory. A residual dielectric permittivity $\epsilon_I$ is assumed in the interfacial region between the metal image plane (located at $z=0$) and the electrolyte boundary seperated by a charge-separation distance $a$. Therefore, the overall dielectric profile of the system in the region $z>0$ is given by
\begin{align}\label{DielectricPermittivity}
    \epsilon(r)=
\begin{cases}
\epsilon_I\,\quad\text{for $0<z<a$},\\
\epsilon_B\,\quad\text{for $z>a$},\\
\end{cases}
\end{align}
where $z$ is the coordinate perpendicular to the metal surface. For the purpose of clarity we will abbreviate the 3D spatial coordinate as $r\equiv\vec{r}$. 

The potential energy of the system of ion species and their image charges interacting by Coulomb interaction $V(r,r')=1/(4\pi\epsilon(r)\abs{r-r'}) $ is
\begin{align}\label{Hamiltonian}
    U=\frac{1}{4}\int_{r,r'}\rho(r)V(r,r')\rho(r') \ ,
\end{align}
where the integration domain comprises both the electrolyte and metal image-charge regions. The one-quarter prefactor (instead of the usual one-half \cite{netzPoissonBoltzmannFluctuationEffects2000}) is due to the fact that the electrostatic energy of the real system of electrolyte and metal surface charge (electrons) is equal to half of the coulombic energy of the system of ions and image charges \cite{taddeiSubtletiesEnergyCalculations2009}. The charge density, which comprises both ionic and image charges, is given by
\begin{align}
    \rho(r)&=\rho_{\text{ion}}(r)+\rho_{\text{image}}(r) \nonumber \\
    &=\sum_{i=a/c}\sum_{k=1}^{N_i}q_{i}\left(\delta(r-r_{i,k})-\delta(r-r_{i,k}^{\text{im}})\right),
\end{align}
where $i = a, c$ denote anions and cations, and $q_i$ and $r_{i,k}$, respectively, are the charge and position of ion $k$ of type $i$. Its image charge is located at $r_{i,k}^\text{im}$, which is given by $r_{i,k}$ but with the sign of the $z$ coordinate reversed, $z_{i,k}^\text{im} = -z_{i,k}$. Unlike the canonical ensemble with fixed ionic particle numbers \(N_i\), the grand-canonical ensemble fixes the ionic chemical potentials \(\mu_i\) via external reservoirs, determining \(N_i\) and thus the metal charge due to electroneutrality. Here, the grand-canonical description is adopted to derive a theory for interface charging as a function of the applied electrode potential. 

In order to calculate the thermodynamic quantities of a grand-canonical system, it is first necessary to determine the grand-canonical partition function $\zgc$. The grand-canonical partition function for a system comprising $N_i$ ionic charges with thermal wavelength $\Lambda_i$, fugacity $\lambda_i=\exp(\beta\mu_i)$, and chemical potential $\mu_i$ is the configuration integral over all states,
\begin{align}
    \zgc(\left\{\lambda_i \right\})=\prod_{i=a/c}\sum_{N_{i}=0}^{\infty}\lambda_i^{N_i}\frac{1}{N_{i}!(\Lambda_i^{3})^{N_{i}}}\prod_{j=1}^{N_{i}}\int d^{3}r_{i,k} \nonumber\\
    \Theta(z_{i,k}-a)\exp(-\beta U).
\end{align}
The ions are free to move in the region of the electrolyte solution but constrained by a minimal distance $a$ to the metal image plane, as described by the Heaviside step function $\Theta$. A Hubbard-Stratonovich (HS) transformation, as outlined in Refs. \cite{hubbardCalculationPartitionFunctions1959,bruchVariationalFunctionalTheory2024}, introduces the HS field, $\psi$, which yields the field-theoretic formulation of the partition function,
\begin{align}\label{PartitionFunction}
    \zgc(\left\{\lambda_i \right\})=\int D\psi\exp\left(-\beta S[\psi]\right),
\end{align}
as a functional over a non-linear action,
\begin{align}\label{PsiAction}
    &S[\psi] =\int_{r}\epsilon(r)\left(\nabla\psi(r)\right)^{2} \nonumber \\
    &-\sum_{i=a/c}\beta^{-1}\lambda_{i}\Lambda_i^{-3}\int_{r}\Theta(z-a)e^{-i\beta q_{i}\left(\psi(r)-\psi(r^{\text{im}})\right)}.
\end{align}
The action functional is non-local as $\psi$ is evaluated both at $r$ and $r^\text{im}$. The problem is significantly simplified by utilization of the mirror (anti)symmetry with respect to the image plane at $z=0$, and restricting the integral measure to HS fields in the solution region,
\begin{align}
D\psi&=\Pi_{r\in\text{(Electrode+Solution)}} d\psi(r) \nonumber\\
    &=\Pi_{r\in\text{(Solution)}} d\psi(r)d\psi(r^\text{im})\equiv D(\psi,\psi^\text{im}),\label{NewMeasure}
\end{align}
effectively doubling the functional degrees of freedom. This freedom allows to define a linear field transformation to the symmetric, $\psi_s$, and anti-symmetric, $\psi_a$, components of $\psi$,
\begin{align}
    \psi_s(r)=\psi(r)+\psi(r^\text{im}),\label{LinearTrafo1} \\
    \psi_a(r)=\psi(r)-\psi(r^\text{im})\label{LinearTrafo2}.
\end{align}
Inserting Eq. \eqref{NewMeasure}, and applying the variable transformations \eqref{LinearTrafo1} and \eqref{LinearTrafo2} in \eqref{PartitionFunction} one arrives at
\begin{align}\label{PartitionFunction2}
    \zgc(\left\{\lambda_i \right\})=\int D(\psi_a,\psi_s)\exp\left(-\beta S[\psi_a,\psi_s]\right),
\end{align}
with
\begin{align}\label{PsiAction2}
    &S[\psi_a,\psi_s] =\int_{r>0}\frac{\epsilon(r)}{2}\left(\nabla\psi_s(r)\right)^{2}+\int_{r>0}\frac{\epsilon(r)}{2}\left(\nabla\psi_a(r)\right)^{2} \nonumber \\
    &+(-1)\sum_{i=a/c}\beta^{-1}\lambda_{i}\Lambda_i^{-3}\int_{r>0}\Theta(z-a)e^{-i\beta q_{i}\psi_a(r)},
\end{align}
where $\int_{r>0}$ stands symbolically for the integration over the electrolyte solution region $(z>0)$. The value of the Jacobian determinant of the linear transformation above has been absorbed into the measure $D(\psi_a,\psi_s)$. The integrals over $\psi_s$ and $\psi_a$ are fully decoupled. Performing the Gaussian functional integral over $\psi_s$ yields a simple multiplicative constant that can be absorbed into the functional measure \footnote{Multiplicative constants do not play a role for thermodynamic averages and are therefore neglected. This is due to the fact that averages are normalized with respect to the partition function. Therefore, multiplicative constants simply cancel out and can therefore be omitted.}. What is left is the integral over $\psi_a$,
\begin{align}\label{FinalPartitionFunction}
     \zgc(\left\{\lambda_i \right\})=\int D\psi_a \exp\left(-\beta S[\psi_a]\right),
\end{align}
with an effective action
\begin{align}
    &S[\psi_a]=S_\text{free}[\psi_a]+S_\text{int}[\psi_a,\left\{\lambda_i \right\}] =\underbrace{\int_{r>0}\frac{\epsilon(r)}{2}\left(\nabla\psi_a(r)\right)^{2}}_{S_\text{free}} \nonumber \\
    &+\overbrace{(-1)\sum_{i=a/c}\beta^{-1}\lambda_{i}\Lambda_i^{-3}\int_{r>0}\Theta(z-a)e^{-i\beta q_{i}\psi_a(r)}}^{S_\text{int}} \ ,
    \label{Action}
\end{align}
where $S_\text{free}$ is the action of a field theory without interactions and $S_\text{int}$ is the Coulomb interacting part of the theory, controlled by the ionic chemical potentials (fugacities $\lambda_i$). The image-charge contribution to the action is the restriction of the integral to the solution side along with the boundary condition for $\psi_a(z=0)=0$ (because $\psi_a$ is anti-symmetric). From Eq.~\eqref{FinalPartitionFunction}, thermodynamic quantities can be extracted utilizing functional averages,
\begin{align}\label{Defn:FunctionalAverage}
    \avg{...}=\frac{1}{\zgc}\int D\psi_a e^{-\beta S[\psi_a]} (...).
\end{align}
Adding a fictitious term to the potential energy (Eq.~\eqref{Hamiltonian}) of the form $\int_r\;v_i^\text{aux}(r)n_i(r)$ that couples an auxiliary potential $v_i^\text{aux}$ to the ion density $n_i(r)=\sum_k \delta(r-r_{i,k})$ allows to compute the the average ion density (cf. \cite{bruchVariationalFunctionalTheory2024}) through
\begin{align}\label{Densities}
    \avg{n_{i}(r)}&=-\frac{1}{\beta}\frac{1}{\zgc}\frac{\delta\zgc}{\delta v_i^\text{aux}(r)}\at_{v_i^\text{aux}=0}\nonumber \\
    &=\lambda_{i}\Lambda_i^{-3}\Theta(z-a)\avg{e^{-i\beta q_{i}\psi_a(r)}}.
\end{align}
The Schwinger-Dyson equation $\avg{\delta S/\delta\psi}=0$ \cite{lauFluctuationCorrelationEffects2008} for Eq.~\eqref{Action} is the Poisson equation, 
\begin{align}\label{PoissonEquation}
    -\nabla\left(\epsilon(r)\nabla\phi(r)\right)=\sum_{i=a/c}q_i\avg{n_i(r)},
\end{align}
with the electrostatic potential given by
\begin{align}\label{Defn:ElectrostaticPotential}
    \phi(r)=i \avg{\psi_a(r)},
\end{align}
where $i$ is the imaginary unit. Through the electrostatic potential, the surface charge density at the metal phase can be obtained using Gauss's theorem,
\begin{align}\label{MetalChargeDensity}
    \sigma_{M}&=-\epsilon_{B}\frac{\partial\phi(r)}{\partial z}\at_{z=a},
\end{align}
where $\epsilon_B$ is the bulk permittivity of the electrolyte solvent, cf. Eq.~\eqref{DielectricPermittivity}. As a primary objective, we aim to determine the contribution of electron-ion correlations, viz., the image-charge effect, to the interfacial capacitance, $C_d$. The latter is defined as the (partial) derivative of the metallic surface charge with respect to the electrode potential, $\mathcal{E}$,
\begin{align}\label{Defn:DifferentialCapacitance}
    C_d = \frac{\partial \sigma_M}{\partial\mathcal{E}}.
\end{align}

\subsection{Definition of the electrode potential}\label{SubSec:ElectrodePotential}
Experimentally, the electrode potential is defined as the measurable voltage between the working and reference electrodes, which is essentially equal to the difference between the respective electronic chemical potentials. In classical modelling, the electrode potential is typically defined as the difference between the inner electrostatic potentials of the metal electrode and electrolyte, which is consistent with the experimental definition up to an unimportant additive constant. For the present image-charge model, however, the definition of the electrode potential requires clarification. In Eq.~\eqref{Defn:ElectrostaticPotential}, the electrostatic potential is determined by the anti-symmetric HS field, which is zero at the electrode surface (mirror plane) and approaches an inner bulk value that is equal in magnitude and opposite in sign to the inner potential in the electrolyte, see Fig.~\ref{fig:SchemcaticModel}. The question thus arises whether the ``real'' inner potential of the electrode is given by the value of the anti-symmetric electrostatic potential at the electrode surface or in the bulk electrode region. In other words, does the electrode potential correspond to the electrostatic potential difference between solution bulk and metal \emph{surface}, or, due to the presence of the image charges, between solution bulk and metal \emph{bulk}? 

Fundamentally, the electrode potential is defined as the free-energy change for transferring a neutral pair of an electron and ion from a reference state to the working electrode \cite{boettcherPotentiallyConfusingPotentials2021}. Thermodynamically, the partial derivative of the grand-potential  $\Omega$ with respect to the electrode potential is equal to the negative excess charge of the metal surface, $Q_M$ \cite{binningerFirstprinciplesTheoryElectrochemical2023},
\begin{align}\label{Def:ElectrodePotential}
    \frac{\partial\Omega}{\partial\mathcal{E}} = -Q_M \ .
\end{align}
Due to the required overall charge neutrality, the electrode surface charge equals the negative of the net ionic charge of the electrolyte, $Q_M=-q\left(N_{c}-N_{a}\right)$, where $q$ is  the charge number of cations and anions. For single-valent ions, $q$ is equal to the elementary charge, $q = e$. The grand-potential is the Legendre transform of the free energy
\begin{align}\label{DefnOmega}
    \Omega(\mu_a,\mu_c)=\mathcal{F}(N_a,N_c) - \mu_a N_a-\mu_c N_c,
\end{align}
where $\mu_i$ and $N_i$ are the chemical potential and particle number of anions and cations. It will turn out useful to perform a linear transformation of the ionic chemical potentials to define two new chemical potentials, 
\begin{align}
    \mu_{+} = \frac{\mu_{c}+\mu_{a}}{2}, \label{MuBar} \\
     \mu_{-} = \frac{\mu_c-\mu_a}{2}\label{MuTilde}.
\end{align}
Inserting Eqs.~\eqref{MuBar} and \eqref{MuTilde} into Eq.~\eqref{DefnOmega} leads to
\begin{align}
    \Omega(\mu_{+},\mu_{-}) &=\mathcal{F}(N_a,N_c) -\mu_{+}\left(N_{c}+N_{a}\right)-\mu_{-}\left(N_{c}-N_{a}\right), \nonumber\\
    &=\mathcal{F}(N_a,N_c) -\mu_{+}\;N_\text{tot}-\mu_{-}\; N_\text{diff} \label{GrandPotentialChemicalPotentials} \ ,
\end{align}
with the total number of ions, $N_\text{tot}=N_{c}+N_{a}$, and the number difference between cations and anions, $N_\text{diff}=N_{c}-N_{a}$, corresponding to the net charge $Q_\text{sol}=q N_\text{diff}$ of the electrolyte solution, which is the negative of the net metallic surface charge $Q_\text{M}=-Q_\text{sol}$. Computing the partial derivatives of $\Omega$ with respect to $\mu_{+}$ and $\mu_{-}$, shown in Appendix \ref{Appendix:ComputingPartialDerivatives}, 
\begin{align}
    \frac{\partial\Omega(\mu_{+},\mu_{-})}{\partial \mu_{+}} &=-N_\text{tot},\label{PartialDerivative1} \\
    \frac{\partial\Omega(\mu_{+},\mu_{-})}{\partial \mu_{-}} &=-N_\text{diff} \,=\, Q_\text{M}/q,\label{PartialDerivative2}
\end{align}
shows that $\mu_{+}$ and $\mu_{-}$ correspond to the chemical potentials for $N_\text{tot}$ and $N_\text{diff}$, respectively. Clearly, comparing Eq.~\eqref{Def:ElectrodePotential} with Eq.~\eqref{PartialDerivative2} shows that the electrode potential is defined by
\begin{align}\label{ElectrodePotential}
    \mathcal{E}  = -\frac{\mu_{-}}{q}\ .
\end{align}

\subsection{Perturbative expansion}\label{SubSec:PerturbativeExpansion}
Computing quantities such as the electrostatic potential (Eq.~\eqref{Defn:ElectrostaticPotential}) with the full functional average (Eq.~\eqref{Defn:FunctionalAverage}) is not possible to achieve analytically. The only functional averages that can be performed analytically are Gaussian averages, i.e., where the action is quadratic in $\psi_a$ \cite{zinn-justinQuantumFieldTheory2021}. This work employs a perturbative expansion of the HS field \cite{netzPoissonBoltzmannFluctuationEffects2000,podgornikInhomogeneousCoulombFluid1988} using the loop parmeter $\ell$, corresponding to a mean-field theory in zeroth order, and then systematically incorporating coulombic correlation effects, including image-charge correlations, with increasing orders.

The perturbative expansion
\begin{align}\psi_a(r)=\psi_c(r)+\ell^{-1/2}\delta\psi(r)
\label{eq_expansion_HS_field}
\end{align}
is performed around a specific field configuration $\psi_c$ that is specified later. In this context, the loop parameter $\ell$ serves the purpose of distinguishing between terms of different orders and is set to $\ell=1$ at the end of the calculation. In addition $\ell$ must be multiplied in front of the action (Eq.~\eqref{Defn:FunctionalAverage}) to systematically organize corrections in even powers of that parameter \cite{netzPoissonBoltzmannFluctuationEffects2000}. Expanding the electrostatic potential (Eq.~\eqref{Defn:ElectrostaticPotential}) yields
\begin{align}\label{PerturbativeElectrostaticPotential}
    \phi(r)=i \frac{1}{\zgc} \int D\delta\psi \left(\psi_c+\ell^{-1/2}\delta\psi\right)e^{-\beta \ell S[\psi_c+\ell^{-1/2}\delta\psi]}.
\end{align}
The action $S$, as defined in Eq.~\eqref{Action}, expanded around $\psi_c$, takes the following form
\begin{align}\label{ActionExpansion}
   & S[\psi_c+\ell^{-1/2}\delta\psi,\left\{\lambda_i \right\}]=S[\psi_{c},\left\{\lambda_i \right\}]
    \nonumber \\
    &+\ell^{-1/2}\int_{r,r'}\delta\psi(r)\frac{\delta S[\psi,\left\{\lambda_i \right\}]}{\delta\psi(r)}\at_{\psi=\psi_{c}}\nonumber \\
    &+\frac{\ell^{-1}}{2}\int_{r,r'}\delta\psi(r)\frac{\delta^{2}S[\psi,\left\{\lambda_i \right\}]}{\delta\psi(r)\delta\psi(r')}\at_{\psi=\psi_{c}}\delta\psi(r') \nonumber \\
    &+\sum_{n=3}^{\infty} \ell^{-n/2}\Gamma^{(n)}[\left\{\lambda_i \right\}],
\end{align}
with the so-called bare vertex,
\begin{align}\label{TheoryVertex}
    \Gamma^{(n)}[\left\{\lambda_i \right\}] = \frac{1}{n!}&\int_{r_1,\dots,r_n}\delta\psi(r_1)\ldots\delta\psi(r_n) \times \nonumber \\
    &\times\Gamma^{(n)}(\left\{\lambda_i\right\},r_1,\dots,r_n),
\end{align}
and its corresponding vertex function,
\begin{align}\label{TheoryVertexFunction}
    \Gamma^{(n)}(\left\{\lambda_i \right\},r_1,\dots,r_n) = \frac{\delta^n S[\psi,\left\{\lambda_i \right\}]}{\delta\psi(r_1)\,...\,\delta\psi(r_n)}\at_{\psi=\psi_{c}}.
\end{align}
Once the perturbative expansion exceeds the zeroth-order term, the chemical potentials must be adjusted in order to yield a well-defined theory without un-physical divergences \cite{lauFluctuationCorrelationEffects2008,brownEffectiveFieldTheory2001}. We choose an appropriate gauge chemical potential $\mu_\text{g}$ for the ion chemical potentials $\mu_i\to\mu_i+\ell^{-1}\mu_\text{g}$, which is the same for both ion species. For the fugacities, this results in another expansion
\begin{align}\label{FugacityExpansion}
    \lambda_i=e^{\beta \left(\mu_i+\ell^{-1}\mu_\text{g}\right)}=\lambda^0_i\left(1+\ell^{-1}\omega+\frac{\ell^{-2}}{2}\omega^2+\mathcal{O}(\ell^{-3})\right),
\end{align}
where $\omega=\beta\mu_\text{g}$. Inserting Eq.~\eqref{FugacityExpansion} into Eq.~\eqref{ActionExpansion} one arrives at
\begin{align}\label{ExpansionAction}
    &S[\psi,\left\{\lambda_i \right\}]=S[\psi_{c},\left\{\lambda^0_i \right\}] \nonumber\\
    &+\ell^{-1}S_{\text{int}}[\psi_c,\left\{\lambda^0_i \omega\right\}] + \ell^{-1/2}\int_r\delta\psi(r)\frac{\delta S[\psi,\left\{\lambda^0_i \right\}]}{\delta\psi(r)}\at_{\psi=\psi_c}  \nonumber\\
    &+\frac{\ell^{-1}}{2}\int_{r,r'}\delta\psi(r)\frac{\delta^{2}S[\psi,\left\{\lambda^0_i \right\}]}{\delta\psi(r)\delta\psi(r')}\at_{\psi=\psi_{c}}\delta\psi(r') \nonumber \\
    &+\ell^{-3/2}S_I^{(3)}[\psi_c] + \mathcal{O}(\ell^{-2})\ ,
\end{align}
with $S_{\text{int}}$ as defined in Eq.~\eqref{Action} and a renormalized vertex function of order three,
\begin{align}\label{FirstOrderIneracting}
    &S_I^{(3)}[\psi_c]=\int_{r,r'}\frac{\delta S_{\text{int}}[\psi,\left\{\lambda^0_i \omega \right\}]}{\delta\psi(r)}\at_{\psi=\psi_{c}}\delta\psi(r) \nonumber \\
    &+\frac{1}{3!}\int_{r,r',r''}\Gamma^{(3)}(\left\{\lambda^0_i \right\},r,r',r'')\delta\psi(r)\delta\psi(r')\delta\psi(r'').
\end{align}
The saddle-point configuration of the action functional of Eq.~\eqref{Action} is the field configuration with the largest statistical weight in the partition function of Eq.~\eqref{FinalPartitionFunction} \cite{podgornikInhomogeneousCoulombFluid1988}. Therefore, it is selected as the expansion point $\psi_c$ in the loop expansion of the HS field, Eq.~\eqref{eq_expansion_HS_field}. The saddle-point field satisfies
\begin{align}\label{SaddlePointEquation}
\frac{\delta S[\psi,\left\{\lambda^0_i \right\}]}{\delta\psi(r)}\at_{\psi=\psi_c} = 0.
\end{align}
Inserting Eq.~\eqref{ExpansionAction} along with Eq.~\eqref{SaddlePointEquation} into the expansion for the electrostatic potential (Eq.~\eqref{PerturbativeElectrostaticPotential}) yields
\begin{align}
    &\phi(r)=i\psi_c(r)+\nonumber\\
    &i\ell^{-1/2}\left\langle\delta\psi(r) \,\exp\left(-\beta\ell^{-1/2}S_I^{(3)}[\psi_c] + \mathcal{O}(\ell^{-2})\right)\right\rangle_G\ ,\label{FullExpansionElectrostaticPotential}
\end{align}
where 
\begin{align}\label{GaussianAverage}
    \avgG{...} = \frac{1}{\zg}\int\delta\psi\, (...)\,e^{-\frac{\beta}{2}\int_{r,r'}\delta\psi(r)G^{-1}(r,r')\delta\psi(r')},
\end{align}
with $\zg=\int\delta\psi\,\exp\left(-\frac{\beta}{2}\int_{r,r'}\delta\psi(r)G^{-1}(r,r')\delta\psi(r')\right)$, is a Gaussian functional average with respect to the second-order functional derivative of the action,
\begin{align}\label{DiffeqGreensFunction}
    G^{-1}(r,r') = \frac{\delta^2S}{\delta\psi(r)\delta\psi(r')}\at_{\psi=\psi_c}.
\end{align} 
In particular, since the average of Eq.~\eqref{GaussianAverage} is Gaussian, the average of a product of two HS field fluctuations is exactly given by
\begin{align}\label{Defn:CorrelationFunction}
    \avgG{\delta\psi(r)\delta\psi(r')}=\beta^{-1}G(r,r'),    
\end{align}
which is defined by
\begin{align}\label{ComputationOfCorrelationFunction}
    \int_{r''}G^{-1}(r,r'')G(r'',r')=\delta(r,r').
\end{align}
The correlation function  $G(r,r')$ (Eq.~\eqref{Defn:CorrelationFunction}) quantifies correlations in $\delta\psi$ by linking its amplitude at $r$ to that at $r'$. In a similar fashion, inserting the perturbative expansion of $\psi$ (Eq.~\eqref{eq_expansion_HS_field}) into the formula for the ion densities (Eq.~\eqref{Densities}) leads to 
\begin{align}\label{DensityExpansion}
     &\avg{n_{i}(r)}=\lambda_{i}\Lambda_i^{-3}\Theta(z-a)e^{-i\beta q_{i}\psi_c(r)}\nonumber \\
     &\avgG{e^{-\ell^{-1/2}i\beta q_{i}\delta\psi(r)}e^{-\beta\ell^{-1/2}S_I^{(3)}[\psi_c] + \mathcal{O}(\ell^{-2})}}.
\end{align}
The expansions of the electrostatic potential (Eq.~\eqref{FullExpansionElectrostaticPotential}) and ion densities (Eq.~\eqref{DensityExpansion}) are exact but also impossible to calculate analytically. The following sections present approximations corresponding to different orders in the perturbative expansion, i.e., truncating the expansions at different orders of $\ell$.

\subsection{Mean-field theory}\label{Sec:MFTheory}
At the mean-field (MF) level, only terms of zeroth order in $\ell$ are kept, i.e., $\delta\psi=0$. At this level of theory, the field $\psi_a(r)$ is replaced in all functional averages by the saddle-point field
$\psi_c(r)$. We will see at the end of this section, that this theory at MF level coincides with the traditional GCS theory with a homogeneous surface charge density, corresponding to an in-plane averaged (smeared-out) image-charge distribution. In physical terms, this means that the MF approach does not account for ion-ion and ion-electron correlations. 

\paragraph{Electrostatic potential.} The electrostatic potential (Eq.~\eqref{FullExpansionElectrostaticPotential}) at the MF level is simply given by 
\begin{align}
    \phi^\text{MF}(r)=i\psi_c(r).    
\end{align}
The zero potential reference is set at the metal surface (image plane) rather than in the bulk electrolyte. This choice is required by the asymmetry of $\psi_c$ upon reflection at the metal surface, which arises from the asymmetry of $\psi_a$ in Eq.~\eqref{LinearTrafo2} and the inherent asymmetry of the action in Eq.~\eqref{FinalPartitionFunction}. The saddle-point equation (Eq.~\eqref{SaddlePointEquation}) of the action (Eq.~\eqref{Action}) defining $\psi_c(r)$ (and thus $\phi^\text{MF}(r)$) within the electrolyte solution region $(z>a)$ is given by
\begin{align}\label{EqnSaddlePointSolution}
&-\epsilon_{B}\nabla^{2}\phi^\text{MF}(r)-\sum_{i=a/c}q_{i}\lambda_{i}\Lambda_i^{-3}e^{-\beta q_{i}\phi^\text{MF}(r)}=0
\end{align}
and within $(0<z<a)$ by
\begin{align}
  -\epsilon_{I}\nabla^{2}\phi^\text{MF}(r) = 0.
\end{align}
Far from the interface ($z\gg a$), where $\phi^\text{MF}$ is constant, the saddle-point equation is satisfied for
\begin{align}\label{DonnanPotential}
    \phi^\text{MF}(\infty)\equiv\phi_{0}=-\frac{1}{2\beta q}\log\left(\frac{\lambda_{a}\Lambda_{a}^{-3}}{\lambda_{c}\Lambda_{c}^{-3}}\right) \ ,
\end{align}
where a symmetric electrolyte with $q = q_{c} = -q_{a}$ was assumed. Inserting the redefined chemical potentials Eqs.~\eqref{MuBar} and \eqref{MuTilde} into Eq.~\eqref{DonnanPotential}, the electrolyte bulk value of $\phi^\text{MF}$ is determined by $\mu_{-}$ alone, $\phi_0  =\mu_{-}/q$. In other words, the electrolyte bulk value of $\phi^\text{MF}$ depends only on the difference between the ionic chemical potentials and thus directly corresponds to the electrode potential according to Eq.~\eqref{ElectrodePotential},
\begin{align}\label{ElectrostaticPotentialBulk}
    \phi_0= \frac{\mu_{-}}{q} \,=\, -\mathcal{E}.
\end{align}
This relation resolves the ambivalence discussed in Sec.~\ref{SubSec:ElectrodePotential} and shows that even in presence of image charges, the common definition of the electrode potential in terms of the electrostatic potential difference between solution bulk and metal surface still holds \cite{schmicklerInterfacialElectrochemistry2010}. Knowing that the saddle-point field $\phi^\text{MF}$ converges to a constant value $\phi_0$ in the electrolyte bulk, we insert $\phi^\text{MF}(r)=\phi_{0}+\Delta\phi^\text{MF}(r)$ into Eq.~\eqref{EqnSaddlePointSolution} and find the approximate solution for the MF electrostatic potential for a weakly charged surface (with details shown in Appendix~\ref{DerivationMFElectrostaticPotential}),
\begin{widetext}
\begin{align}\label{ElectrostaticPotentialMF}
    \phi^\text{MF}(r)& =-\mathcal{E}+\frac{4}{\beta q}\text{arctanh}\left(\tanh\left(\frac{\lambda_{D}}{(\epsilon_{B}/\epsilon_{I})a+\lambda_{D}}\frac{\beta q\mathcal{E}}{4}\right)e^{-\lambda_{D}^{-1}(z-a)}\right),
\end{align}
\end{widetext}
which coincides with the GCS solution. The Debye length
\begin{align}
    \lambda_D=\sqrt{\frac{\epsilon_B}{2\beta q^{2}n_{\text{ion}}^{\text{b}}}}
\end{align}
is determined by the bulk ion density $n_{\text{ion}}^{\text{b}}$ discussed in the following. 

\paragraph{Electroneutrality in the solution bulk.} 
Ion densities (Eq.\eqref{DensityExpansion}) at the MF level are simply given by the Boltzmann-type relation
\begin{align}\label{DensityMF}
    \avg{n_i(r)}^\text{MF}=\lambda_{i}\Lambda_i^{-3}\Theta(z-a)e^{-\beta q_{i}\phi^\text{MF}(r)}.
\end{align}
Inserting $\phi^\text{MF}(r)=\phi_{0}+\Delta\phi^\text{MF}(r)$ yields the same relationship with the zero potential reference in the bulk electrolyte,
\begin{align}
    \avg{n_i(r)}^\text{MF}=\Theta(z-a)n_\text{ion}^\text{b}e^{-\beta q_{i}\Delta\phi(r)} \ ,
\end{align} 
with an identical bulk ion concentration for cations and anions,
\begin{align}
    \avg{n_i(\infty)}\equiv n_\text{ion}^\text{b} =\lambda_{i}\Lambda_i^{-3}e^{-\beta q_{i}\phi_0} = \Lambda_i^{-3}e^{\beta\mu_{+}},
\end{align}
where
\begin{align}\label{MFEffectiveChemicalPotential}
    \mu_i-q_{i}\phi_0 = \frac{\mu_a+\mu_c}{2} = \mu_{+}
\end{align}
was used in the last step.
The equal bulk concentrations of anions and cations mean that electroneutrality in the solution bulk is automatically satisfied. Thus, regardless of the values of the individual ion chemical potentials $\mu_a$ and $\mu_c$ as defined by respective external reservoirs, the inner electrostatic potential in the electrolyte bulk will always adjust to a value such that electroneutrality is satisfied. This confirms the consistency of the present grand-canonical approach where both the cation and anion chemical potentials are free and independent parameters. 

\paragraph{Capacitance.} The metallic surface charge density (Eq.~\eqref{MetalChargeDensity}) is given by 
\begin{align}
    \sigma_{M}^\text{MF}&=-\epsilon_{B}\frac{\partial\phi^\text{MF}}{\partial z}\at_{z=a} \nonumber\\
    &=\frac{2}{\beta q}\frac{\epsilon_{B}}{\lambda_{D}}\sinh\left(\frac{\lambda_{D}}{a(\epsilon_{B}/\epsilon_{I})+\lambda_{D}}\frac{\beta q\mathcal{E}}{2}\right)
\end{align}
and the (inverse) differential capacitance is calculated according to Eq. ~\eqref{Defn:DifferentialCapacitance},
\begin{align}\label{MFCapacitance}
    C_{\text{MF}}^{-1}&=\left(\frac{\partial\sigma_M^\text{MF}}{\partial \mathcal{E}}\right)^{-1} \nonumber \\
    &=\frac{\lambda_{D}}{\epsilon_{B}}\frac{1}{\cosh\left(\frac{q\beta}{2}\left(\mathcal{E}-\frac{a}{\epsilon_{I}}\sigma_M^\text{MF}(\mathcal{E})\right)\right)}+\frac{a}{\epsilon_{I}} \nonumber \\
    &=C_{\text{GC}}^{-1}+C_{\text{H}}^{-1}=C_{\text{GCS}}^{-1}.
\end{align}
This expression is precisely equal to the inverse Gouy-Chapman-Stern (GCS) capacitance, which is the sum of inverse Gouy-Chapman and Helmholtz capacitances, the latter given by $a/\epsilon_{I}$ in the present model. This result proves that the image-charge theory reduces to GCS theory when restricted to the MF level. At this level, the instantaneous inhomogeneous distribution of image charges, and thus (electronic) surface charge, gets completely smeared out, resulting in a constant in-plane averaged surface charge density as assumed in GCS theory. In particular, this shows that the image charges consistently represent the entire charge of the electrode, and no additional surface charge must be applied in this model.

\subsection{First-order perturbation theory}\label{Sec:DHTheory}\label{SubSec:Theory1}
It is generally difficult to make systematic improvements over classical GCS theory, and extensions are thus usually phenomenological in nature \cite{bikermanStructureCapacityElectrical1942,bazantDoubleLayerIonic2011}. After showing that the MF level of our theoretical model corresponds to GCS theory, we are now in a position to improve the theory in a systematic way by going beyond the saddle-point approximation. We thereby focus on the capacitive response around the potential of zero charge (PZC), which allows for analytical calculations of perturbative corrections to the MF result. Accordingly, this section presents the development of a first-order extended GCS theory that incorporates coulombic correlation effects, specifically ion--ion and ion--image-charge correlations, with the latter representing interfacial ion--electron correlations. The first-order perturbative expressions are derived for the electrostatic potential, the ion densities, and the differential capacitance by keeping terms up to order $\ell^{-1}$ of the respective loop expansion in powers of $\ell$. The obtained first-order results will be denoted by ``1L'', short for ``one loop'', due to the fact that the 1L level diagrammatically contains one self-energy loop  \cite{netzPoissonBoltzmannFluctuationEffects2000}.

\paragraph{Correlation function.}
The correlation function, $G(r,r')$, is the primary object of any perturbative analysis of the partition function and all its derived quantities. It was already explained that at the MF level, particles interact only via the average (mean) charge distributions. The correlation function effectively captures all interactions due to instantaneous ``out-of-average'' density fluctuations that are ignored at the MF level. 

By inserting Eq.~\eqref{DiffeqGreensFunction} into Eq.~\eqref{ComputationOfCorrelationFunction} one finds that, near the PZC, the correlation function $G$ satisfies
\begin{align}\label{LinearizedFMinusEquation}
    \left(-\nabla_r\epsilon(r)\nabla_r+2\beta q^{2}n_{\text{ion}}^{\text{b}}\Theta(z-a)\right)G(r,r')=\delta(r-r') \ ,
\end{align}
as shown in greater detail in Appendix~\ref{Appendix:CorrelationFunction}. Utilizing the in-plane translational invariance, $G(r,r') = G(r-r')$, the solution reads
\begin{widetext}
\begin{align}\label{CorrelationFunction}
    G(r,r')= \frac{1}{4\pi\epsilon_B}\int_{0}^{\infty}dk\;J_{0}(k\,\rho)\;\frac{k}{\gamma(k)}\left(e^{-\gamma(k)\abs{z'-z}}+e^{-\gamma(k)(z+z'-2a)}\left(1-\frac{2k}{k+(\epsilon_{B}/\epsilon_{I})\gamma(k)\tanh(ka)}\right)\right),
\end{align}
\end{widetext}
where $J_0$ is the zeroth Bessel function of the first kind, and $\gamma(k)=\sqrt{k^2+\lambda_D^{-2}}$. Further details are given in Appendix~\ref{Appendix:SolutionCorrelationFunction}. 

\paragraph{Interface correlation energy.} The correlation function (Eq.~\eqref{CorrelationFunction}) evaluated at equal point (or coincident point), $r=r'$, is divergent. This divergence can be corrected by subtracting the (equally divergent) limiting expression for the equal-point correlation function in the bulk dielectric solvent (here water), which is given by 
\begin{align}
    G_0(r,r)=\frac{1}{4\pi\epsilon_B}\int_0^\infty dk.
\end{align}
The renormalized (corrected) self energy
\begin{align}\label{Defn:SelfEnergy}
    V(z)=\frac{q^2}{2}\left(G(r,r)-G_0(r,r)\right)
\end{align} 
(in case of Eq.~\eqref{CorrelationFunction} only a function of $z$) quantifies the correlation energy of a single point charge at $r$ compared to the self energy of a single ion in a bulk dielectric solvent \cite{netzDebyeHUckelTheory1999,lauFluctuationCorrelationEffects2008,hedleyWhatDoesIon2025}. \begin{figure*}[t]
	\centering
	\includegraphics[width=\textwidth]{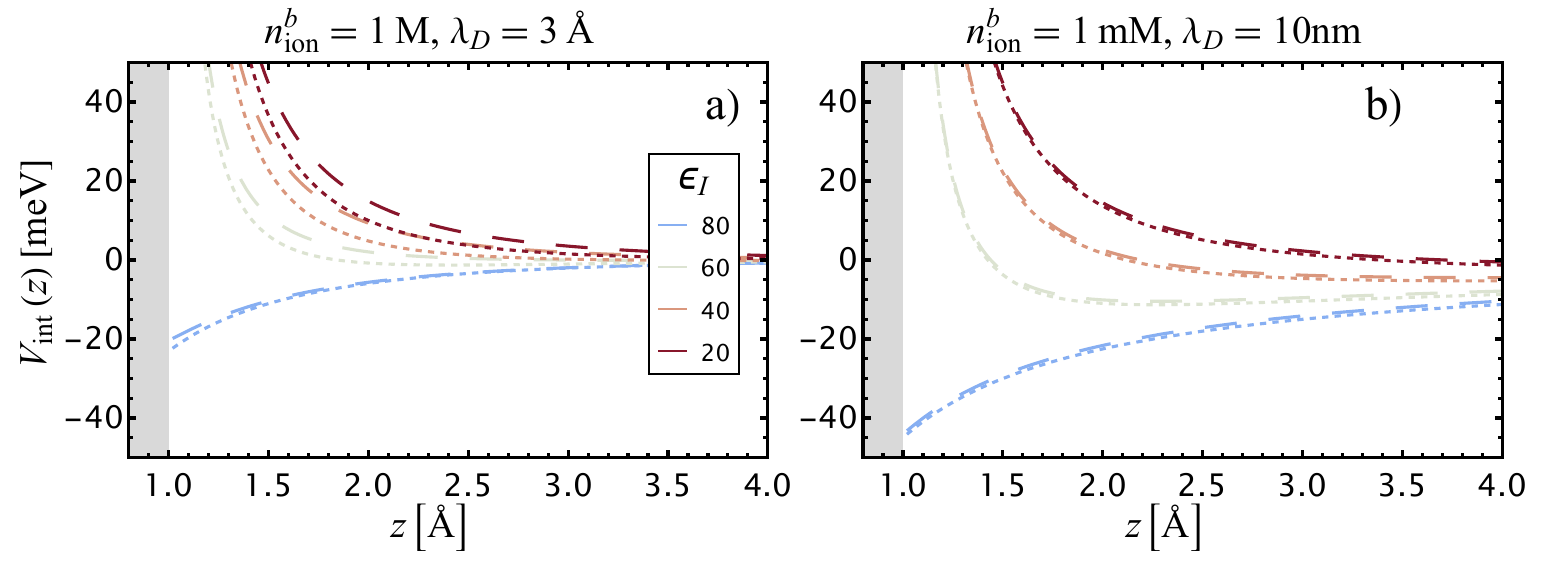}
	\caption{The numerical solution (long-dashed) of the interface correlation function in Eq.~\eqref{FullSelfEnergy} of an aqueous electrolyte solution with $\epsilon_B=80$ and a charge-separation distance of $a=1\,\text{\AA}$  for ionic concentrations 1M (a) and 1mM (b). For comparison, the analytical approximation  Eq.~\eqref{SelfEnergyApprox} is shown as short-dashed curves. Shown are results for $\epsilon_I=80$ (blue) down to $\epsilon_I=10$ (red).}
 \label{fig:SelfEnergyApprox}
\end{figure*} 
An ion energetically favors moving from the bulk to position $z$ if $V(z)<0$, and, vice versa, requires energy if $V(z)>0$. Inserting Eq.~\eqref{CorrelationFunction} into Eq.~\eqref{Defn:SelfEnergy},  $V(z)$ splits into two finite parts,
\begin{align}\label{SelfEnergy}
    V(z)=V_{\text{DH}}+V_{\text{int}}(z),
\end{align}
where $V_\text{DH}$ is the known Debye-Hückel self energy of a charged particle in a bulk electrolyte solution \cite{netzDebyeHUckelTheory1999,lauFluctuationCorrelationEffects2008},
\begin{align}\label{DHSelf-Energy}
    V_{\text{DH}}=-\frac{q^2}{8\pi\epsilon_{B}\lambda_{D}}.
\end{align}
The additional component,
\begin{align}\label{FullSelfEnergy}
    V_{\text{int}}(z)&=\frac{q^2}{8\pi\epsilon_B}\int_{0}^{\infty}dk\;\frac{k}{\gamma(k)}e^{-2\gamma(k)(z-a)/\lambda_D} \nonumber \\
    &\left(1-\frac{2k}{k+(\epsilon_{B}/\epsilon_{I})\gamma(k)\tanh(ka)}\right),
\end{align}
will be referred to as the \emph{interface correlation energy} in the following, that arises due to correlations between ions in the solution and the electrode surface, i.e. it describes electron-ion correlations. The integral determining the interface correlation function in Eq.~\eqref{FullSelfEnergy} is not amenable to an analytical solution. However, in the limit of infinite dilution ($\lambda_D\to\infty$), an analytical solution is given by
\begin{widetext}
\begin{align}\label{SelfEnergyApprox}
    V_{\text{int}}(z)= \frac{q^2}{8\pi\epsilon_{B}\lambda_D} \left(\frac{e^{-2(z-a)/\lambda_D}}{2(z-a)/\lambda_D}\left( 1- \frac{{}_2F_1(1,z/a-1,z/a;\frac{(\epsilon_B/\epsilon_I)-1}{(\epsilon_B/\epsilon_I)+1})}{((\epsilon_B/\epsilon_I)+1)/2}\right)-\frac{e^{-2z/\lambda_D}}{2z/\lambda_D}\frac{{}_2F_1(1,z/a,1+z/a;\frac{(\epsilon_B/\epsilon_I)-1}{(\epsilon_B/\epsilon_I)+1})}{((\epsilon_B/\epsilon_I)+1)/2}\right),
\end{align}
\end{widetext}
with ${}_2F_1$ being the hypergeometric function. 

Figure~\ref{fig:SelfEnergyApprox} compares the full numerical solution for the interface correlation function (Eq.\eqref{FullSelfEnergy}) with its analytical approximation (Eq.\eqref{SelfEnergyApprox}) for a charge-separation distance of $a=1\,\text{\AA}$ at various values for $\epsilon_I$ and two different concentrations of an aqueous electrolyte solution characterized by $\epsilon_B=80$. In concentrated solutions (a), i.e., for a short Debye length, the approximate analytical solution (short-dashed) of Eq.~\eqref{SelfEnergyApprox} agrees only at a qualitative level with the numerical solution (long-dashed), because the former was derived under the assumption of a large Debye length. Accordingly, in the regime of a large Debye length (b), i.e., in the dilute limit,  the analytical and numerical solutions align very well independent of the values of other parameters. Further in-depth analysis of $V_\text{int}$ is postponed to the results section (Sec.~\ref{SubSec:Summary1}). 

\paragraph{Electrostatic potential.}\label{SubSec:ElectrostaticPotential}
Expanding the exponential function in Eq.~\eqref{FullExpansionElectrostaticPotential} in powers of $\ell$ and keeping only the terms up to (including) $\mathcal{O}(\ell^{-1})$ and setting $\ell=1$ yields the electrostatic potential to 1L order,
\begin{align}\label{ElectrostaticPotentialFirstOrder}
    \phi^\text{1L}(r)=\phi^\text{MF}(r)-i\beta\avgG{\delta\psi(r)S_I^{(3)}},
\end{align}
where $\avgG{\delta\psi(r)}=0$ since odd moments in $\delta\psi(r)$ vanish for Gaussian distributions. The first term on the r.h.s. is the MF result (Eq.~\eqref{ElectrostaticPotentialMF}) and the second term is the first-order correction, which depends on $S_I^{(3)}$ (defined in Eq.~\eqref{FirstOrderIneracting}). As shown in the Appendix~\ref{SubSubSec:ThirdVariation}, inserting $S_I^{(3)}$ into Eq.~\eqref{ElectrostaticPotentialFirstOrder} gives the electrostatic potential as an integral expression
\begin{widetext}
    \begin{align}\label{ElectrostaticPotential1L}
    &\phi^\text{1L}(r)=\phi^\text{MF}(r)+\Delta\phi^\text{1L}(r) \nonumber \\
    &=\phi^\text{MF}(r)+\frac{\epsilon_B}{\beta q\lambda_{D}^{2}}
    \int_{r'>a}\sinh\left(4\,\text{arctanh}\left(\tanh\left(\frac{\lambda_{D}}{(\epsilon_{B}/\epsilon_{I})a+\lambda_{D}}\frac{\beta q \mathcal{E}}{4}\right)e^{-\lambda_{D}^{-1}\abs{z'-a}}\right)\right) 
   G(r,r')\left(\frac{\beta q^{2}}{2}G(r',r')-\omega\right).
\end{align}
\end{widetext}
An important fact here is that due to the factor $\sim\sinh$ the correlation contribution $\Delta\phi^\text{1L}$ vanishes in the electrolyte bulk $z\to\infty$. This means that in the electrolyte bulk there is \textit{no} correction of the electrostatic potential. This also means that the relation between electrode potential and the bulk electrostatic potential, Eq.~\eqref{ElectrostaticPotentialBulk}, remains the same.

Clearly, the 1L correction directly depends on the correlation function $G(r,r')$ but also on the equal-point correlation function $G(r,r)$, which is infinite. The gauge constant $\omega=\beta\mu_\text{g}$ in Eq.~\eqref{ElectrostaticPotential1L} can be used to renormalize the equal-point correlation function to a finite value. To choose an appropriate gauge chemical potential it is necessary to consider the ion densities.

\paragraph*{Ion densities.}\label{SubSec:ElectrolyteDensities}
Truncating the expansion of the ion densities, Eq.~\eqref{DensityExpansion}, beyond $\mathcal{O}(\ell^{-1})$ yields,
\begin{align}
    \avg{n_i(r)}^{\text{1L}} = \lambda_{i}\Lambda_{i}^{3}e^{-q_{i}\beta \phi^\text{1L}(r)}\exp\left(-\left(\frac{\beta q_{i}^{2}}{2}G(r,r)-\omega\right)\right),
\end{align}
which depends on the electrostatic potential in Eq.~\eqref{ElectrostaticPotential1L} and $G(r,r)$. As discussed before, in the electrolyte bulk, the 1L electrostatic potential correction vanishes ($\phi^\text{1L}=\phi^\text{MF}$). Also, far from the interface the influence of the electrode vanishes, $V_\text{int}=0$ (see Figure~\ref{fig:SelfEnergyApprox}). Hence, according to Eq.~\eqref{SelfEnergy}, in the solution bulk the equal-point correlation function reduces to $\frac{q_i^2}{2} G(r,r)=V_\text{DH}+\frac{q_i^2}{2}G_0$. The ion density in the solution bulk is then
\begin{align}
    \avg{n_i(\infty)}^{\text{1L}} = e^{\beta \mu_i^\text{eff}} \Lambda_{i}^{-3},
\end{align}
with an ``effective'' chemical potential, 
\begin{align}
    \mu_i^\text{eff} = \mu_i - q_{i}\phi_0-\beta^{-1}\log \gamma_i = \mu_{+} -\beta^{-1}\log \gamma_i,
\end{align}
that is shifted compared to the MF result in Eq.~\eqref{MFEffectiveChemicalPotential} by the activity coefficient
\begin{align}\label{BulkActvityCoefficient}
    \log\gamma_i = -\frac{\beta q_{i}^{2}}{8\pi\epsilon_{B}\lambda_{D}}+\frac{\beta q_i^2}{2} G_0(r,r) -\beta \mu_\text{g}.
\end{align}
The gauge chemical potential, $\mu_\text{g}$, thus enables tuning the bulk ion density in the 1L theory to the same value as for the MF theory, which is required for a consistent comparison of the interfacial capacitance results between the MF and 1L level. Selecting
\begin{align}\label{ChosenReferenceState}
    \omega=\beta\mu_\text{g}=-\frac{\beta q_{i}^{2}}{8\pi\epsilon_{B}\lambda_{D}}+\frac{\beta q_i^2}{2} G_0(r,r)\ ,
\end{align}
the reference activity coefficients in the electrolyte bulk become $\gamma_i=1$ and the bulk ion densities at the MF and 1L levels of theory coincide \footnote{If the gauge is chosen to be $\mu_\text{g}=q_i^2V_0$, then the the 1L expansion has an activity coefficient of DH form $\log\gamma_i = -\frac{\beta q_{i}^{2}}{8\pi\epsilon_{B}\lambda_{D}}$. This shows that the 1L level that is being discussed here is, in fact, equivalent to the DH theory \cite{Debye_1923}. In the case of symmetric electrolytes, both anion and cation densities are equivalently modified, so electroneutrality is still satisfied in the bulk. However, this means that the 1L ion densities in the electrolyte bulk are different from the MF densities.},
\begin{align}\label{Density1L}
    \avg{n_i(\infty)}^\text{MF}=\avg{n_i(\infty)}^\text{1L}.
\end{align}
With the chosen gauge chemical potential (Eq.~\eqref{ChosenReferenceState}), the ion densities become
\begin{align}\label{IonDensity1L}
    \avg{n_i(r)}^\text{1L} = n_\text{ion}^\text{b}e^{-  \beta q_i (\Delta\phi^{\text{MF}}(r)+\Delta\phi^{\text{1L}}(r))} e^{-\beta V_{\text{int}}(r)}.
\end{align}
Clearly, interfacial Coulomb correlations affect the ionic densities through both changes in electrostatic potential and the additional interface correlation energy. Note that the latter can also be interpreted as a spatially dependent activity coefficient
\begin{align}\label{EDLActivity}
\log\gamma_i(z)=\beta V_{\text{int}}(z).
\end{align}

\paragraph*{Capacitance.}\label{SubSec:TheInterfaceCapacitance}
Calculating the electrode surface charge density $\sigma_M$, from Eq.~\eqref{MetalChargeDensity} yields
\begin{align}\label{Theory1LChargeDensity}
    \sigma^\text{1L}_{M}&=-\epsilon_{B}\frac{\partial\phi^\text{1L}(r)}{\partial z}\at_{z=a} \nonumber \\
    &=-\epsilon_{B}\left(\frac{\partial\Delta\phi^{\text{MF}}(r)}{\partial z}\at_{z=a}+\frac{\partial\Delta\phi^{\text{1L}}(r)}{\partial z}\at_{z=a}\right),
\end{align}
with a contribution to the charge density from the MF solution $\Delta\phi^\text{MF}$ and one from the 1L correction $\Delta\phi^\text{1L}$. As mentioned before, this article is entirely focused on the conditions around the PZC. Therfore, the $\cosh$ and $\sinh$ terms of the MF (Eq.~\eqref{ElectrostaticPotentialMF}) and 1L (Eq.~\eqref{ElectrostaticPotential1L}) contributions are linearized to
\begin{align}
    \Delta\phi^\text{MF}(z) &\,\approx\, \mathcal{E}\,\frac{\lambda_{D}}{(\epsilon_{B}/\epsilon_{I})a+\lambda_{D}}e^{-(z-a)/\lambda_D}\ , \label{TheoryMFPhiPZC}\\[0.2cm]
    \Delta\phi^\text{1L}(z)&\,\approx\,  \frac{1}{2}\,\frac{ \beta \mathcal{E}}{\lambda_{D}}\,\frac{\lambda_{D}}{(\epsilon_{B}/\epsilon_{I})a+\lambda_{D}}\int_{z'>a} e^{-(z'-a)/\lambda_{D}}
   \nonumber \label{Theory1LPhiPZC}\\
   &\left( e^{-\vert z-z'\vert/\lambda_D}-e^{-(z+z')/\lambda_D}\right) V_\text{int}(z')\ ,
\end{align}
yielding
\begin{align}
    -\epsilon_B \frac{\partial\Delta\phi^{\text{MF}}(r)}{\partial z}\at_{z=a}&\,\approx\, \frac{\epsilon_B}{(\epsilon_{B}/\epsilon_{I})a+\lambda_{D}}\mathcal{E}\ , \\[0.2cm]
    -\epsilon_B \frac{\partial\Delta\phi^{\text{1L}}(r)}{\partial z}\at_{z=a}&
    \,\approx\,\frac{-\epsilon_B}{(\epsilon_{B}/\epsilon_{I})a+\lambda_{D}}\mathcal{E}\nonumber \\
    &\frac{\beta}{\lambda_D}\int_{z'>a} e^{-2z'/\lambda_{D}}
V_{\text{int}}(z').
\end{align}
Thus, around the PZC, the metallic surface charge of Eq.~\eqref{Theory1LChargeDensity} is given by
\begin{align}\label{1LChargeDensity}
    \sigma^\text{1L}_{M} = \frac{\epsilon_{B}}{(\epsilon_{B}/\epsilon_{I})a+\lambda_{D}}\,\mathcal{E}\,\left(1-\frac{\beta}{\lambda_D}\int_{z'>a} e^{-2z'/\lambda_{D}}
V_{\text{int}}(z')\right).
\end{align}
Inserting Eq.~\eqref{1LChargeDensity} into the definition of the differential capacitance, Eq.~\eqref{Defn:DifferentialCapacitance}, the result for the capacitance at 1L level can be expressed in terms of the MF GCS capacitance scaled by an additional 1L factor,
\begin{align}\label{Capacitance1Loop}
    C_\text{1L} = R_\text{1L}\times C_{\text{GCS}} \ .
\end{align}
This shows that interface correlations modify the capacitance mathematically akin to a geometrical electrode surface roughness, as often invoked to explain discrepancies between experimental capacitance data and GCS predictions at the PZC \cite{forestiModelEffectRoughness1993, parsonsInterphaseMercuryAqueous1965, daikhinDoublelayerCapacitanceRough1996}. Here, this factor has an entirely different physical origin in interfacial electron-ion correlations, and we therefore denote it as the \emph{interface correlation factor},
\begin{align}\label{Roughness1Loop}
    R_\text{1L}= 1-\frac{\beta}{\lambda_D}\int_{a}^\infty dz \;e^{-2z/\lambda_{D}}
V_{\text{int}}(z) \ .
\end{align}

\subsection{Clover approximation}\label{Sec:InfiniteOrderTheory} 
\begin{figure*}[t]
	\centering
	\includegraphics[width=2\columnwidth]{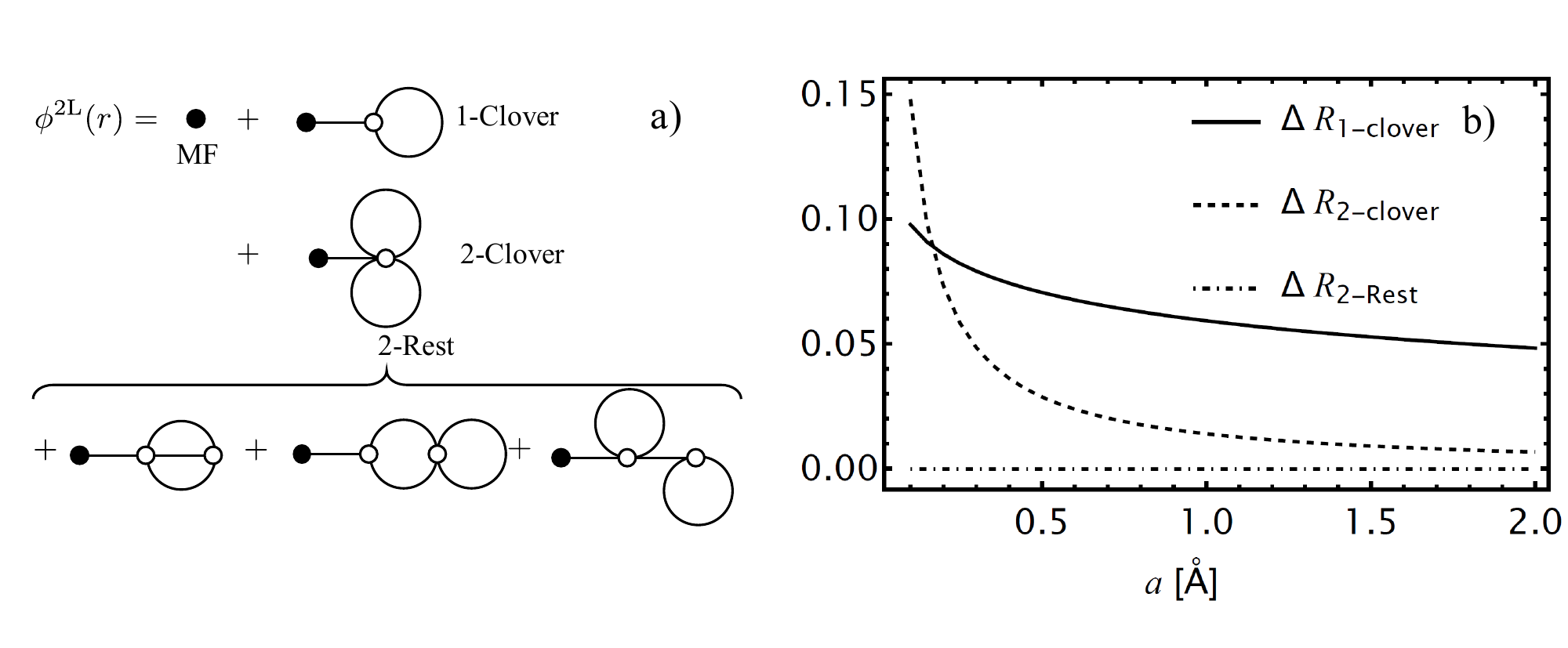}
	\caption{a) Diagrams contributing to the second-order pertubation theory of $\phi^{\text{2L}}$. (b) Contributions to the interface correlation factor $R_\text{2L}$ (Eq.~\eqref{RoughnessTwoLoopResult}), evaluated at the PZC, as a function of the charge-separation distance $a$, for a fixed residual permittivity of $\epsilon_I=20$ and ionic bulk concentration of $n^b_\text{ion}=100\,\text{mM}$.}
 \label{fig:2LCorrections}
\end{figure*}
In the previous section, the influence of electron-ion correlation effects on electrical double layer properties such as interfacial capacitance at PZC have been calculated based on first-order (1L) perturbation theory. In the first part of this section, the second-order (two-loop) perturbation theory (2L) of the interfacial capacitance at PZC is calculated and compared in its magnitude to the 1L result. It is demonstrated that the 1L expansion becomes inaccurate for $a<1\,\text{\AA}$, where the 2L order becomes signifcant in magnitude comapred to the 1L order. The 2L expansion of the capacitance consists of several different contributions of which one is clearly more important than the others. In fact, when represented as a Feynman diagram, cf. Appendix~\ref{Section:ExplanationDiagrams}, the 1L and the dominant 2L contributions share a common structure, resembling a clover leaf with one and two leaves, respectively. In the second part of this section all possible Feynman diagrams of clover structure are added up, which leads to a capacitance formula of infinite order in pertubation theory, which we refer to as the \emph{clover approximation}.

\paragraph{Second-order perturbation theory.}
To evaluate the accuracy of the first-order perturbation theory presented in Sec.~\ref{Sec:DHTheory}, in this section the interface correlation factor (cf. Eq.~\eqref{Roughness1Loop}) is derived to second order and compared with the first-order result. Expanding the electrostatic potential in Eq.~\eqref{FullExpansionElectrostaticPotential} to second order similarly to Eq.~\eqref{ElectrostaticPotentialFirstOrder} yields 
\begin{align}\label{ElectricPotential2L}
    \phi^\text{2L}(r)&=\phi^\text{1L}(r)-i\beta\ell^{-2}\Bigg(\avgG{\delta\psi(r)S_{I}^{(5)}}\nonumber \\
    &-\beta\avgG{\delta\psi S_{I}^{(3)}S_{I}^{(4)}} -\frac{\beta^{2}}{3!}\avgG{\delta\psi(r)\left(S_{I}^{(3)}\right)^{3}}\Bigg),
\end{align}
where the first term on the r.h.s. is the 1L result (Eq.~\eqref{ElectrostaticPotential1L}) and the remaining three Gaussian averages are the 2L correction to the electrostatic potential. In Appendix~\ref{Sec:SecondOrder}, details on the calculation of the Gaussian averages are presented, yielding a lengthy expression for the electrostatic potential to second order shown in Eq.~\eqref{ElectricPotential2LResult}. 

In higher-order perturbation theory, many terms share common structures. A more compact way to keep track of all possible terms is to use diagrammatic methods to efficiently represent the perturbative expansion. For our theory, diagrammatic rules are presented in Appendix~\ref{Section:ExplanationDiagrams}. Figure~\ref{fig:2LCorrections}a diagrammatically depicts the second-order expansion of the electrostatic potential, including all five contributing diagrams from Eq.~\eqref{DiagramAsIntegrals}. Clover-1 represents the first-order contribution of $\phi^\text{1L}$, while Clover-2 and the remaining three diagrams account for the second-order contributions.

Using the solution for the electrostatic potential (Eq.~\eqref{ElectricPotential2LResult}) and the definition of the capacitance (Eq.~\eqref{Defn:DifferentialCapacitance}), it is shown in Appendix \ref{Sec:SecondOrder} that to second order, the capacitance can be written as
\begin{align}\label{2LCapacitance}
    C_\text{2L}=R_\text{2L}\times C_\text{GCS},
\end{align}
where 
\begin{align}\label{RoughnessTwoLoopResult}
    R_\text{2L} = 1 +\Delta R_{\text{1-clover}}+\Delta R_{\text{2-clover}}+ \Delta R_{\text{2-rest}}
\end{align}
is the second-order interface correlation factor consisting of three contributions $\Delta R_{\text{1-clover}},\;\Delta R_{\text{2-clover}}\;\text{and}\;\Delta R_{\text{2-rest}}$ that are directly linked to the diagrammatic contributions shown in Fig.~\ref{fig:2LCorrections}a. Their mathematical form is given in Eq.~\eqref{Roughness2Loop}. In Fig.~\ref{fig:2LCorrections}b, the three contributions to the interface correlation factor from Eq.~\eqref{RoughnessTwoLoopResult}, evaluated at the PZC, are shown as functions of the charge-separation distance for a residual permittivity of $\epsilon_I=20$ and a bulk ion concentration of $n^b_\text{ion}=100\,\text{mM}$, for which the analytical approximate expression for the interface correlation energy, Eq.~\eqref{SelfEnergyApprox}, is accurate. The 1-clover contribution, $\Delta R_{\text{1-clover}}$, is larger than all second-order contributions for $a>0.2\,\text{\AA}$. Among the second-order contributions, the 2-clover diagram corresponding to $\Delta R_\text{2-clover}$ is significantly more important than all the remaining ones combined, $\Delta R_{\text{2-rest}}$. Physically, this may be explained by the fact that in the dilute limit considered here, interactions involving multiple ions become increasingly unlikely due to large interionic distances. In the diagrammatic representation (see Appendix~\ref{Section:ExplanationDiagrams}), such interactions are indicated by straight lines. In Fig.~\ref{fig:2LCorrections}a, the 2-clover graph has only one pairwise interaction, while the graphs contributing to 2-rest have four, three, and two, from left to right. By lowering $a$, image-charge correlation effects become stronger, which can be seen by the increase in $\Delta R_\text{1-clover}$ and $\Delta R_\text{2-clover}$. Below $a\approx1\,\text{\AA}$, the first-order expansion becomes inaccurate as the $\Delta R_\text{2-clover}$ contribution becomes a significant magnitude relative to $\Delta R_\text{1-clover}$. 

Of the four second-order diagrams in Fig.~\ref{fig:2LCorrections}a, the 2-clover diagram clearly dominates. We therefore assume that the same holds at higher expansion orders and that systematic improvements in accuracy can be achieved by accounting for clover-type diagrams of any order.

\paragraph{Infinite-order clover expansion.}
\begin{figure}[t]
	\centering
	\includegraphics[width=\columnwidth]{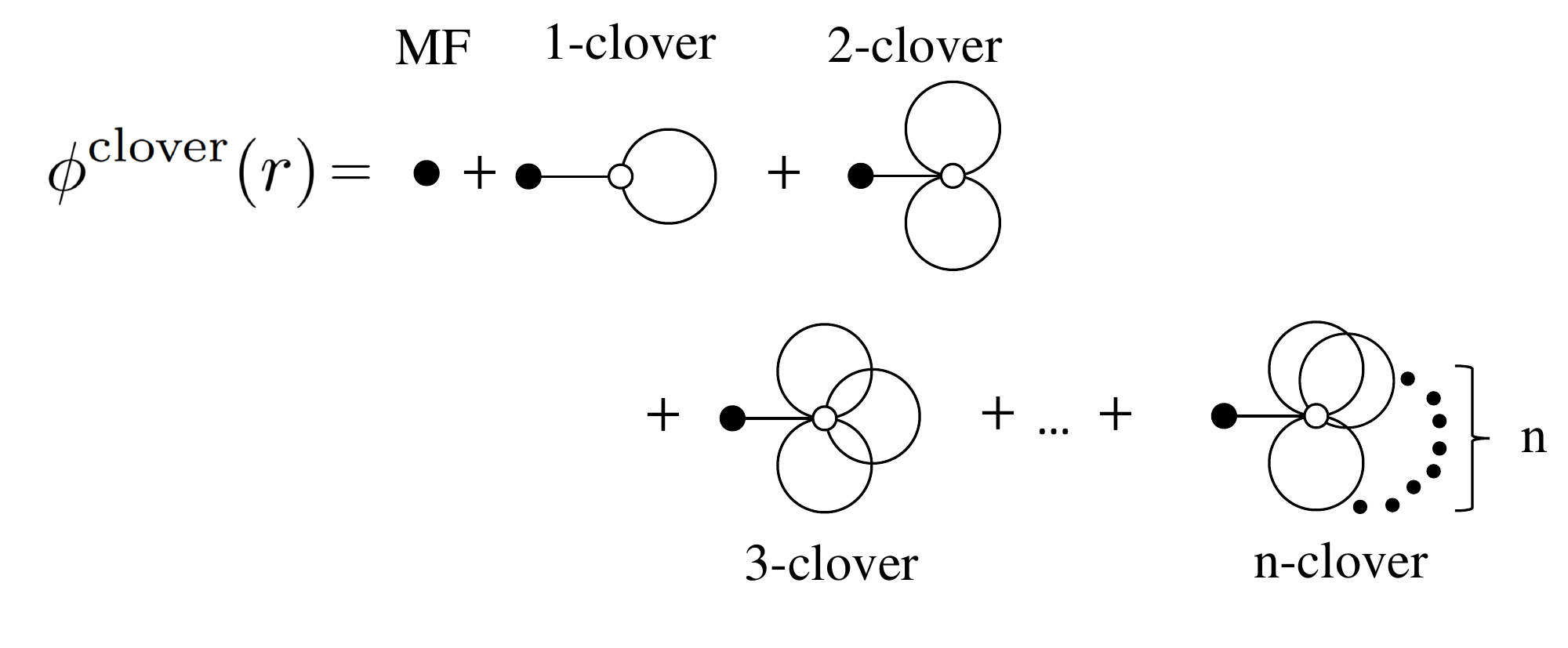}
	\caption{Diagrammatic representation of the electrostatic potential in clover approximation, cf. Eq.~\eqref{ElectrostaticPotentialSeries}.}
 \label{fig:InfniteOrderExpansion}
\end{figure}
As demonstrated in the preceding section, diagrams depicting the clover structure, characterized by a single black internal line with one external leg on one side and self-energy loops on the other, have been identified as being of paramount significance. In this section we derive a more accurate formula for the interface correlation factor accounting for all possible clover diagrams. This allows us to significantly improve the applicability of the theory and compare our model results with experimental data in Sec.~\ref{SubSec:Summary1}.

Diagrams of clover structures have in common that they only contain one renormalized vertex of odd order (i.e., $S_I^{(3)}$ or $S_I^{(5)}$), as seen in the first-order and second-order clover diagrams in Fig.~\ref{fig:2LCorrections}a. The expansion of the electrostatic potential (Eq. ~\eqref{PerturbativeElectrostaticPotential}) in clover diagrams is shown in Fig.~\ref{fig:InfniteOrderExpansion}. Mathematically, the clover expansion of the electrostatic potential is given by
\begin{align}
    \phi^\text{clover}(r)=\phi^\text{MF}(r)-i\sum_{n=1}^{\infty}\beta\ell^{-n}\avgG{\delta\psi(r)S_I^{(2n+1)}}.\label{ElectrostaticPotentialSeries}
\end{align}
All averages of $\delta\psi$ with $S_I^{(2n+1)}$ are readily computed since they follow a repetitive pattern, as shown in Appendix ~\ref{Appendix:Clover}. Upon inserting Eq.~\eqref{nThOrderWick} into Eq.~\eqref{ElectrostaticPotentialSeries} we find
\begin{widetext}
\begin{align}\label{ElectrostaticPotentialClover}
    \phi^\text{clover}(r) = \phi^\text{MF}(r) - \frac{\epsilon_B}{\beta q \lambda_{D}^{2}}\int_{r'>a} 
    \sinh\left(4\,\text{arctanh}\left(\tanh\left(\frac{\lambda_{D}}{(\epsilon_{B}/\epsilon_{I})a+\lambda_{D}}\frac{\beta q\mathcal{E}}{4}\right)e^{-\lambda_{D}^{-1}\abs{z'-a}}\right)\right)G(r,r')\left(e^{-\beta V_\text{int}(z')}-1\right).
\end{align}
\end{widetext}

\begin{figure*}[t]
	\centering
	\includegraphics[width=\textwidth]{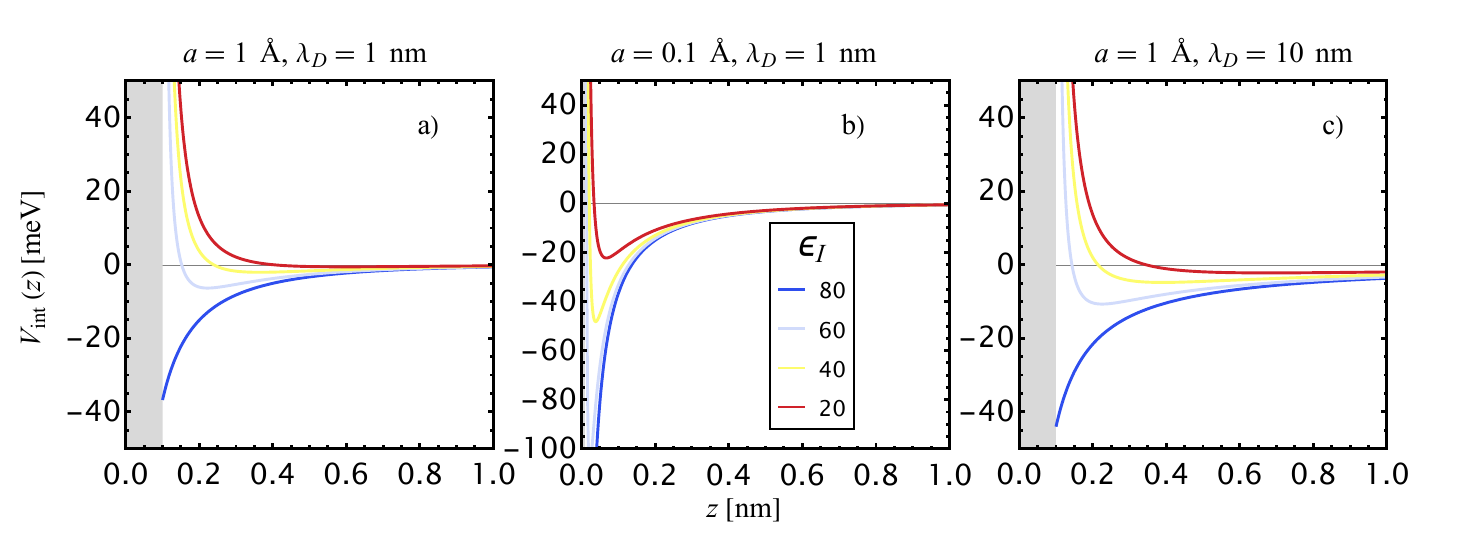}
	\caption{Interface correlation energy $V_{\text{int}}$ for an aqueous solution ($\epsilon_B=80$) as function of distance to the electrode surface for (a) $a=1\,\text{\AA}$, $\lambda_D=1\,\text{nm}$,  (b) $a=0.1\,\text{\AA}$, $\lambda_D=1\,\text{nm}$, and (c) $a=1\,\text{\AA}$, $\lambda_D=10\,\text{nm}$ and several values of $\epsilon_I$. The region closer than the charge-seperation distance to the electrode surface is shown in light gray. It defines the minimum distance between the ionic charge and electronic counter-charge.}
 \label{fig:ActivityCoefficient}
\end{figure*}

Following Eqs.~\eqref{MetalChargeDensity} and \eqref{Defn:DifferentialCapacitance}, the capacitance in clover approximation is then given by
\begin{align}\label{CapacitanceClover}
    C_\text{clover} = R_\text{clover}\times C_\text{GCS},
\end{align}
with the interface correlation factor in clover approximation,
\begin{align}\label{RoughnessInfinity}
    R_\text{clover}=1+\frac{1}{\lambda_{D}}\int_{a}^{\infty}dz\,e^{-2z/\lambda_{D}}\left(\exp\left(-\beta V_{\text{int}}(z)\right)-1\right).
\end{align}
\section{Results}\label{SubSec:Summary1}
In the following, the image-charge theory developed in the previous section is used to examine how electron-ion correlation effects affect the local properties of the EDL. It is thereby found to provide a novel and consistent explanation for the previously established discrepancies between GCS theory and experimental differential capacitance data.

\subsection{Electron-ion correlations as attractive interaction}\label{ResultsCorrelations}
It is well established that image charges cause an attractive interaction between ions of an electrolyte solution and metallic surfaces \cite{kornyshevImagePotentialDielectric1977,gabovichExcessNonspecificCoulomb2006,sonImagechargeEffectsIon2021,zhouImageChargeEffects2024,hedleyWhatDoesIon2025}. However, the image-charge interaction is often considered an additional surface polarization effect that acts on top of the electrostatic interaction caused by the net (excess) surface charge of the metal electrode~\cite{ntimMolecularDynamicsSimulations2023, hedleyWhatDoesIon2025,zhouImageChargeEffects2024}. In the present theory, besides the image charges, no additional excess surface charge was applied. In fact, it was found that the image charges comprise the net surface-charge electrostatics, precisely reproducing GCS theory at the mean-field level (cf. Sec.~\ref{Sec:MFTheory}). Image-charge interactions that go beyond MF electrostatics are described by the interface correlation energy $V_\text{int}(z)$ (Eq.~\eqref{SelfEnergyApprox}), which corresponds to the change in energy of an ion at a distance $z$ from the metal surface caused by image-charge \emph{correlations}, i.e., interfacial electron--ion correlations. The interface correlation energy is shown in Figure~\ref{fig:ActivityCoefficient}a as a function of distance from the electrode surface. For $\epsilon_I = \epsilon_B $ (blue), $V_\text{int}(z)$ is strictly negative, indicating that ions gain energy by approaching the electrode due to attraction to their image charges. When $\epsilon_I$ is decreased, $V_\text{int}(z)$ turns to positive values near the electolyte solution edge as ions must expend energy to approach a lower-permittivity region, which effectively corresponds to a (partial) desolvation of the ion. This demonstrates that $V_{\text{int}}$ simultaneously accounts for the competing effects of electron--ion correlations and desolvation. 

By decreasing the charge-separation distance $a$ (Figure~\ref{fig:ActivityCoefficient}b), the influence of electron--ion correlations becomes significantly enhanced as evident by the negative shift of $V_{\text{int}}$. Similarly, by decreasing the bulk  electrolyte concentration from $100\,\text{mM}$ (Fig.~\ref{fig:ActivityCoefficient}a) to $1\,\text{mM}$ (Fig.~\ref{fig:ActivityCoefficient}c), the interfacial correlation energy becomes more negative, indicating a stronger attraction in more dilute solutions, which is due to the fact that the interaction between electronic and ionic charges is less screened. 

\begin{figure*}[t]
	\centering
	\includegraphics[width=\textwidth]{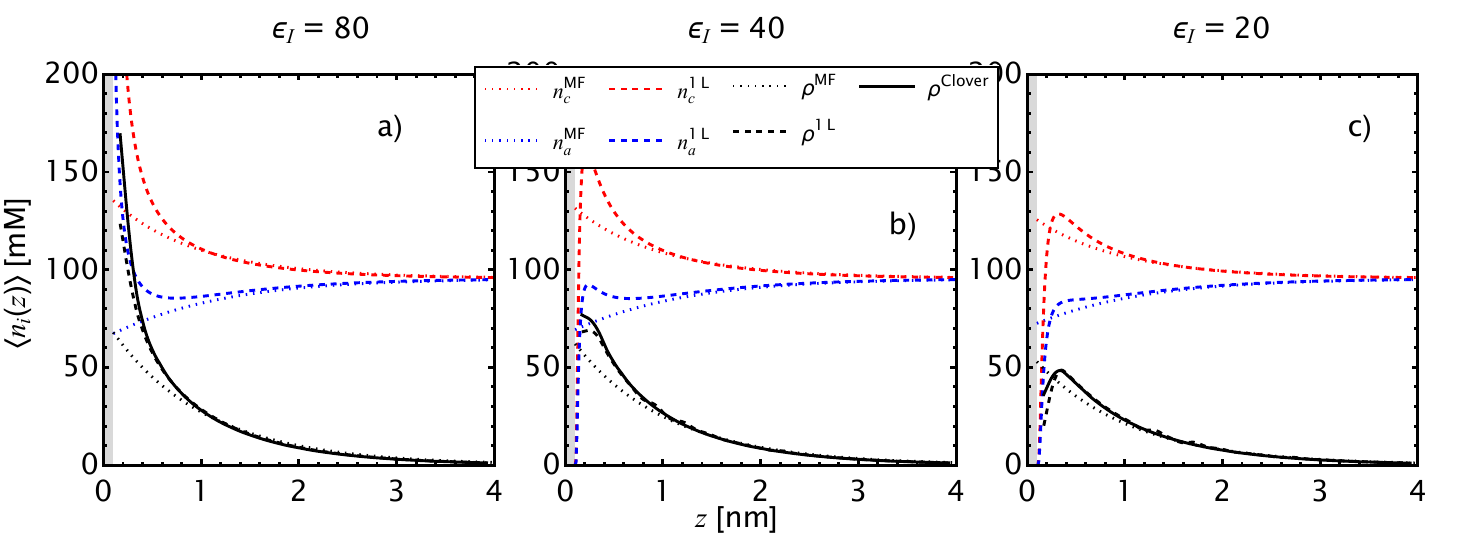}
	\caption{Ionic and total charge densities in MF, 1L, and clover approximation for a $100\,$mM electrolyte solution with an applied electrode potential of $\mathcal{E}=-10\,$mV. From left (a) to right (c), the residual permittivity $\epsilon_I$ is varied from a value of $80$ down to $20$, at a constant charge-separation distance of $a=0.1\,\text{nm}$.}
 \label{fig:Densities}
\end{figure*}

In Fig.~\ref{fig:Densities}, we compare ion densities in the clover approximation with those from the MF and 1L models, using three different values for the residual permittivity, $\epsilon_I$, while maintaining a fixed electrode potential of $-10\,\text{mV}\;\text{vs.}\;\mathcal{E}_\text{pzc}$, for $a=1\,\text{\AA}$. For the clover approximation, the charge density,
\begin{align}
    \rho(r) = q \left( n_c(r) - n_a(r) \right),
\end{align}
was calculated from the second derivative of the electrostatic potential (Eq.~\eqref{PoissonEquation}). In the electrolyte bulk, both anion and cation densities are equal in all models, a consequence of electroneutrality. For the negative electrode potential, anion densities are lower than cation densities near the interface in the MF model, with 1L densities matching the MF solution at distances $>1\,\text{nm}$ from the metal surface, corresponding to about one Debye length at $100\,\text{mM}$ concentration. This is because the interfacial correlation energy (Fig.~\ref{fig:ActivityCoefficient}) becomes negligible at more than about one Debye length from the metal surface.

\begin{figure*}[t]
	\centering
	\includegraphics[width=\textwidth]{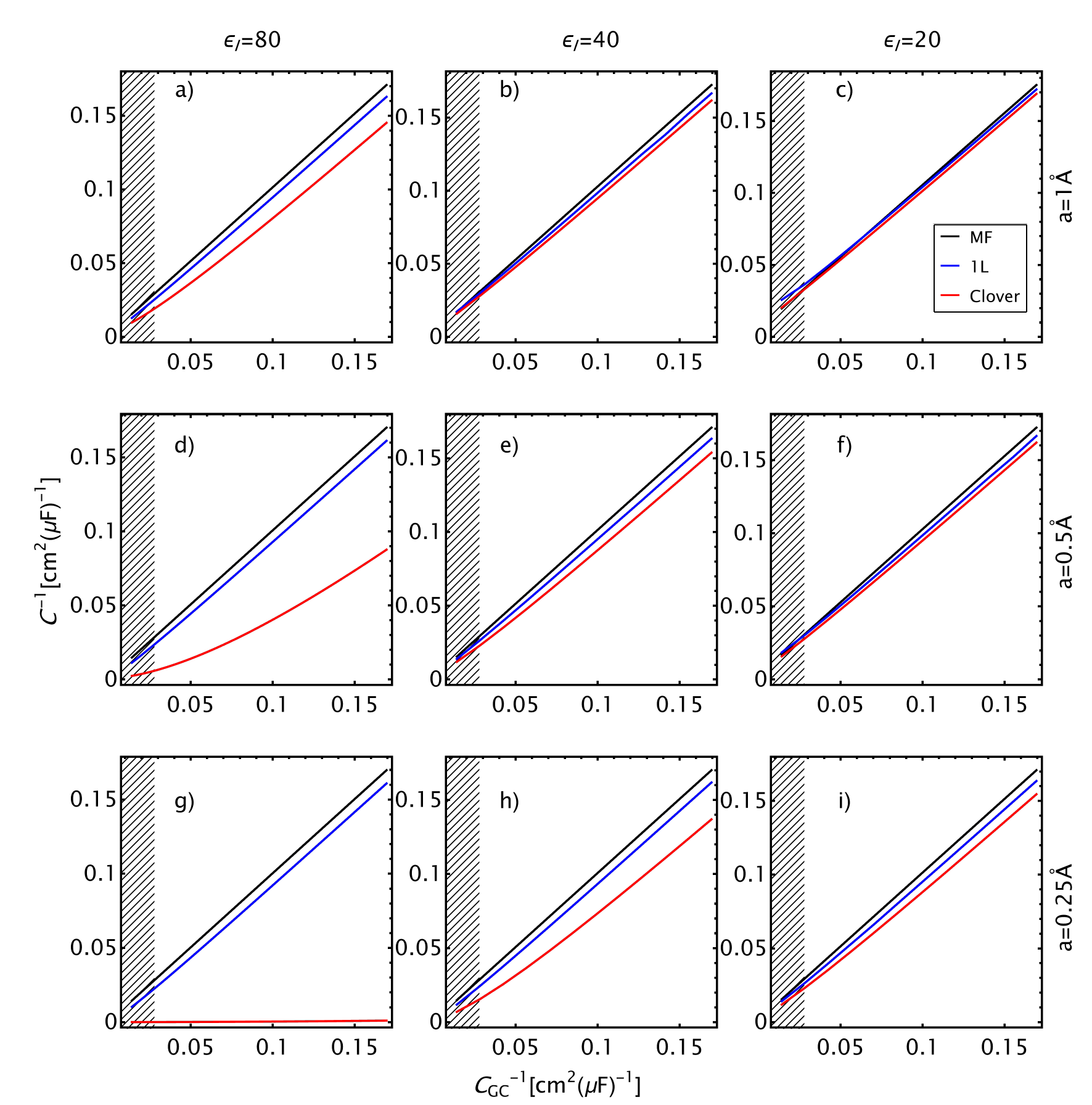}
	\caption{Parsons-Zobel plots of the differential capacitance in MF (GCS), 1L, and clover approximation for various combinations of the residual dielectric permittivity $\epsilon_I$ and charge-seperation distance $a$, while keeping $\epsilon_B=80$ fixed. The shaded area indicates the range of higher concentrations where the approximate analytical solution of the interface correlation energy $V_\text{int}$, Eq.~\eqref{SelfEnergyApprox}, employed in the 1L and clover models, becomes inaccurate.}
 \label{fig:CapacitanceModelComparison}
\end{figure*}

Strong deviations from the MF results are observed within a distance of one Debye length of the metal surface. For $\epsilon_I = \epsilon_B$ (Fig.~\ref{fig:Densities}a), the image-charge interaction strongly increases both ion densities near the surface, leading to a divergence of the densities as $a$ approaches zero. As the value of $\epsilon_I$ is decreased (Fig.~\ref{fig:Densities}b), the competition between the attractive image charge and the repulsive desolvation effects results in a distinct maximum in the charge density near the metal. Very close to $z = a$, the repulsive bound charge of the dielectric discontinuity at the solution edge causes the densities to decrease. Further decreasing $\epsilon_I$ increases the strength of this repulsive interaction and reduces the ion densities near $z=a$ compared to the MF solution, as shown in Fig.~\ref{fig:Densities}c.

\subsection{Capacitance at the PZC: Parsons-Zobel analysis}
In a Parsons-Zobel (PZ) plot, the inverse of the measured/simulated capacitance is plotted against the inverse of the GC model capacitance ($C_\text{GC}$), the latter being effectively varied by changing the (bulk) electrolyte concentration, i.e., Debye length. The MF-GCS capacitance formula (Eq.~\eqref{MFCapacitance}) predicts a linear relationship between $C^{-1}$ and $C_\text{GC}^{-1}$ with a slope of one. Figure~\ref{fig:CapacitanceModelComparison} shows PZ plots of the capacitance predicted by the present theory for six different combinations of the residual dielectric permittivity $\epsilon_I$ and charge-seperation distance $a$ for an aqueous solution ($\epsilon_B=80$). All curves in the same row correspond to the same charge-separation distance, while all PZ plots in the same column correspond to the same residual dielectric permittivity. The MF curve (black) corresponds to GCS theory. The shaded region indicates the range of higher concentrations where the approximate analytical solution of the interface correlation energy $V_\text{int}$, Eq.~\eqref{SelfEnergyApprox}, employed in the 1L and clover models, becomes inaccurate, see Fig.~\ref{fig:SelfEnergyApprox}a. 

It is apparent in Figure~\ref{fig:CapacitanceModelComparison} that the 1L and clover models, which include electron--ion correlation effects, exhibit PZ slopes of less than one across all investigated parameter combinations, corresponding to larger values of the interfacial capacitance than predicted by GCS theory. Thereby, the capacitance in clover approximation is larger than at the 1L level for the same set of parameters. Notably, significant deviations are observed between the predictions from the clover-level theory in comparison to (mean-field) GCS theory even for low electrolyte concentrations (right-hand part of the PZ plot), where GCS theory is typically assumed to be accurate. This indicates that electron--ion correlations can play a role in experimentally established discrepancies to GCS predictions around the potential of zero charge in the dilute limit.

The PZ slope of the capacitance curve is effectively tuned by the charge-separation distance. When $a$ is very small, electron-ion correlation effects are significantly affecting the interface correlation energy, cf. Fig.~\ref{fig:ActivityCoefficient}b. In panel (g) of Fig.~\ref{fig:CapacitanceModelComparison}, the slope of the PZ plot of the clover approximation is nearly zero, a behavior typically attributed to pseudocapacitive charging due to ion adsorption that increases the capacitance beyond the GCS prediction~\cite{schmicklerEffectWeakAdsorption2021,gonzalezReviewSupercapacitorsTechnologies2016}. Conversely, for larger $a$ (Fig.~\ref{fig:CapacitanceModelComparison}a--c), capacitance curves, which include electron-ion correlation effects, converge to GCS predictions. This shows that less-than-unity slopes in PZ plots, as routinely encountered in experimental data, could be a measure for the strength of interfacial correlations between ions and electrons.

Interestingly, PZ plots with similar $\epsilon_I/a$ ratios, such as those shown in panels (a), (e), and (i), or (b) and (f) of Fig.~\ref{fig:CapacitanceModelComparison}, display similar shapes. This suggests that $a$ and $\epsilon_I$ are not independent parameters for the interfacial capacitance characteristics, similar to the Helmholtz capacitance in GCS theory. However, while in the latter, $C_{\text{H}}$ only determines the y-offset of the PZ plots, the parameters $\epsilon_I/a$ in the image-charge theory affect both the y-intercept and the slope of the curves.

\subsection{Quantitative agreement with experimental data}
In the following, the theory of interfacial electron-ion correlations presented in this article is shown to provide a consistent explanation for the puzzling discrepancies between experimental capacitance data and GCS theory over a wide spectrum of electrode materials and electrolyte systems.

\begin{figure*}[t]
	\centering
	\includegraphics[width=\textwidth]{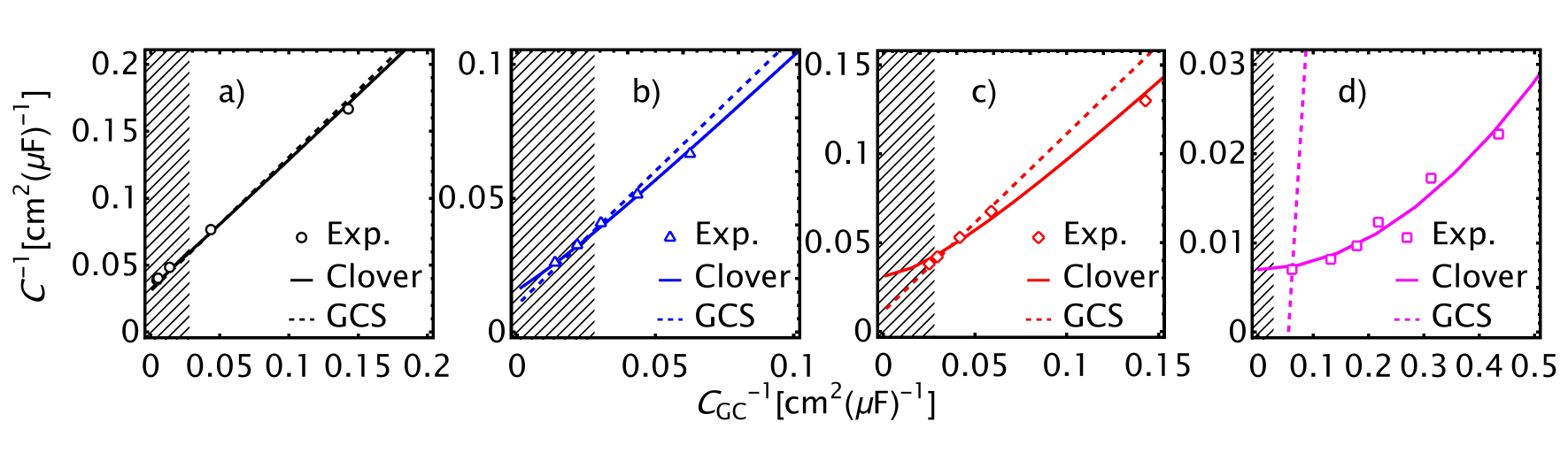}
	\caption{Experimental PZ plots (symbols) for (a) Mercury--NaF \cite{grahameDifferentialCapacityMercury1954}, (b) Ag(111)--KPF$_6$ \cite{valetteDoubleLayerSilver1989}, (c) Au(111)--KPF$_6$ \cite{hamelinStudyGoldLow1987} and (d) Pt(111)--HClO$_4$ \cite{ojhaDoubleLayerPt111Aqueous2020} interfaces, along with model results from GCS theory (dashed lines) and clover theory (solid lines), Eq.~\eqref{FitModel}, with respective model parameters given in Tab.~\ref{tab:Data}.}
 \label{fig:CapacitanceAqueous}
\end{figure*}

\begin{table}[t]
    \caption{Values of model parameters used to reproduce experimental capacitance data in Figs.~\ref{fig:CapacitanceAqueous} and \ref{fig:CapacitanceNonAqueous}. Literature references for the experimental data are indicated. The PZ slope refers to a linear fit of the experimental PZ plot. Fitted parameters for the capacitance model in Eq.~\eqref{FitModel} include the charge-separation distance $a$, interfacial permittivity $\epsilon_I$, and external capacitance $C_\text{ext}$. The value of the bulk permittivity $\epsilon_B$ corresponds to the known permittivity of the respective solvent.}  
	\begin{ruledtabular}
		\begin{tabular}{lc|cccc}\label{tab:Data}
			System & PZ slope & $a\,[\text{\AA}]$  &\makecell{$C_\text{ext}^{-1}$ \\ $[\text{cm}^2/\mu\text{F}]$} & $\epsilon_B$ &$\epsilon_I$\\
			\hline
            \multicolumn{6}{l}{Aqueous Electrolytes} \\
            Mercury-NaF \cite{grahameDifferentialCapacityMercury1954} & $0.97$ & $1.8$  & $0.021$ & $80$&$20$\\
            Ag(111)-KPF$_6$ \cite{valetteDoubleLayerSilver1989} & $0.85$ & $0.3$  & $0.013$ & $80$&$20$\\
            Au(111)-KPF$_6$ \cite{hamelinStudyGoldLow1987}& $0.79$ & $0.1$  & $0.03$ & $80$&$20$\\
            Pt(111)-HClO$_4$ \cite{ojhaDoubleLayerPt111Aqueous2020} & $0.04$ & $0.035$   & $0.007$ & $80$&$20$ \\[0.2cm]
            \multicolumn{6}{l}{Organic Electrolytes} \\
            Au(111)-KPF$_6$ DMSO \cite{shatlaDifferentialCapacitancePotential2021}& $0.57$ & $0.2$ & $0.04$ & $46.7$&$12$ \\
            Au(111)-KPF$_6$ ACN \cite{shatlaDifferentialCapacitancePotential2021}& $0.28$ & $0.25$ & $0.02$  & $37.4$&$10$ \\
            Au(111)-KPF$_6$ DG \cite{shatlaDifferentialCapacitancePotential2021}& $0.22$ & $1.4$ & $-0.08$  & $7.23$& $2$
		\end{tabular}
	\end{ruledtabular}
\end{table}
Using Eq.~\eqref{CapacitanceClover} for the differential capacitance at the potential of zero charge (PZC), we can directly test whether the theory is able to reproduce previously published experimental data in PZ plots. The fitting to the experimental capacitance data was performed manually with the residual permittivity ($\epsilon_I$) and charge-seperation distance ($a$) as the only adjustable parameters determining the PZ slope. An additional (concentration-independent) series capacitance $C_\text{ext}$ is introduced to allow for vertical adjustments of the PZ plots,
\begin{align}\label{FitModel}
    C^{-1} = C_\text{ext}^{-1} + R_\text{clover}^{-1} C_\text{GCS}^{-1} \ .
\end{align}
Unlike in GCS theory (Eq.~\eqref{MFCapacitance}), where the compact layer capacitance $C_\text{H}$ sets the y-intercept and can be tuned independent from the slope, our image-charge theory (Eq.~\eqref{CapacitanceClover}) has no concentration independent part, because $R_\text{clover}^{-1}$ affects both the compact layer and the diffuse layer contributions in $C_\text{GCS}$. However, since our primary interest lies in the explanation of anomalous PZ slopes, the ad hoc introduction of $C_\text{ext}$ has no bearings on the further analysis.  

\begin{figure*}[t]
	\centering
	\includegraphics[width=0.8\textwidth]{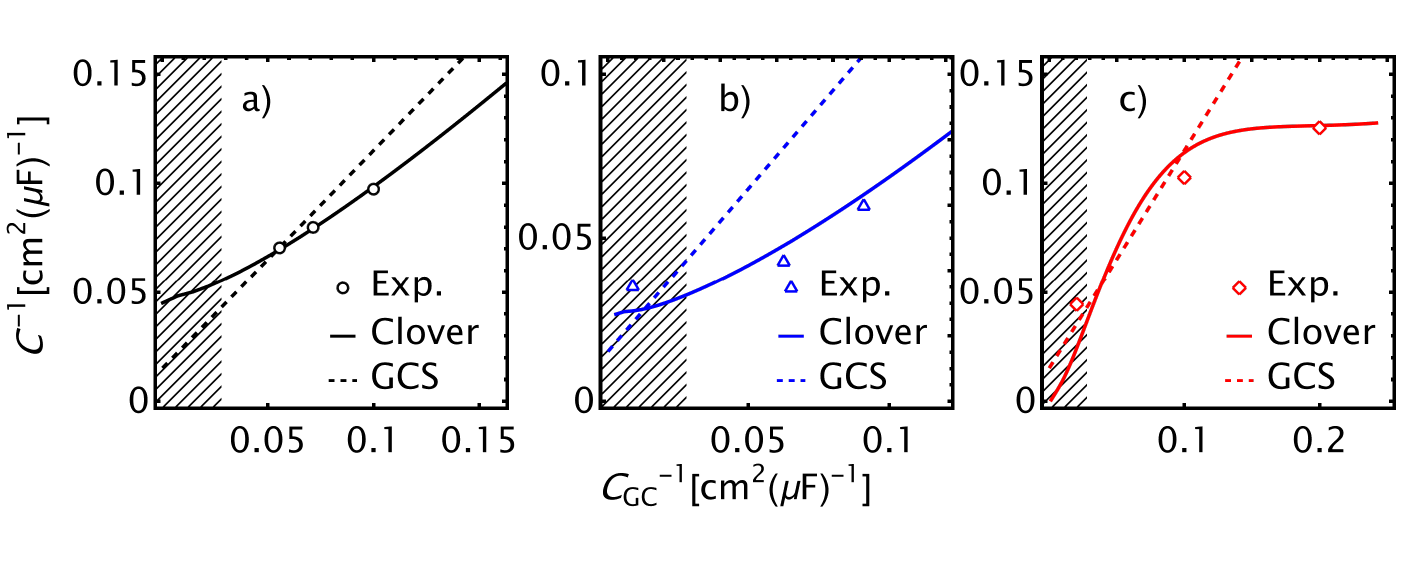}
	\caption{Experimental PZ plots \cite{shatlaDifferentialCapacitancePotential2021} (symbols) for Au(111) electrodes in contact with different KPF$_6$ electrolyte solutions with (a) DMSO, (b) ACN, and (c) DG as solvents, along with model results from GCS theory (dashed lines) and clover theory (solid lines), Eq.~\eqref{FitModel}, with respective model parameters given in Tab.~\ref{tab:Data}.}
 \label{fig:CapacitanceNonAqueous}
\end{figure*}

The fitted parameter values for various experimental data sets are summarized in Table~\ref{tab:Data}, with corresponding comparisons between theory and experiment shown in Figs.~\ref{fig:CapacitanceAqueous} and \ref{fig:CapacitanceNonAqueous}. Bulk dielectric permittivities in the theoretical models were fixed based on literature values for the respective electrolyte solvents. As noted in Sec.~\ref{ResultsCorrelations}, the charge-separation distance ($a$) and its dielectric permittivity ($\epsilon_I$) are not independent parameters. Various combinations can yield almost identical capacitance curves. Therefore, $\epsilon_I$ was fixed to $25\%$ of the bulk dielectric permittivity, i.e, for the aqueous solutions to a value of $\epsilon_I=20$, corresponding to $25\%$ of the bulk water permittivity. Accordingly, the model effectively had only two fit parameters: $a$ and $C_\text{ext}$. Alternative combinations of values for $\epsilon_I$ and $a$ yielding similar results are listed in Table~\ref{tab:fitValues} of the Appendix. Generally, smaller values of $\epsilon_I$ also require smaller values of $a$.

Four different electrode materials are compared: mercury, silver, gold, and platinum, in combination with aqueous electrolytes that are expected to exhibit negligible specific chemisorption, see Refs.~\cite{grahameDifferentialCapacityMercury1954,parsonsInterphaseMercuryAqueous1965,valetteDoubleLayerSilver1989,eberhardtImpedanceStudiesReconstructed1996,hamelinStudyGoldLow1987,ojhaDoubleLayerPt111Aqueous2020}. Figure~\ref{fig:CapacitanceAqueous} presents the corresponding PZ plots, demonstrating excellent quantitative agreement between the theoretical model and experimental data across all electrode materials. The model reproduces the decreased PZ slopes entirely through variations in the distance of closest approach of the ionic charge density to the image plane of the metal surface. For example, to achieve quantitative agreement with experimental data for a mercury electrode, a distance of $1.8\,\text{\AA}$ is obtained, corresponding to weak image-charge interactions and resulting in nearly perfect GCS behavior. In contrast, for the Pt(111) electrode, this distance is dramatically decreased to only $0.035\,\text{\AA}$, where strong attractive interactions between ions and their image charges lead to significantly enhanced capacitance and a suppressed PZ slope. Of course, such a small value of $a$ at a sub-atomic scale does not appear reasonable from the perspective of classical GCS theory, where $a$ would be interpreted as the distance of closest approach of an ion to the metal surface atoms. The present theory, however, does not impose such picture. Instead, $a$ corresponds to the minimum distance between ions and their electronic counter (image) charges of the metal surface. Values smaller than typical interatomic distances are thus possible, as suggested by recent simulation results of Pt(111) interfaces where the surface electron density was found to penetrate far into the solution region \cite{liElectronSpilloverWater2025}. In Sec.~\ref{Sec:Discussion}, the meaning and interpretation of such small values of $a$ are discussed in greater detail, with profound consequences for our perspective on interfacial charging processes.

Interestingly, the values of the charge-separation distance $a$ appear to correlate with the wettability of the electrode materials. Specifically, more hydrophobic materials tend to exhibit larger charge-separation distances, while more hydrophilic ones show smaller ones. A quantitative measure of wettability, i.e., hydrophilicity, is provided by the work of adhesion ($W_\text{ad}$), which reflects the strength of interaction between the electrode surface and water. In Ref.~\cite{gimStructureDynamicsWettability2019}, computed values of $W_\text{ad}$ for single-crystal surfaces show the trend Ag $<$ Au $<$ Pt, with platinum being the most hydrophilic. Mercury, although not included in this ranking, is well established as highly hydrophobic \cite{valetteDoubleLayerSilver1982}, and accordingly requires the largest value of the charge-separation distance in our model. This trend suggests a clear relationship: as hydrophilicity increases (from mercury to platinum), the charge-separation distance decreases, accompanied by stronger electron--ion correlations at the metal--water interface.

The present theory not only explains capacitance data for aqueous electrolytes, but equally holds for non-aqueous solvents with much smaller dielectric constants. We collected published experimental capacitance data for Au(111) electrodes in KPF$_6$ electrolyte solutions involving three different organic solvents, dimethyl sulfoxide (DMSO), acetonitrile (ACN), and diethylene glycol dimethyl ether (DG), with $\epsilon_B$ values ranging from $46.7$ down to $7.23$ (in addition to the case of water solvent discussed before) to examine the role of the dielectric solvent in tuning the strength of interfacial electron--ion correlations. Similarly to the models for aqueous electrolytes, the value of $\epsilon_I$ was fixed to $25\%$ of the bulk dielectric permittivity for each solvent, while the charge-separation distance $a$ was manually adjusted to best match the experimental data, cf. Tab.~\ref{tab:Data}. Figure~\ref{fig:CapacitanceNonAqueous} presents the PZ plots of the Au(111)--KPF$_6$ capacitance data for the DMSO, ACN, and DG solvents. The quantitative agreement between the theoretical model and experimental data is very good, underscoring the robustness of the approach. The model explains the trend that as the bulk dielectric constant decreases, the slope of the PZ plot also decreases. This occurs because lower dielectric constants result in weaker screening of electrostatic interactions compared to water, which increases the interaction between ions and their electronic image charges. It is interesting that the model values of the Au(111)--KPF$_6$ charge-separation distance are very similar for water, DMSO, and ACN solvents. Therefore, the different PZ slopes of the respective capacitance curves can be directly related to the differences in dielectric permittivity of the solvents. Remarkably, while nearly linear PZ plots with a slightly positive curvature are predicted (and observed) for water (Fig.~\ref{fig:CapacitanceAqueous}c) and DMSO (Fig.~\ref{fig:CapacitanceNonAqueous}a) solvents with relatively large permittivities, the theory predicts a non-linear PZ plot with a negative curvature for the case of DG solvent with a very low dielectric permittivity, in good agreement with the available experimental data (Fig.~\ref{fig:CapacitanceNonAqueous}c). These findings highlight the need for dedicated experiments, in particular, accurate capacitance measurements in dilute, non-aqueous electrolytes. Non-linear PZ plots with a negative curvature in low-$\epsilon$ solvents are suggested as an experimentally accessible ``smoking gun'' of interfacial electron--ion correlations.

\section{Discussion}\label{Sec:Discussion}

\begin{figure*}[t]
	\centering
	\includegraphics[width=2\columnwidth]{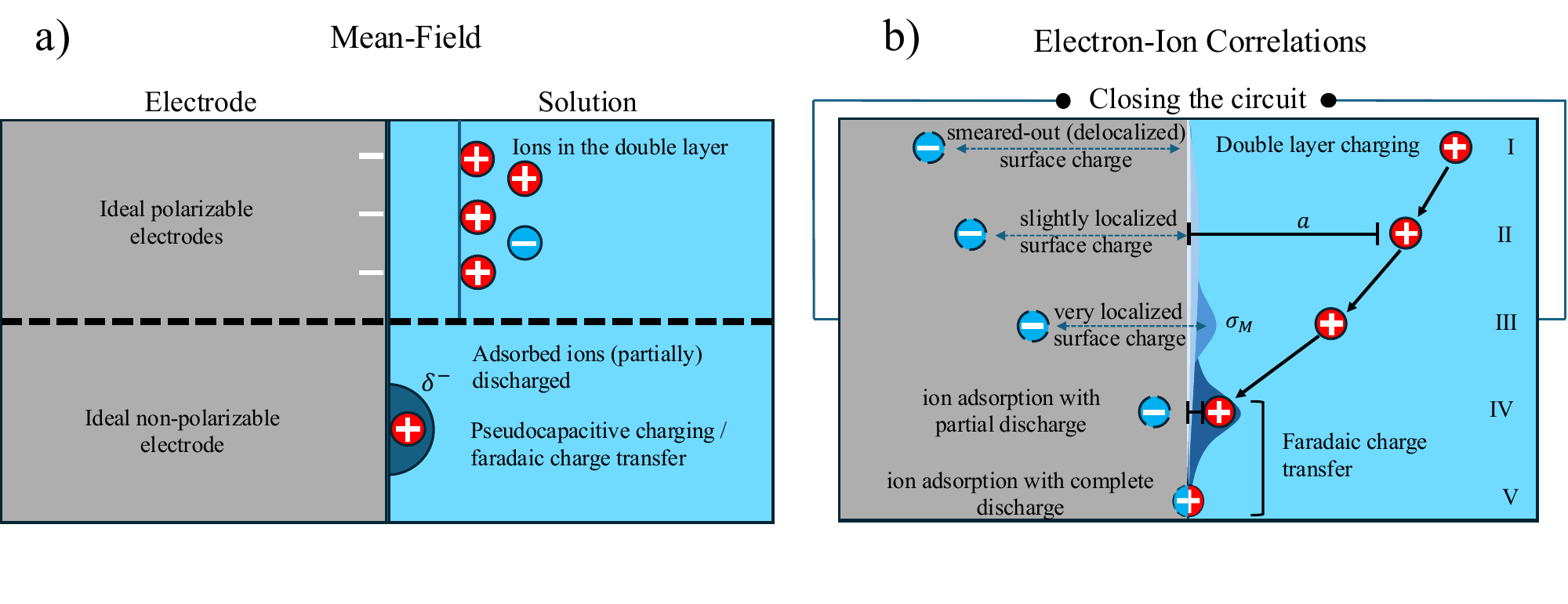}
	\caption{(a) Comparison of the distinct models of an ideal polarizable electrode with double-layer charging (top) and an ideal non-polarizable electrode with pseudocapacitive charging due to ion adsorption and discharge, i.e., faradaic charge transfer (bottom). (b) Unified model from double-layer charging to ion adsorption. A representative cation is shown as it approaches the metal surface. As the charge-separation distance $a$ gets smaller from (I) to (V), the electronic image charge (dashed circles) gets closer to the surface and the corresponding physical surface charge gets more and more localized and contracted to the place of the approaching ion. In the limiting case (V) of $a \to 0$, the electronic (image) and ionic charge centers coincide and exactly cancel each other, corresponding to complete ionic discharge (integer charge transfer).} 
 \label{fig:MFChargingPhenomena}
 \label{fig:SchematicUnification}
\end{figure*}

Traditional GCS theory predicts a unity slope in Parsons-Zobel (PZ) capacitance plots. Experimental capacitance data usually presents smaller-than-one PZ slopes, meaning that the measured differential capacitance is higher than GCS predictions even in the limits of dilute electrolyte concentrations and weak surface charges (i.e., around the PZC) where MF-GCS theory is considered to be accurate. Such deviations have often been attributed to surface roughness, i.e., the difference between the geometric surface area (the area used for the normalization of the surface charge density) and the (generally unknown) real surface area of the electrode \cite{parsonsInterphaseMercuryAqueous1965,forestiModelEffectRoughness1993,daikhinDoublelayerCapacitanceRough1996}. However, similar discrepancies were observed for single-crystal electrode surfaces, where relevant surface roughness can be excluded  \cite{forestiModelEffectRoughness1993,schmicklerEffectWeakAdsorption2021}. As summarized in Table~\ref{tab:Data}, the extent of the deviations strongly depends on the electrode material, with a close-to-unity PZ slope for the case of mercury \cite{grahameDifferentialCapacityMercury1954} to the extreme case of Pt(111) with a slope close to zero, corresponding to a strongly enhanced double-layer capacitance \cite{ojhaDoubleLayerPt111Aqueous2020}. Coinciding with the centenary of GCS theory, the latter recent findings have triggered a vivid debate on the current understanding of the EDL properties in the GCS limit \cite{doblhoff-dierElectricDoubleLayer2023}. Several different explanations for the very low PZ slopes of Pt(111) have been proposed, involving the residual adsorption of OH$^{-}$~\cite{huangZoomingInnerHelmholtz2023} or electrolyte (an)ions~\cite{schmicklerEffectWeakAdsorption2021} on the electrode surface, or an additional weak surface--ion attractive potential~\cite{doblhoff-dierModelingGouyChapman2021}, the origin of which however remained ambiguous. Image-charge interactions were discussed as one possible origin~\cite{doblhoff-dierModelingGouyChapman2021}, since the experimental observations indicated that this attractive potential was acting on both anion and cation species, while being insensitive to their detailed chemical nature (e.g., Na$^+$ or Li$^+$). At the same time, however, it was questioned whether image-charge effects would be sufficiently strong to explain the results. Two central questions were emphasized in a recent article \cite{doblhoff-dierElectricDoubleLayer2023}: First, why is the attractive interaction similar for all investigated electrolytes? Second, why is this interaction apparently stronger on Pt(111) than on other single-crystal metal surfaces or mercury?

The present theory offers a compelling answer to these questions, indeed demonstrating that image-charge (electron--ion) correlations provide a very likely explanation for the puzzling recent results on Pt(111). Moreover, it was shown that the latter represent a limiting case in the context of the general long-standing question in electrochemistry for the origin of non-unity PZ slopes. Quantified by the \textit{interface correlation factor}, Eq.~\eqref{RoughnessInfinity}, electron--ion correlations were found to play an important role in shaping the electric double layer at the PZC by effectively attracting ions toward the electrode surface. This mechanism enhances the differential capacitance in a way that is mathematically akin to the impact of surface roughness, but entirely different in physical nature. 

Because the electron--ion correlations considered here are electrostatic in nature, the proposed mechanism is agnostic to the chemical nature of the electrolyte ions, meaning that the effective interaction is unspecific to the ion species. The interaction strength is primarily controlled by the charge-separation distance $a$ between the ionic and electronic (image) charges, as well as the (bulk) dielectric constant of the electrolyte solvent. In aqueous electrolytes, the trend in PZ slopes across different metals was found to correlate with the hydrophilicity of the metal surfaces, reflected by the respective variations in the value of the charge-separation distance parameter within this theory, cf. Table~\ref{tab:Data}. The more hydrophobic metals, such as mercury, correspond to a larger effective separation of the electronic and ionic charge distributions, and vice versa for the more hydrophilic cases. Besides the results for aqueous systems, we consider it a remarkable success of our theory to also explain trends in PZ slopes for various non-aqueous electrolytes based on the dielectric permittivity of the respective solvents, cf. Fig.~\ref{fig:CapacitanceNonAqueous}. These results strongly support the theory of electron--ion correlations as explanation for the experimental trends in PZ slopes across a variety of electrode--electrolyte systems.
 
The case of the aqueous Pt(111) system deserves further discussion. To fit the respective experimental data, a nearly zero value of the charge-separation distance parameter ($a$) was required. This cannot be purely explained by Pt(111) being rather hydrophilic, allowing for a closer approach of electrolyte ions to the metal surface. In fact, recent DFT simulations showed that the electronic charge distribution of the Pt(111) surface penetrates to a significant depth into the interfacial water region \cite{liElectronSpilloverWater2025}. While this certainly contributes to pushing the effective image plane closer to the electrolyte charge distribution, the effect would likely not be sufficient to explain the essentially zero value of $a$ in the present theory. Instead, in the following we argue that the limit of $a\to0$ corresponds to the transition from traditional EDL charging to (unspecific) ion adsorption. The theory thereby overcomes the conceptual distinction between double-layer and pseudocapacitive charging phenomena~\cite{huangVariantsSurfaceCharges2024}.

Fig.~\ref{fig:MFChargingPhenomena}a shows the traditional distinction between double-layer charging and ionic discharge during adsorption (pseudocapacitive charging), which correspond to different mechanisms of interfacial charge storage and transfer as described within the ideal polarizable and ideal non-polarizable electrode models, respectively. Double-layer charging, shown in the top part of Fig.~\ref{fig:MFChargingPhenomena}a, arises from the electrostatic accumulation of charge at the electrode--electrolyte interface. This process is commonly referred to as non-faradaic, meaning that no electron transfer or electrochemical reaction occurs. The respective charging process is fundamentally described by GCS-based theories. On the other hand, pseudocapacitive charging due to ion adsorption, shown in the bottom part of Fig.~\ref{fig:MFChargingPhenomena}a, involves faradaic charge transfer across the electrode--electrolyte interface due to partial or complete discharge of the adsorbing ions. Such charge storage mechanism is characterized by a large apparent (pseudo)capacitance as compared to pure double-layer charging. 

Within a mean-field description, such as GCS theory, double-layer and pseudocapacitive charging must be treated as two separate mechanisms. At the mean-field level, an ion in the electrolyte solution, regardless of its distance from the electrode surface, only interacts with the fully delocalized (in-plane averaged) electronic surface charge density, which is a consequence of the neglect of coulombic correlations. As a consequence, the MF-GCS capacitance, Eq.~\eqref{MFCapacitance}, is bounded from above by the GC contribution ($C_{\text{GC}}$) that is independent of the charge-separation distance $a$. Letting $a\to 0$ only affects the Helmholtz contribution ($C_{\text{H}}$) that is acting in series to $C_{\text{GC}}$. Accordingly, pseudocapacitance must be incorporated into a MF model in a heuristic way either by adding an additional attractive potential for electrolyte ions to the electrode surface~\cite{doblhoff-dierModelingGouyChapman2021,doblhoff-dierElectricDoubleLayer2023}, or by explicitly introducing a species of adsorbed ions that form localized dipole moments at the electrode surface~\cite{schmicklerEffectWeakAdsorption2021,huangZoomingInnerHelmholtz2023,huangVariantsSurfaceCharges2024}. 

In our theory, pseudocapacitive charging and ion adsorption phenomena are naturally described at the beyond-MF level in the form of local electron--ion (image-charge) correlations. As such, the processes of double-layer charging and (unspecific) ion adsorption are conceptually unified, representing limiting cases of the same interfacial charging phenomenon. A continuous transition between these limits is provided by tuning the effective charge-separation distance $a$. In particular, the local electron--ion interactions captured by the interface correlation factor lead to a divergence of the overall interfacial capacitance (Eq.~\eqref{CapacitanceClover}) in the limit of $a \to 0$. This directly corresponds to the massive pseudocapacitive enhancement in the faradaic limit of complete ionic discharge. 

This conceptual unification is schematically visualized in Fig.~\ref{fig:SchematicUnification}b, tracking a representative cation as it approaches the metal surface. In case (I), the charge-separation distance $a$ is large. Consequently, the image charge is distant from the interface and the corresponding physical metallic surface charge is smeared out (very light blue distribution). In this limit, which is representative for the ideal polarizable electrode, the theory effectively becomes equivalent to a MF description, where the electronic charge density is constant over the surface and the EDL behaves as described by classical GCS theory. Moving to cases (II) and  (III), the ionic charges and image-charges approach each other at closer distances. The surface charge density becomes more localized the closer the ions get to the metal (light blue and blue distributions). In this limit the local interaction/correlation with the image charge becomes important. In case (IV), where $a$ approaches the atomic scale, the electronic surface charge gets contracted to the position of the approaching ion, forming a local surface dipole as in ion adsorption with partial discharge. This again highlights that the model parameter $a$ must be understood as an effective separation distance between the electronic and ionic charges, and \emph{not} as a minimum distance between ions and metal surface atoms as in GCS theory. Therefore, $a$ can assume values of sub-atomic scale and, in the case of complete ionic discharge, turn to zero, corresponding to coinciding centers of charge of the local electronic and ionic charge densities. This limit of $a\to0$ is shown as case (V) in Fig.~\ref{fig:SchematicUnification}b and represents ion adsorption with complete charge neutralization and integer charge transfer. The present theory thus provides a continuous transition from double layer charging to ion adsorption and conceptually unifies both types of processes. Interfacial electron--ion correlations are thus shown to be the link for ``closing the circuit'' from EDL charging to electrosorption. This unified interpretation highlights the continuity between these charging mechanisms rather than treating them as separate phenomena. The present theory is specifically developed to describe the effects of electron--ion correlations in the weak charging limit around the PZC and in dilute electrolyte solutions (GCS limit). Under such conditions, the effective surface coverages of electrosorbed ions (cases (IV) and (V) in Fig.~\ref{fig:SchematicUnification}b) are small, and coverage effects due to finite ion sizes, as heuristically included in another recent electrosorption model~\cite{hedleyWhatDoesIon2025}, can be neglected.

Of course, the limit of $a\to0$ must be taken with a grain of salt in a classical theory like the one presented here that describes ions and image electrons as point charges. At (sub-)atomic scales, the spatial distribution of the electronic image charge would require a quantum-mechanical treatment to capture its orbital structure. An accurate description of (ion-specific) chemisorption therefore goes beyond the scope of the present theory. Ion adsorption processes described herein are independent of the chemical nature of the ion species, because they are solely driven by classical electrostatic interactions between ions and their electronic image charges. This unspecific character is in agreement with experimental results for Pt(111), where PZ slopes were found to be largely insensitive to the ion species in the electrolyte~\cite{doblhoff-dierModelingGouyChapman2021}. The present insights thus reconcile some of the previously proposed explanations for the drastically decreased PZ slopes on Pt(111) in terms of either a weak attractive ion--surface interaction or an effect of residual ion adsorption, both of which are identified as manifestations of interfacial electron--ion correlations.

\section{Conclusion}

In summary, a systematic extension of Gouy–Chapman–Stern (GCS) theory for the electric double layer (EDL) at metal--electrolyte interfaces was developed incorporating electron--ion correlations beyond the mean-field (MF) level. The theory is applicable in the weak charging limit and for dilute electrolyte solutions. This framework provides compelling explanations for long-standing discrepancies between experimental capacitance data and GCS predictions, thereby overcoming previous conceptual boundaries in the classification of interfacial charging phenomena. The key developments and findings include:

\begin{itemize} 
    \item Correlations shape the EDL: Beyond mean-field theory, two competing correlation effects, attractive electron--ion (i.e., image-charge--ion) and repulsive dielectric boundary interactions govern the ion accumulation near the electrode surface around the potential of zero charge (PZC).
    \item Image-charge effects explain anomalous slopes in Parsons-Zobel (PZ) plots: Decreased slopes in PZ plots result from ion--image-charge attraction, viz., electron--ion correlations. While, even for single-crystal electrodes, surface roughness was often postulated to explain discrepancies between experimental PZ slopes and GCS theory, the present findings show that this can be explained purely on the basis of electrostatic phenomena.
    \item To explain the extremely low PZ slopes observed for Pt(111), a sub-atomic value of the effective charge-separation distance $a$ is required within the present theory, representing the case of ion electrosorption. In the limit of vanishing $a\to0$, the electronic (image) charge completely neutralizes the ionic charge. Interfacial electron--ion correlations are thus connecting previous explanatory schemes of weak additional ion--surface interactions and ion adsorption.
    \item The theory explains why hydrophobic metals such as mercury yield a near-unity PZ slope, corresponding to weak electron--ion correlations, while more hydrophilic metals exhibit decreased PZ slopes, representing stronger correlation effects. The PZ slope is thus identified as a fingerprint of interfacial electron--ion correlations. 
    \item Non-aqueous electrolytes with a lower solvent permittivity than water enhance coulombic interactions and correlation effects, resulting in non-linear PZ plots. These findings suggest that additional measurements in non-aqueous electrolytes are required to gain further insights into electrostatic electron--ion correlations at electrochemical interfaces.
    \item The theory conceptually unifies the mechanisms of double layer charging and ion adsorption (electrosorption), commonly treated as different phenomena at the MF level. A continuous transition from double-layer charging to ion adsorption is provided by tuning the charge-separation distance $a$ to zero, demonstrating that both processes can be regarded as the two sides of the same coin.
\end{itemize}

\section*{Acknowledgments}
This work was carried out within the framework of the research program Materials and Technologies for the Energy Transition (MTET) funded by the Helmholtz Association of German Research Centres.

\paragraph*{Author Contributions}
\textbf{Nils Bruch: }Conceptualization, Methodology, Formal analysis, Writing – original draft, Writing – review \& editing. \textbf{Tobias Binninger: }Conceptualization, Methodology, Formal analysis, Writing – original draft, Writing – review \& editing. \textbf{Michael Eikerling: }Conceptualization, Writing – review \& editing.
\onecolumngrid
\appendix

\section{Intermediate steps of the derivations in the theory section}

\subsection{Computing partial derivatives leading to  Eqs.~\eqref{PartialDerivative1} and \eqref{PartialDerivative2}}\label{Appendix:ComputingPartialDerivatives}
For calculating the partial derivatives of $\Omega(\mu_a,\mu_c)$ in Eq.~\eqref{GrandPotentialChemicalPotentials} with respect to $\mu_{+}$ and $\mu_{-}$, the thermodynamic definition of the chemical potentials $\mu_i=\partial\mathcal{F}/\partial N_i$ is needed along with the inverse transformations $\mu_a=\mu_{+}-\mu_{-}$ and $\mu_c=\mu_{+}+\mu_{-}$, which follow immediately from Eqs.~\eqref{MuBar} and \eqref{MuTilde}. Then the partial derivative of $\Omega(\mu_a,\mu_c)$ with respect to $\mu_{-}$ is given by
\begin{align}
    \frac{\partial\Omega(\mu_{+},\mu_{-})}{\partial\mu_{-}} &= \frac{\partial \mathcal{F}}{\partial N_a}\frac{\partial N_a}{\partial \mu_{-}}+\frac{\partial \mathcal{F}}{\partial N_c}\frac{\partial N_c}{\partial \mu_{-}}-\mu_+ \frac{\partial N_\text{tot}}{\partial \mu_{-}}- N_\text{diff} - \mu_{-} \frac{\partial N_\text{diff}}{\partial \mu_{-}} \nonumber \\
    &= \mu_a\frac{\partial N_a}{\partial \mu_{-}}+\mu_c\frac{\partial N_c}{\partial \mu_{-}}-\mu_+ \frac{\partial (N_c+N_a)}{\partial \mu_{-}}- N_\text{diff} - \mu_{-} \frac{\partial (N_c-N_a)}{\partial \mu_{-}} \nonumber \\
    &= \frac{\partial N_a}{\partial \mu_{-}}\left(\mu_a-(\mu_{+}-\mu_{-})\right)+\frac{\partial N_c}{\partial \mu_{-}}\left(\mu_c-(\mu_{+}+\mu_{-})\right) - N_\text{diff} \nonumber \\
    &=- N_\text{diff} \ ,
\end{align}
where in the second line the definitions of $\mu_i$, $N_\text{tot}$, and $N_\text{diff}$ were inserted. In the third line the inverse transformations were inserted into $\mu_+$ and $\mu_{-}$.

\subsection{Computing the MF electrostatic potential in Eq.~\eqref{ElectrostaticPotentialMF}}\label{DerivationMFElectrostaticPotential}

Utilizing the ansatz $\phi^\text{MF}(r)=\phi_0+\Delta\phi^\text{MF}(r)$, in the electrolyte region, $z>a$, the differential equation Eq.~\eqref{SaddlePointEquation} simplifies to
\begin{align}
    -\epsilon_{B}\nabla^{2}\Delta\phi^\text{MF}(r)+2qn_{\text{ion}}^{\text{b}}\sinh\left(\beta q_{i}\Delta\phi^\text{MF}(r)\right) = 0,
\end{align}
which is the well-known Poisson-Boltzmann equation that is solved by \cite{schmicklerInterfacialElectrochemistry2010}
\begin{align}\label{SaddlePointNonLinear}
    \phi^\text{MF}(r)=\phi_{0}
    +\frac{4}{\beta q}\text{arctanh}\left(\tanh\left(\frac{\beta q\left(\phi^\text{MF}(a)-\phi_{0}\right)}{4}\right)e^{-\lambda_{D}^{-1}(z-a)}\right),
\end{align}
with the Debye length
\begin{align}
    \lambda_D=\sqrt{\frac{\epsilon_B}{2\beta q^{2}n_{\text{ion}}^{\text{b}}}}.
\end{align}
Within the charge-separation distance ($0<z<a$), Eq.~\eqref{SaddlePointEquation} is solved by the ansatz $\phi^\text{MF}(r)=Az + B$, where $B=0$ due to the asymmetry of $\phi^\text{MF}$ (due to the asymmetry boundary condition of $\psi_a$ at $z=0$), and $A$ is fixed by the boundary conditions at $z=a$,
\begin{align}
    \epsilon_{B}\frac{\partial\phi^\text{MF}}{\partial z}\at_{z=a^{+}}-\epsilon_{I}\frac{\partial\phi^\text{MF}}{\partial z}\at_{z=a^{-}} = 0, 
\end{align}
which leads to
\begin{align}
    A=\frac{\epsilon_{B}}{\epsilon_I}\frac{\partial\phi^\text{MF}}{\partial z}\at_{z=a^{+}}.
\end{align}
The continuity requirement at $z=a$ leads to a self-consistent equation for the value of $\phi^\text{MF}$ at $z=a$,
\begin{align}\label{SelfConsistentEq}
   \phi^\text{MF}(a)=\frac{\epsilon_{B}}{\epsilon_{I}}\frac{\partial\phi^\text{MF}}{\partial z}\at_{z=a}a  = \frac{2\epsilon_{B}}{\epsilon_{I}}\frac{a}{q\beta\lambda_{D}}\sinh\left(\frac{ q\beta}{2}\left(\phi_{0}-\phi^\text{MF}(a)\right)\right),
\end{align}
which can not be solved analytically. For weakly charged surfaces, $\phi^\text{MF}$ varies only slightly between the bulk and the interface. As a result, the argument of the $\sinh$ function ($\phi^\text{MF}(a)-\phi_{0}$), remains small. In this limit the $\sinh$ function can be linearized and Eq.~\eqref{SelfConsistentEq} is solved by $\phi^\text{MF}(a)=\frac{\epsilon_{B}a}{\epsilon_{B}a+\epsilon_{I}\lambda_{D}}\phi_{0}$. Substituting this result into Eq.~\eqref{SaddlePointNonLinear} yields the electrostatic potential on a MF level near PZC, shown in Eq.~\eqref{ElectrostaticPotentialMF}.

\subsection{Deriving the correlation function in Eq.~\eqref{LinearizedFMinusEquation}}\label{Appendix:CorrelationFunction}
The second-order variational derivative is given by
\begin{align}\label{AppendixSecondDerivative}
    &\frac{\delta S}{\delta\psi(r)\delta\psi(r')}=\frac{\delta}{\delta\psi(r)}\Bigg(-\nabla\epsilon(r')\nabla\psi_a(r')+i\sum_{i=a/c}q_{i}\lambda_{i}\Lambda_{i}^{-3} \Theta(z'-a)e^{-i\beta q_{i}\psi_a(r')}\Bigg)\nonumber\\
    &=-\nabla\epsilon(r')\nabla\delta(r,r')+\beta\sum_{i=a/c}q_{i}^{2}\lambda_{i}\Lambda_{i}^{-3}\Theta(z'-a)e^{-i\beta q_{i}\psi_a(r)}\delta(r,r').
\end{align}
The inverse correlation function (Eq.~\eqref{DiffeqGreensFunction}) is obtained by replacing $\psi_a$ in Eq.~\eqref{AppendixSecondDerivative} with the saddle-point field $\psi_c$, which is related to the electrostatic potential on a MF level via $\phi^\text{MF}=i\psi_c$ as shown in Eq.~\eqref{ElectrostaticPotentialMF}. Substituting Eq.~\eqref{ElectrostaticPotentialMF} into Eq.~\eqref{AppendixSecondDerivative} then yields
\begin{align}\label{InvseGreen}
    G^{-1}(r,r')&=-\nabla\epsilon(r')\nabla\delta(r,r') +2\beta z^{2}e_{0}^{2}n_{\text{ion}}^{\text{b}}\cosh\left(4\text{arctanh}\left(\tanh\left(\frac{\lambda_{D}}{(\epsilon_{B}/\epsilon_{I})a+\lambda_{D}}\frac{\beta q \mathcal{E}}{4}\right)e^{-\lambda_{D}^{-1}\abs{z'-a}}\right)\right) \Theta(z'-a)\delta(r,r').
\end{align}
Using the definition, Eq.~\eqref{Defn:CorrelationFunction}, the correlation function satisfies the differential equation
\begin{align}\label{DGLForG}
    -\nabla\epsilon(r')\nabla G(r,r')  +2\beta z^{2}e_{0}^{2}n_{\text{ion}}^{\text{bulk}} \cosh\left(4\text{arctanh}\left(\tanh\left(\frac{\lambda_{D}}{(\epsilon_{B}/\epsilon_{I})a+\lambda_{D}}\frac{\beta q \mathcal{E}}{4}\right)e^{-\lambda_{D}^{-1}\abs{z'-a}}\right)\right) G(r,r')=\delta(r-r').
\end{align}
This work focuses on the region close to the potential of zero charge (PZC), which is characterized by $\mathcal{E}\ll1$. For $\mathcal{E}\ll1$ the nonlinearity in the equation above can be linearized, leading to Eq.~\eqref{LinearizedFMinusEquation}.

\subsection{Derivation of  Eq.~\eqref{CorrelationFunction} with in-plane Fourier transformation of the correlation function}\label{Appendix:SolutionCorrelationFunction}
We follow the same procedure as Lau \cite{lauFluctuationCorrelationEffects2008} or Netz \cite{netzDebyeHUckelTheory1999} who provided the basis for the following calculations. Through the translational symmetry in $xy$ direction, Fourier-transforming the differential equation for the correlation function (Eq.~\eqref{LinearizedFMinusEquation}) yields
\begin{align}
    G(r,r')=\frac{1}{\sqrt{2\pi}^{2}}\int d^{2}k\;g(k,z,z')\;e^{ik_{x}(x-x')}e^{ik_{y}(y-y')}=\frac{1}{2\pi}\int d^{2}k\;g(k,z,z')\;e^{i\boldsymbol{k}\cdot\boldsymbol{\rho}},
\end{align}
which can be simplified by utilizing polar coordinates,
\begin{align}\label{GFInPolarCoordinates}
    G(r,r') = \int_{0}^{\infty}dk\;k\;g(k,z,z')\;J_{0}(k\,\rho),
\end{align}
where $J_0$ is the zeroth Bessel function of first kind. Similarly, the Fourier-transform of the delta function is
\begin{align}
    \delta(r-r')&=\frac{1}{2\pi}\int dk\;\delta(z-z')\;k\, J_{0}(k\,\rho).
\end{align}
Lastly, rewriting the gradients in Eq.~\eqref{LinearizedFMinusEquation} in polar coordinates, yields an equation for $g$,
\begin{align}
    \left(-\partial_{z'}^{2}+k^{2}+\lambda_{D}^{-2}\right)g(k,z,z')&=l_{B}\delta(z-z')\quad\text{for \ensuremath{\abs z>a}},\\
    \left(-\partial_{z'}^{2}+k^{2}\right)g(k,z,z')&=l_{I}\delta(z-z')\quad\text{for \ensuremath{\abs z<a}},
\end{align}
where
\begin{align}
    l_{B}=\frac{1}{2\pi \epsilon_{B}},\quad l_{I}=\frac{1}{2\pi \epsilon_{I}}.
\end{align}
For a fixed $z'>a$, there are three regions to be distinguished,
\begin{align}
    \left(-\partial_{z}^{2}+k^{2}+\lambda_{D}^{-2}\right)h_{\pm}(k,z)=0&\text{ für \ensuremath{z>a}},\\\left(-\partial_{z}^{2}+k^{2}\right)g_{\pm}(k,z)=0&\text{ für \ensuremath{-a<z<a}},\\\left(-\partial_{z}^{2}+k^{2}+\lambda_{D}^{-2}\right)h_{\pm}(k,z)=0&\text{ für \ensuremath{z<-a}},
\end{align}
with solutions
\begin{align}
    h_{\pm}(k,z)&=\exp\left(\pm\sqrt{k^{2}+\lambda_{D}^{-2}}z\right)\equiv\exp\left(\pm\gamma(k)z\right),\\
    g_{\pm}(k,z)&=\exp\left(\pm kz\right).
\end{align}
The boundary conditions are continuity in $g$ across the three regions and  discontinuity in the partial derivative of $g$ with repsect to $z$ due to the delta funtions at $z=z'$ and $z=-z'$, cf. Ref.~\cite{markovichIonicProfilesClose2016}. This leads to a linear system of ten equations for ten unknown amplitude functions of $z$. For the purposes of this work, only the region $z>a$ is of interest, where the function $g$ takes the form:
\begin{align}\label{AppendixSmallG}
    g(k,z,z')=\frac{l_{B}}{2\gamma(k)}\left(e^{-\gamma(k)\abs{z'-z}}+e^{-\gamma(k)(z+z'-2a)}\left(1-\frac{2k}{k+(\epsilon_{B}/\epsilon_{I})\gamma(k)\tanh(ka)}\right)\right)\quad\forall\;z,z'>a.
\end{align}
Inserting Eq.~\eqref{AppendixSmallG} into Eq.~\eqref{GFInPolarCoordinates} yields Eq.~\eqref{CorrelationFunction}.

\subsection{Computing the 1L electrostatic potential $\phi^\text{1L}$}\label{SubSubSec:ThirdVariation}
In this section the derivation of the first-order correction to the electrostatic potential is discussed in detail. 
\paragraph{Computation of $S_I^{(3)}$}
For sake of clarity, the first non-vanishing part of the expanded action (Eq.~\eqref{FirstOrderIneracting}) is repeated here,
\begin{align}\label{AppendixFirstOrderInteracting}
    S_I^{(3)}[\psi_c]=\int_{r,r'}\frac{\delta S_{\text{int}}[\psi,\left\{\lambda^0_i \omega \right\}]}{\delta\psi(r)}\at_{\psi=\psi_{c}}\delta\psi(r)+\frac{1}{3!}\int_{r,r',r''}\Gamma^{(3)}(\left\{\lambda^0_i \right\},r,r',r'')\delta\psi(r)\delta\psi(r')\delta\psi(r'').
\end{align}
Clearly, computing Eq.~\eqref{AppendixFirstOrderInteracting} requires computing the first and third functional derivative of the action Eq.~\eqref{Action}. The first variational derivative of the action is given by
\begin{align}
  \frac{\delta S_{\text{int}}[\psi,\left\{\lambda^0_i \omega \right\}]}{\delta\psi(r)}\at_{\psi=\psi_{c}}=i\sum_{i=a/c}q_{i}\lambda_i\omega\Lambda_{i}^{-3}\Theta(z'-a)e^{-i\beta q_{i}\psi_{c}(r')},
\end{align}
and insertion of the saddle-point $\psi_c=i^{-1}\phi^\text{MF}$ (Eq.~\eqref{ElectrostaticPotentialMF}) simplifies the expression to
\begin{align}\label{ResultFirstFunctionalDerivative}
\frac{\delta S_{\text{int}}[\psi,\left\{\lambda^0_i \omega \right\}]}{\delta\psi(r)}\at_{\psi=\psi_{c}}=-\frac{i\epsilon_{B}\omega}{\beta q \lambda_{D}^{2}}\sinh\left(4\text{arctanh}\left(\tanh\left(\frac{\lambda_{D}}{(\epsilon_{B}/\epsilon_{I})a+\lambda_{D}}\frac{\beta q \mathcal{E}}{4}\right)e^{-\lambda_{D}^{-1}\abs{z'-a}}\right)\right)\Theta(z'-a).
\end{align}
In a  similar fashion, the third variational derivative yields
\begin{align}\label{AppendixThirdVariation}
    \frac{\delta^3 S}{\delta\psi(r)\delta\psi(r')\delta\psi(r'')} = -i\beta^{2}\sum_{i=a/c}q_{i}^{3}\lambda_{i}\Lambda_{i}^{-3}\Theta(z'-a)e^{-i\beta q_{i}\psi_a(r)}\delta(r',r'') \delta(r,r'),
\end{align}
and evaluating at the saddle-point gives
\begin{align}
    \sum_{i=a/c}q_{i}^{3}\lambda_{i}\Lambda_{i}^{-3}e^{-i\beta q_{i}\psi_c(r)}=2q_{i}^{3}n_{\text{ion}}^{\text{b}}\sinh\left(4\text{arctanh}\left(\tanh\left(\frac{\lambda_{D}}{(\epsilon_{B}/\epsilon_{I})a+\lambda_{D}}\frac{\beta q \mathcal{E}}{4}\right)e^{-\lambda_{D}^{-1}\abs{z-a}}\right)\right),
\end{align}
which yields the bare vertex of order three,
\begin{align}\label{ThirdVariation}
    \Gamma^{(3)}(\left\{\lambda^0_i \right\},r,r',r'') &= i\beta q\epsilon_B\lambda_{D}^{-2}\sinh\left(4\text{arctanh}\left(\tanh\left(\frac{\lambda_{D}}{(\epsilon_{B}/\epsilon_{I})a+\lambda_{D}}\frac{\beta q \mathcal{E}}{4}\right)e^{-\lambda_{D}^{-1}\abs{z-a}}\right)\right)\Theta(z'-a)\delta(r',r'')\delta(r,r').
\end{align}
Inserting Eq.~\eqref{ThirdVariation} into the second part on the r.h.s. of Eq.~\eqref{AppendixFirstOrderInteracting} and integrating over $r''$ and $r'''$ leads to 
\begin{align}\label{AppendixGamma3}
    \int_{r,r',r''}\Gamma^{(3)}(\left\{\lambda^0_i \right\},r,r',r'')\delta\psi(r)&\delta\psi(r')\delta\psi(r'')=i\beta q\epsilon_{0}\lambda_{D}^{-2}\int_{r>a}\sinh\left(4\text{arctanh}\left(\tanh\left(\frac{\beta q \mathcal{E}}{4}\right)e^{-\abs{z}/\lambda_{D}}\right)\right)\delta\psi(r)^{3}.
\end{align}
In summary, inserting Eq.~\eqref{AppendixGamma3} and Eq.~\eqref{ResultFirstFunctionalDerivative} into Eq.~\eqref{AppendixFirstOrderInteracting} one arrives at the expression for the renormalized vertex of order three,
\begin{align}\label{InteractingAction}
    &S^{(3)}_{I}[\psi_c] = -\int_{r>a}i\epsilon_B\lambda_{D}^{-2} \sinh\left(4\text{arctanh}\left(\tanh\left(\frac{\lambda_{D}}{(\epsilon_{B}/\epsilon_{I})a+\lambda_{D}}\frac{\beta q \mathcal{E}}{4}\right)e^{-\lambda_{D}^{-1}\abs{z-a}}\right)\right)\left(\frac{\omega}{\beta q}\delta\psi(r)-\frac{1}{3!}\beta q\delta\psi(r)^{3}\right).
\end{align}

\paragraph{Computation of the Gaussian averages}
Inserting Eq.~\eqref{InteractingAction} into Eq.~\eqref{ElectrostaticPotentialFirstOrder} shows that obtaining $\phi^\text{1L}$ requires computing the average
\begin{align}\label{AppendixComputationOfElectrostaticPotential}
    \avgG{\delta\psi(r)S_I^{(3)}} =&-\int_{r'>a}i\epsilon_{0}\lambda_{D}^{-2}\sinh\left(4\text{arctanh}\left(\tanh\left(\frac{\lambda_{D}}{(\epsilon_{B}/\epsilon_{I})a+\lambda_{D}}\frac{\beta q \mathcal{E}}{4}\right)e^{-\lambda_{D}^{-1}\abs{z'-a}}\right)\right) \nonumber\\&\left(\frac{\omega}{\beta q}\avgG{\delta\psi(r)\delta\psi(r')}-\frac{1}{3!}\beta q\avgG{\delta\psi(r)\delta\psi(r')^{3}}\right).
\end{align}
The average $\avgG{\delta\psi(r)\delta\psi(r'))}$ is defined by Eq.~\eqref{Defn:CorrelationFunction}.  Since the averages are with respect to a Gaussian (quadratic) action, higher order moments than two can be computed using Wicks theorem \cite{altlandCondensedMatterField2010}, 
\begin{align}\label{WickContraction1}
    \avgG{\delta\psi(r)\delta\psi(r')^3}=3\,\beta^{-1}G(r,r')\, 1\,\beta^{-1} G(r',r').
\end{align}
The `3' on the r.h.s is due to the fact that there are three possible combinations to pair $\delta\psi(r)$ with $\delta\psi(r')$. Once $\delta\psi(r)$ is paired to $\delta\psi(r')$, there is only `1' possibility to pair two $\delta\psi(r')$. Inserting Eq.~\eqref{WickContraction1} into Eq.~\eqref{AppendixComputationOfElectrostaticPotential} yields
\begin{align}\label{ComputationClover1}
    \avgG{\delta\psi(r)S_{I}^{(3)}}=\int_{r'>a}\frac{1}{\beta^{2}q}&\left(i\epsilon_{B}\lambda_{D}^{-2}\sinh\left(4\text{arctanh}\left(\tanh\left(\frac{\lambda_{D}}{(\epsilon_{B}/\epsilon_{I})a+\lambda_{D}}\frac{\beta q \mathcal{E}}{4}\right)e^{-\lambda_{D}^{-1}\abs{z'-a}}\right)\right)\right) \nonumber \\
    &G(r,r')\left(\frac{\beta q^{2}}{2}G(r',r')-\omega\right),
\end{align}
which when inserted into Eq.~\eqref{ElectrostaticPotentialFirstOrder} leads to Eq.~\eqref{ElectrostaticPotential1L}.
\subsection{Diagrammatic representation of Gaussian averages}\label{Section:ExplanationDiagrams}
\begin{figure}[t]
	\centering
	\includegraphics[width=0.5\columnwidth]{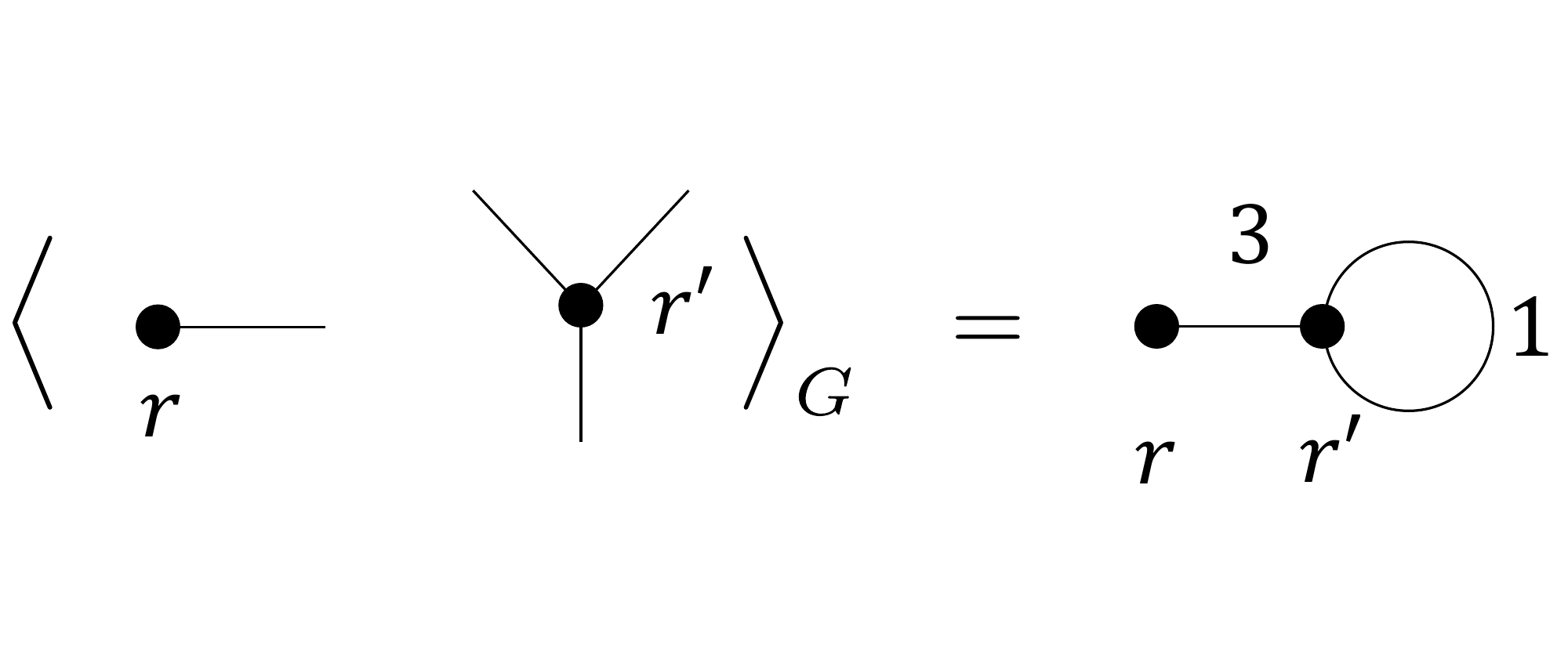}
	\caption{Graphical representation of the Gaussian average $\avgG{\delta\psi(r)\delta\psi^3(r')}=3\,\beta^{-1}G(r,r')\,1\,\beta^{-1} G(r',r')$.}
 \label{fig:1stOrder}
\end{figure}
At higher orders in perturbation theory, additional terms appear in the expansion of any observable quantity of interest, e.g. the electrostatic potential. To systematically identify the relevant corrections, it is helpful to represent them diagrammatically, following conventions from quantum field theory \cite{altlandCondensedMatterField2010}. This section outlines the diagrammatic approach adopted in this work and introduces the basic rules for simplifying such expansions. It is generally distinguished between two types of diagrams: those involving simple polynomials (e.g., $\delta\psi(r_i)^m$) and those built from renormalized vertices (e.g., $S^{(3)}_I$). Renormalized in this context means that the bare vartices yield unphysical infinite corrections, for instance, to the electrostatic potential. Through the choice of a gauge chemical potential in Eq.~\eqref{FugacityExpansion}, the vertex is renormalized to a finite result.

\paragraph{Diagramatic rules} First, the individual components of the diagrams are introduced. A polynomial $\delta\psi(r_i)^m$ is represented by dot (•) (with label $r_i$) connected to $m$ lines called external legs. The correlation function $G(r_i,r_j)$ acts as a propagator between the dots at $(r_i)$ and $(r_j)$, depicted as a line connecting $(r_i)$ and $(r_j)$. 

Second, the Gaussian average of diagrams is defined. Wick's theorem states that a Gaussian expectation value is given by the sum of all possible pairings \cite{altlandCondensedMatterField2010}
\begin{align}\label{WicksTheorem}
    \avgG{\delta\psi(r_1)\delta\psi(r_2)...\delta\psi(r_{2n})}=\sum_{\text{pairings of }{r_1,...,r_{2n}}} \avgG{\delta\psi(r_{k_1})\delta\psi(r_{k_2})}...\avgG{\delta\psi(r_{k_{2n-1}})\delta\psi(r_{k_{2n}})}.
\end{align}
A pairing is nothing but a propagator, i.e. $\avgG{\delta\psi(r_n)\delta\psi(r_m)}=\beta^{-1}G(r_n,r_m)$ (cf. Eq.~\eqref{Defn:CorrelationFunction}), and is therefore symbolized by a line connecting the dot at $r_n$ with a dot at $r_m$. Figure~\ref{fig:1stOrder} diagrammatically represents the Gaussian average in Eq.~\eqref{WickContraction1}. The dot $ r $ has one leg for $\delta\psi(r)$, while the dot at $ r' $ has three for $ \delta\psi(r')^3 $. The only possible pairing configuration is shown on the right-hand side, where $\delta\psi(r)$ pairs with one leg at $ r' $ (three options), and the remaining two legs at $ r' $ pair uniquely. This graphical representation corresponds to the mathematical expression on the r.h.s. of Eq.~\eqref{WickContraction1}.

\begin{figure}[t]
	\centering
	\includegraphics[width=0.5\columnwidth]{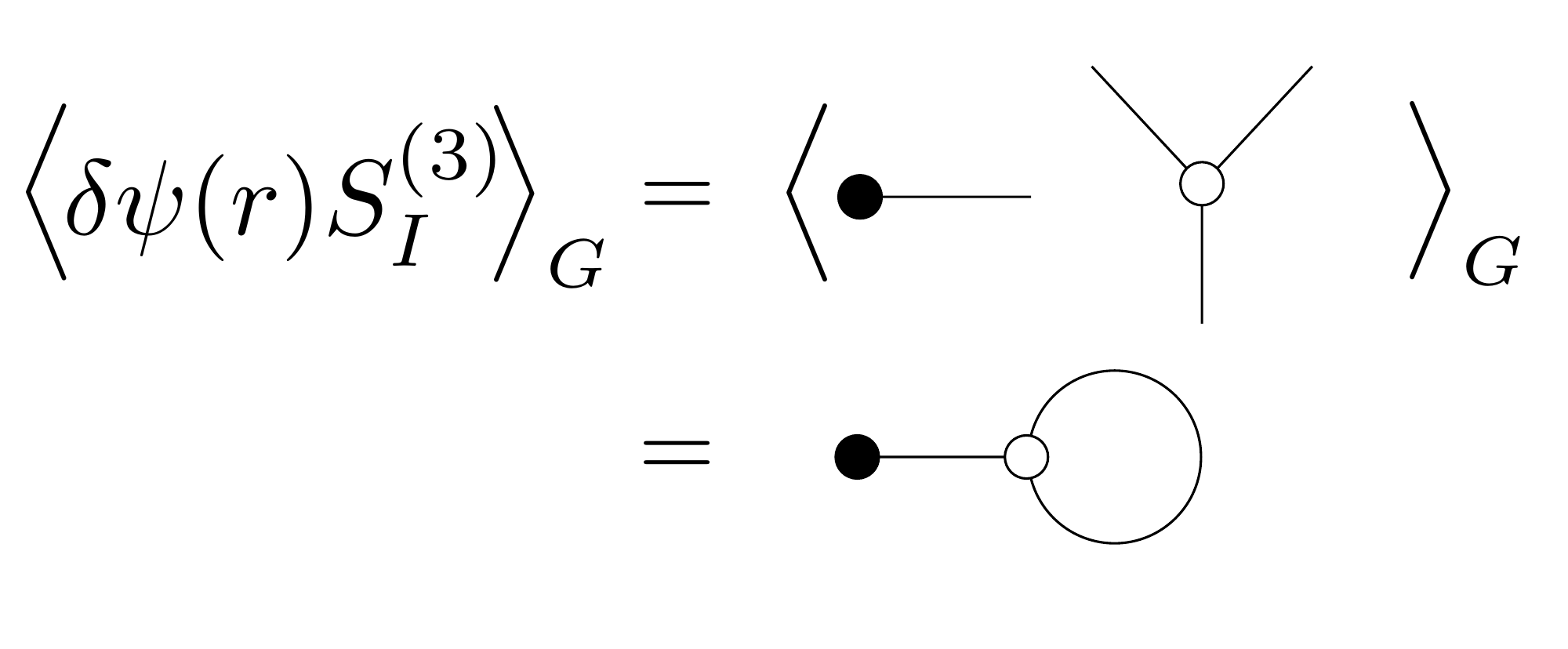}
	\caption{Graphical representation of the Gaussian average in Eq.~\eqref{ElectrostaticPotentialFirstOrder} and the result Eq.~\eqref{AppendixComputationOfElectrostaticPotential}.}
 \label{fig:Feynman1LElectricPotential}
\end{figure}

Using the diagrammatic rules, one can readily compute, for instance, the Gaussian average involving a bare vertex
\begin{align}\label{AppendixBareVertex}
    \avgG{\delta\psi(r)\Gamma^{(3)}[\left\{\lambda_i \right\}]}=&\int_{r'>a}i\epsilon_B\lambda_{D}^{-2} \sinh\left(4\text{arctanh}\left(\tanh\left(\frac{\lambda_{D}}{(\epsilon_{B}/\epsilon_{I})a+\lambda_{D}}\frac{\beta q \mathcal{E}}{4}\right)e^{-\lambda_{D}^{-1}\abs{z'-a}}\right)\right)\left(\frac{1}{3!}\beta q\avgG{\delta\psi(r)\delta\psi(r')^{3}}\right)\nonumber \\ 
    =\int_{r'>a}\frac{1}{\beta^{2}q}&\left(i\epsilon_{B}\lambda_{D}^{-2}\sinh\left(4\text{arctanh}\left(\tanh\left(\frac{\lambda_{D}}{(\epsilon_{B}/\epsilon_{I})a+\lambda_{D}}\frac{\beta q \mathcal{E}}{4}\right)e^{-\lambda_{D}^{-1}\abs{z'-a}}\right)\right)\right) G(r,r')\left(\frac{\beta q^{2}}{2}G(r',r')\right).
\end{align}
\paragraph{Diagrams of a renormalized vertex.}
The diagrammatics as presented above help calculating Gaussian averages involving polynmials $\delta\psi(r)^m$ explicitly. However, as shown in the calculation of $\avgG{\delta\psi(r)S_I^{(3)}}$ in Eq.~\eqref{AppendixComputationOfElectrostaticPotential} a computation of a renormalized vertex requires several Gaussian averages such as $\avg{\delta\psi(r)\delta\psi(r')}$ and $\avg{\delta\psi(r)\delta\psi(r')^3}$, which makes computation of higher order renormalized vertices cumbersome. However, it was shown in Eq.~\eqref{AppendixBareVertex} that the result of computing a bare (Eq.~\eqref{AppendixBareVertex}) or renormalized vertex Eq.~\eqref{ComputationClover1}, in the end, just differs by a shift of the equal-point correlation function by $\omega$. In other words, mathematically a Gaussian average involving a renormalized vertex can be treated as a Gaussian average of a bare vertex but with replacing all equal-point correlation functions by
\begin{align}
    \frac{\beta q^2}{2}G(r',r')\to\frac{\beta q^2}{2}G(r',r')-\omega.
\end{align}
With this knwoledge, we choose a diagrammatic representation of a renormalized vertex. We diagrammatically represent $S^{(n)}_I$ by an empty dot with $n$ external legs, see for instance $S_I^{(n)}$ in Fig.~\eqref{fig:Feynman1LElectricPotential}. To translate back to an integral expression, the result of the Wick contraction in terms of Green's functions is multiplied by the vertex function and then integration is performed over the internal spatial coordinates. The final vertex determines the spatial variable of the overall expression. For instance, for Fig.~\ref{fig:Feynman1LElectricPotential}, the integral expression is shown in Eq.~\eqref{ComputationClover1}.

\subsection{Details of second-order perturbation theory}\label{Sec:SecondOrder}
To evaluate the accuracy of the 1L approximation, it is useful to examine the next order in the perturbation series. This serves two purposes. First, it shows when the perturbation theory begins to break down (typically when second-order contributions become larger than the first-order term). Second, if the second-order term consists of several parts, we can identify which contribution is dominant.
\paragraph{Computing $\phi^\text{2L}$} To compute the electrostatic potential to second order in pertubation theory, one must expand the exponential in Eq.~\eqref{FullExpansionElectrostaticPotential} up to second order,
\begin{align}\label{AverageHSField}
     \phi^\text{2L}(r) &=i\psi_{c}(r)+i\ell^{-1/2}\avg{\delta\psi(r)} \nonumber \\
     &\approx i\psi_{c}(r)+i\ell^{-1/2}\avgG{\delta\psi(r)e^{-\beta\ell^{-1/2}S_{I}^{(3)}-\beta\ell^{-1}S_{I}^{(4)}-\beta\ell^{-3/2}S_{I}^{(5)}-\beta\ell^{-2}S_{I}^{(6)}}} \nonumber \\
     &=\phi^\text{1L}(r)+i\ell^{-3/2}\left(-\beta\avgG{\delta\psi(r)S_{I}^{(4)}}+\frac{\beta^{2}}{2}\avgG{\delta\psi S_{I}^{(3)}S_{I}^{(3)}}\right)+\nonumber \\
     &+\ell^{-2}\left(-\beta\avgG{\delta\psi(r)S_{I}^{(5)}}+\beta^{2}\avgG{\delta\psi S_{I}^{(3)}S_{I}^{(4)}}+\frac{\beta^{3}}{3!}\avgG{\delta\psi(r)\left(S_{I}^{(3)}\right)^{3}}\right)+\mathcal{O}(\ell^{-5/2}),
\end{align}
with
\begin{align}
    &S_{I}^{(3)}=\Gamma^{(3)}[\left\{\lambda_i^0 \right\}]+\int_{r}\frac{\delta S_{\text{int}}[\psi,{\lambda^0_i \omega}]}{\delta\psi(r)}\at_{\psi=\psi_{c}}\delta\psi(r), \label{RenormalizedVertex1}\\
&S_{I}^{(4)}=\Gamma^{(4)}[\left\{\lambda_i^0 \right\}]+\frac{1}{2}\int_{r,r'}\delta\psi(r)\frac{\delta^{2}S_{\text{int}}[\psi,\left\{\lambda_i^0\omega \right\}]}{\delta\psi(r)\delta\psi(r')}\at_{\psi=\psi_{c}}\delta\psi(r'), \label{AppendixRenormalizedVertex2}\\
&S_{I}^{(5)}=\Gamma^{(5)}[\left\{\lambda_i^0 \right\}]+\Gamma^{(3)}[\left\{\lambda_i^0\omega \right\}]+\int_{r}\delta\psi(r)\frac{\delta S_{\text{int}}[\psi,\lambda^0_i \omega^2/2]}{\delta\psi(r)}\at_{\psi=\psi_{c}}.\label{AppendixRenormalizedVertex3}
\end{align}
All Gaussian averages with an odd number of $\delta\psi$ fields vanish, e.g. $\avgG{\delta\psi(r)S_{I}^{(4)}}=0$, which yields Eq.~\eqref{ElectricPotential2L}.

\paragraph{Computing the capacitance $C^\text{2L}$}
To compute the capacitance, one needs an explicit expression for the electrostatic potential (Eq.~\eqref{RenormalizedVertex1}) first. Starting from Eq.~\eqref{AppendixThirdVariation}, subsequent functional differentiations with respectect to $\delta\psi$ yield higher order vertex functions, cf. Eqs.~\eqref{TheoryVertex} and \eqref{TheoryVertexFunction}. For the second-order expansion of the electrostatic potential one needs $\Gamma^{(3)},\Gamma^{(4)}$ and $\Gamma^{(5)}$, which are given by
\begin{align}
    \Gamma^{(3)}[\left\{\lambda_i^0 \right\}]&=\int_{r>a}(+1)(\beta q)\left(i\epsilon_B\lambda_{D}^{-2}\sinh\left(4\text{arctanh}\left(\tanh\left(\frac{\lambda_{D}}{(\epsilon_{B}/\epsilon_{I})a+\lambda_{D}}\frac{\beta q \mathcal{E}}{4}\right)e^{-\lambda_{D}^{-1}\abs{z-a}}\right)\right)\right) \delta\psi(r)^3,\label{AppendixGamma3_2}\\
    \Gamma^{(4)}[\left\{\lambda_i^0 \right\}]&=\int_{r>a}(-i)(\beta q)^{2}\left(i\epsilon_B\lambda_{D}^{-2}\cosh\left(4\text{arctanh}\left(\tanh\left(\frac{\lambda_{D}}{(\epsilon_{B}/\epsilon_{I})a+\lambda_{D}}\frac{\beta q \mathcal{E}}{4}\right)e^{-\lambda_{D}^{-1}\abs{z-a}}\right)\right)\right) \delta\psi(r)^4,\label{AppendixGamma4}\\
    \Gamma^{(5)}[\left\{\lambda_i^0 \right\}]&=\int_{r>a}(-1)(\beta q)^{3}\left(i\epsilon_B\lambda_{D}^{-2}\sinh\left(4\text{arctanh}\left(\tanh\left(\frac{\lambda_{D}}{(\epsilon_{B}/\epsilon_{I})a+\lambda_{D}}\frac{\beta q \mathcal{E}}{4}\right)e^{-\lambda_{D}^{-1}\abs{z-a}}\right)\right)\right)\delta\psi(r)^5.\label{AppendixGamma5}
\end{align}
Inserting the bare vertices (Eqs.~\eqref{AppendixGamma3_2}, \eqref{AppendixGamma4} and \eqref{AppendixGamma5}) into the expressions for the renormalized vertices (Eqs.~\eqref{RenormalizedVertex1} , \eqref{AppendixRenormalizedVertex2} and \eqref{AppendixRenormalizedVertex3}) yields
\begin{align}
    &S_{I}^{(3)}=+\beta q\int_{r>a}i\epsilon_{B}\lambda_{D}^{-2}\sinh\left(4\text{arctanh}\left(\tanh\left(\frac{\lambda_{D}}{(\epsilon_{B}/\epsilon_{I})a+\lambda_{D}}\frac{\beta q \mathcal{E}}{4}\right)e^{-\lambda_{D}^{-1}\abs{z-a}}\right)\right)\nonumber \\
&\hspace{4cm}\left(\frac{1}{3!}\delta\psi(r)^{3}-\frac{\omega}{\beta^2 q^2}\delta\psi(r)\right), \label{AppendixRenormalizedVertex3Result}\\
&S_{I}^{(4)}=+\beta^{2}q^{2}\int_{r>a}\epsilon_{B}\lambda_{D}^{-2}\cosh\left(4\text{arctanh}\left(\tanh\left(\frac{\lambda_{D}}{(\epsilon_{B}/\epsilon_{I})a+\lambda_{D}}\frac{\beta q \mathcal{E}}{4}\right)e^{-\lambda_{D}^{-1}\abs{z-a}}\right)\right) \nonumber \\
&\hspace{4cm}\left(\frac{1}{4!}\delta\psi(r)^{4}-\frac{1}{2}\frac{\omega}{\beta^{2}q^{2}}\delta\psi(r)^{2}\right), \label{AppendixRenormalizedVertex4Result}\\
&S_{I}^{(5)}=(-1)\beta^{3}q^{3}\int_{r>a}\left(i\epsilon_{B}\lambda_{D}^{-2}\sinh\left(4\text{arctanh}\left(\tanh\left(\frac{\lambda_{D}}{(\epsilon_{B}/\epsilon_{I})a+\lambda_{D}}\frac{\beta q \mathcal{E}}{4}\right)e^{-\lambda_{D}^{-1}\abs{z-a}}\right)\right)\right) \nonumber \\
&\hspace{4cm}\left(\frac{1}{5!}\delta\psi(r)^{5}-\frac{\omega}{3!}\frac{1}{\beta^{2}q^{2}}\delta\psi(r)^{3}+\frac{1}{\beta^{4}q^{4}}\frac{\omega^2}{2}\delta\psi(r)\right).\label{AppendixRenormalizedVertex5Result}
\end{align}
Inserting the explicit expressions for the renormalized vertices above into the expression for the 2L electrostatic potential (Eq.~\eqref{AverageHSField}) and applying Wick's theorem (Sec.~\ref{Section:ExplanationDiagrams}), for instance
\begin{align}
\avg{\delta\psi(r)\delta\psi(r_{1})^{3}}	=3\,\beta^{-1}G(r,r_{1})\,\beta^{-1}G(r_{1},r_{1}), \label{WickTheoremResult1}\\
\avg{\delta\psi(r)\delta\psi(r_{1})^{5}}	=5\,3\beta^{-1}G(r,r_{1})\beta^{-2}G(r_{1},r_{1})^{2},\label{WickTheoremResult2}
\end{align}
yields two different terms that contribute to $\phi^\text{2L}$ apart from the 1L contribution (already computed in Eq.~\eqref{ComputationClover1}):
\begin{align}
    \avg{\delta\psi(r)S_{I}^{(5)}}_{G} &= \int_{r'>a}(+1)\frac{1}{\beta^{2}q}\left(i\epsilon_{B}\lambda_{D}^{-2}\sinh\left(4\text{arctanh}\left(\tanh\left(\frac{\lambda_{D}}{(\epsilon_{B}/\epsilon_{I})a+\lambda_{D}}\frac{\beta\mu_{-}}{4}\right)e^{-\lambda_{D}^{-1}\abs{z'-a}}\right)\right) 
    \right)\nonumber \\
    &G(r,r')\frac{1}{2}\left(\frac{\beta q_{i}^{2}}{2}G(r',r')-\omega\right)^{2}\,, \label{RenormalizedVertex3}\\
    \avg{\delta\psi(r) S_{I}^{(3)}S_{I}^{(4)}}&=i\beta^{3}q^{3}\epsilon_{B}^{2}\lambda_{D}^{-4}\int_{r'>a}\int_{r''>a}\sinh\left(4\text{arctanh}\left(\tanh\left(\frac{\lambda_{D}}{(\epsilon_{B}/\epsilon_{I})a+\lambda_{D}}\frac{\beta\mu_{-}}{4}\right)e^{-\lambda_{D}^{-1}\abs{z'-a}}\right)\right) \nonumber \\
    &\cosh\left(4\text{arctanh}\left(\tanh\left(\frac{\lambda_{D}}{(\epsilon_{B}/\epsilon_{I})a+\lambda_{D}}\frac{\beta\mu_{-}}{4}\right)e^{-\lambda_{D}^{-1}\abs{z''-a}}\right)\right)\Bigg(\frac{1}{6}\beta^{-4}G(r,r'')G(r'',r')^{3} \nonumber \\
    &+G(r,r')G(r',r'')^{2}\frac{1}{2\beta^{5}q^{2}}\left(\frac{\beta q^{2}}{2}G(r'',r'')-\omega\right) \nonumber \\
    &+\frac{1}{\beta^{6}q^{4}}G(r,r'')G(r',r'')\left(\frac{\beta q_{i}^{2}}{2}G(r',r')-\omega\right)\left(\frac{\beta q_{i}^{2}}{2}G(r'',r'')-\omega\right)\Bigg).\label{RenormalizedVertex4}
\end{align}
At this point vaccum diagrams have been discarded as they do not contribute to Gaussian averages. The term $\sim\avgG{\delta\psi(r)\left(S_{I}^{(3)}\right)^{3}}$ from Eq.~\eqref{AverageHSField} is omitted because it does not contribute to the differential capacitance. This is due to the fact that $(S_{I}^{(3)})^3$ is proportional to $\sinh^3(\ldots \mathcal{E})$, which vanishes when taking the derivative with respect to the electrode potential $\mathcal{E}$ at $\mathcal{E} = 0$, as required for computing the capacitance. Inserting Eqs.~\eqref{RenormalizedVertex3} and Eq.~\eqref{RenormalizedVertex4} into Eq.~\eqref{AverageHSField} and using the gauge chemical potential in Eq.~\eqref{ChosenReferenceState} yields the electrostatic potential in 2L approximation. Further linearising the $\cosh$ and $\sinh$ functions yields the electrostatic potential in 2L approximation around PZC,
\begin{align}\label{ElectricPotential2LResult}
     &\phi^\text{2L}(r)=\phi^{\text{MF}}(r)+\Delta\phi^{\text{1-clover }}(r)+\Delta\phi^{\text{2-clover}}(r)+\Delta\phi^{\text{2-rest}}(r),
\end{align}
with 
\begin{align}\label{DiagramAsIntegrals}
    &\phi^{\text{MF}}(r)=i\psi_{c}(r),\\
    &\Delta\phi^{\text{1-clover }}(r)=\frac{\epsilon_{B}}{\lambda_{D}^{2}}\frac{\lambda_{D}}{(\epsilon_{B}/\epsilon_{I})a+\lambda_{D}}\mathcal{E}\int_{r'>a}e^{-\lambda_{D}^{-1}\abs{z'-a}}G(r,r')\left(\beta V_\text{int}(r')\right),\\&\Delta\phi^{\text{2-clover}}(r)=-\frac{\epsilon_{B}}{\lambda_{D}^{2}}\frac{\lambda_{D}}{(\epsilon_{B}/\epsilon_{I})a+\lambda_{D}}\mathcal{E}\int_{r'>a}e^{-\lambda_{D}^{-1}\abs{z'-a}}G(r,r')\frac{1}{2}\left(\beta V_\text{int}(r')\right)^{2},\\&\Delta\phi^{\text{2-rest}}(r)=\frac{\epsilon_{B}^{2}}{\lambda_{D}^{4}}\frac{\lambda_{D}}{(\epsilon_{B}/\epsilon_{I})a+\lambda_{D}}\mathcal{E}\int_{r'>a}\int_{r''>a}e^{-\lambda_{D}^{-1}\abs{z'-a}}\Bigg(\frac{1}{6}\beta^{2}q^{4}G(r,r'')G(r',r'')^{3} \nonumber \\
    &+\frac{\beta q^{2}}{2}G(r,r')G(r',r'')^{2}\left(\beta V_\text{int}(r'')\right)+G(r,r'')G(r'',r')\left(\beta V_\text{int}(r')\right)\left(\beta V_\text{int}(r'')\right)\Bigg).\label{AppendixRest}
\end{align}
Applying the same reasoning as in the 1L calculation, the metallic surface charge is evaluated using Eq.~\eqref{MetalChargeDensity}, together with
\begin{align}
    \frac{\partial}{\partial z}\left(e^{-\abs{z-z'}/\lambda_{D}}-e^{-(z+z')/\lambda_{D}}\right)\at_{z=0}=2\frac{e^{-z'/\lambda_{D}}}{\lambda_{D}},
\end{align}
and subsequently the differential capacitance using Eq.~\eqref{Defn:DifferentialCapacitance} yields
\begin{align}
    C_\text{2L}&=R_\text{2L}\times C_\text{GCS}.  
\end{align}
The interface correlation factor on a 2L level,
\begin{align}
    R_\text{2L} = 1+\Delta R_{\text{1-clover}}+\Delta R_{\text{2-clover}}+\Delta R_{\text{2-rest}},
\end{align}
consist of three different contributions $\Delta R_{\text{1-clover}},\;\Delta R_{\text{2-clover}}\;\text{and}\;\Delta R_{\text{2-rest}}$, which are given by
\begin{align}\label{Roughness2Loop}
    R_{\text{1-clover}}&=-\frac{\beta}{\lambda_{D}}\int_{a}^{\infty}dz\,e^{-2z/\lambda_{D}}V_{\text{int}}(z),\\R_{\text{2-clover}}&=\frac{1}{2}\frac{\beta^{2}}{\lambda_{D}}\int_{a}^{\infty}dz\, e^{-2z/\lambda_D}V_{\text{int}}(z)^{2},\\R_{\text{2-rest}}&=-\frac{1}{2}\frac{\beta^2}{\lambda_D^2}\int_{a}^{\infty}d\tilde{z}'\int_{a}^{\infty}d\tilde{z}''\Bigg(\frac{2}{3}\frac{q^{4}\pi \epsilon_B }{\lambda_D}e^{-(z'+z'')/\lambda_D}\int_{0}^{\infty}d\rho'\rho'G(\rho',z',z'')^{3}+\nonumber\\&+2\frac{q^{2}\pi \epsilon_B}{\lambda_D}e^{-2z/\lambda_D'}\int_{0}^{\infty}d\rho''\rho'' G(\rho'',z',z'')^{2}V_{\text{int}}(z'')\nonumber\\&+e^{(-z'+z'')/\lambda_D}\left(e^{-\abs{z'-z''}/\lambda_{D}}-e^{-(z'+z'')/\lambda_{D}}\right)V_{\text{int}}(z')V_{\text{int}}(z'')\Bigg).
\end{align}
\subsection{Computing the electrostatic potential in Clover apprxoximation given in  Eq.~\eqref{ElectrostaticPotentialClover}.}\label{Appendix:Clover}
It is possible to compute all averages in Eq.~\eqref{ElectrostaticPotentialSeries} as they follow a repetitive pattern. The first term to compute ($n=3$) is given by the 1L results that has been computed in Eq.~\eqref{ComputationClover1},
\begin{align}
     \avg{\delta\psi(r)S_{I}^{(3)}}=\int_{r'>a}(+1)\frac{i\epsilon_{B}}{\beta^{2}q\lambda_{D}^{2}}\sinh\left(4\text{arctanh}\left(\tanh\left(\frac{\lambda_{D}}{(\epsilon_{B}/\epsilon_{I})a+\lambda_{D}}\frac{\beta q\mathcal{E}}{4}\right)e^{-\lambda_{D}^{-1}\abs{z'-a}}\right)\right)G(r,r')\left(\beta V_\text{int}(z')\right).
\end{align}
Similarly, from the 2L computations (Eq.~\eqref{RenormalizedVertex3}) it is known that
\begin{align}
    \avg{\delta\psi(r)S_{I}^{(5)}}=\int_{r'>a}(-1)\frac{i\epsilon_{B}}{\beta^{2}q\lambda_{D}^{2}}\sinh\left(4\text{arctanh}\left(\tanh\left(\frac{\lambda_{D}}{(\epsilon_{B}/\epsilon_{I})a+\lambda_{D}}\frac{\beta q\mathcal{E}}{4}\right)e^{-\lambda_{D}^{-1}\abs{z'-a}}\right)\right)G(r,r')\frac{1}{2}\left(\beta V_\text{int}(z')\right)^{2}.
\end{align}
From this one deduces that $n^\text{th}$ order of the clover expansion is given by 
\begin{align}\label{nThOrderWick}
    \avg{\delta\psi(r)S_{I}^{(2n+1)}} = -\int_{r'>a}\frac{i\epsilon_{B}}{\beta^{2}q\lambda_{D}^{2}}\sinh\left(4\text{arctanh}\left(\tanh\left(\frac{\lambda_{D}}{(\epsilon_{B}/\epsilon_{I})a+\lambda_{D}}\frac{\beta q\mathcal{E}}{4}\right)e^{-\lambda_{D}^{-1}\abs{z'-a}}\right)\right)G(r,r')\frac{(-1)^n}{n!}\left(\beta V_\text{int}(z')\right)^{n}.
\end{align}
Upon inserting Eqs.~\eqref{nThOrderWick} into Eq.~\eqref{ElectrostaticPotentialSeries} and using
\begin{align}
    \sum_{n=1}^{\infty}\frac{\left(-x\right)^{n}}{n!}=e^{-x}-1
\end{align}
one arrives at Eq.~\eqref{ElectrostaticPotentialClover}.
\section{Alternative model parameters for the comparison to experimental data}
The model parameters $\epsilon_I$ and $a$ are not independent when determined by comparison to experimental data. Several combinations of the respective parameter values exists that give similar results for the capacitance curves. In addition to the parameter values shown in Tab.~\ref{tab:Data}, which were used for the plots in Figs.~\ref{fig:CapacitanceAqueous} and \ref{fig:CapacitanceNonAqueous}, the parameter values shown in Tab.~\ref{tab:fitValues} yield qualitatively similar results. The parameters presented in Tab.~\ref{tab:Data} correspond to the first column in Tab.~\ref{tab:fitValues}. When the value of $\epsilon_I$ is increased, the required value for $a$ also increases. For instance, for the mercury electrode, the choice of $\epsilon_I=20$ requires a value of $a=1.8\,\text{\AA}$, while $\epsilon_I=80$ leads to $a=8\,\text{\AA}$. This can be explained by the fact that the desolvation effect is stronger at lower $\epsilon_I$, which requires a lower value for $a$ to enhance the attractive effect of electron--ion correlations. Interestingly, an approximately linear relationship between the values of $\epsilon_I$ and $a$ is observed, which means that the ratio of $\epsilon_I/a$ is the key parameter in controlling the model fitting to the experimental data.

\begin{table*}[h]
    \caption{Various combinations of parameter values that achive qualitatively similar model curves as those shown in Figs.~\ref{fig:CapacitanceAqueous} and \ref{fig:CapacitanceNonAqueous}.}  
	\begin{ruledtabular}
		\begin{tabular}{l|llll}\label{tab:fitValues}
			  & $a$ for $\epsilon_I=20$ & $a$ for $\epsilon_I=40$  & $a$ for $\epsilon_I=60$ & $a$ for $\epsilon_I=80$  \\
			\hline
            \multicolumn{5}{l}{Aqueous Electrolytes / Hydrophobic to hydrophilic interfaces from top to bottom} \\
            Mercury-NaF \cite{grahameDifferentialCapacityMercury1954} & $1.8\,\text{\AA}$ & $4\,\text{\AA}$  & $6\,\text{\AA}$ & $8\,\text{\AA}$ \\
            Ag(111)-KPF$_6$ \cite{valetteDoubleLayerSilver1989} & $0.3\,\text{\AA}$ & $0.6\,\text{\AA}$  & $1\,\text{\AA}$ & $1.7\,\text{\AA}$ \\
            Au(111)-KPF$_6$ \cite{hamelinStudyGoldLow1987}& $0.1\,\text{\AA}$ & $0.21\,\text{\AA}$  & $0.36\,\text{\AA}$ & $0.7\,\text{\AA}$ \\
            Pt(111)-HClO$_4$ \cite{ojhaDoubleLayerPt111Aqueous2020} & $0.035\,\text{\AA}$ & $0.07\,\text{\AA}$  & $0.13\,\text{\AA}$ & $0.27\,\text{\AA}$ \\
            \multicolumn{4}{l}{} \\
            & $a$ for $\epsilon_I=0.25\epsilon_B$ & $a$ for $\epsilon_I=\epsilon_B$  & $\epsilon_B$ \\
			\hline
            \multicolumn{5}{l}{Organic Electrolytes } \\
            Au(111)-KPF$_6$ DMSO \cite{shatlaDifferentialCapacitancePotential2021}& $0.2\,\text{\AA}$ & $1.3\,\text{\AA}$ & 46.7  \\
            Au(111)-KPF$_6$ ACN \cite{shatlaDifferentialCapacitancePotential2021}& $0.25\,\text{\AA}$ & $1.5\,\text{\AA}$ &37.4   \\
            Au(111)-KPF$_6$ DG \cite{shatlaDifferentialCapacitancePotential2021}& $1.4\,\text{\AA}$ & $6\,\text{\AA}$ & 7.23  \\
		\end{tabular}
	\end{ruledtabular}
\end{table*}

\twocolumngrid


\begin{thebibliography}{66}%
	\makeatletter
	\providecommand \@ifxundefined [1]{%
		\@ifx{#1\undefined}
	}%
	\providecommand \@ifnum [1]{%
		\ifnum #1\expandafter \@firstoftwo
		\else \expandafter \@secondoftwo
		\fi
	}%
	\providecommand \@ifx [1]{%
		\ifx #1\expandafter \@firstoftwo
		\else \expandafter \@secondoftwo
		\fi
	}%
	\providecommand \natexlab [1]{#1}%
	\providecommand \enquote  [1]{``#1''}%
	\providecommand \bibnamefont  [1]{#1}%
	\providecommand \bibfnamefont [1]{#1}%
	\providecommand \citenamefont [1]{#1}%
	\providecommand \href@noop [0]{\@secondoftwo}%
	\providecommand \href [0]{\begingroup \@sanitize@url \@href}%
	\providecommand \@href[1]{\@@startlink{#1}\@@href}%
	\providecommand \@@href[1]{\endgroup#1\@@endlink}%
	\providecommand \@sanitize@url [0]{\catcode `\\12\catcode `\$12\catcode
		`\&12\catcode `\#12\catcode `\^12\catcode `\_12\catcode `\%12\relax}%
	\providecommand \@@startlink[1]{}%
	\providecommand \@@endlink[0]{}%
	\providecommand \url  [0]{\begingroup\@sanitize@url \@url }%
	\providecommand \@url [1]{\endgroup\@href {#1}{\urlprefix }}%
	\providecommand \urlprefix  [0]{URL }%
	\providecommand \Eprint [0]{\href }%
	\providecommand \doibase [0]{https://doi.org/}%
	\providecommand \selectlanguage [0]{\@gobble}%
	\providecommand \bibinfo  [0]{\@secondoftwo}%
	\providecommand \bibfield  [0]{\@secondoftwo}%
	\providecommand \translation [1]{[#1]}%
	\providecommand \BibitemOpen [0]{}%
	\providecommand \bibitemStop [0]{}%
	\providecommand \bibitemNoStop [0]{.\EOS\space}%
	\providecommand \EOS [0]{\spacefactor3000\relax}%
	\providecommand \BibitemShut  [1]{\csname bibitem#1\endcsname}%
	\let\auto@bib@innerbib\@empty
	\bibitem [{\citenamefont {Stern}(1924)}]{sternZurTheorieElektrolytischen1924}%
	\BibitemOpen
	\bibfield  {author} {\bibinfo {author} {\bibfnamefont {O.}~\bibnamefont
			{Stern}},\ }\bibfield  {title} {\bibinfo {title} {Zur {{Theorie Der
					Elektrolytischen Doppelschicht}}},\ }\href
	{https://doi.org/10.1002/bbpc.192400182} {\bibfield  {journal} {\bibinfo
			{journal} {Z. Elektrochem.}\ }\textbf {\bibinfo {volume} {30}},\ \bibinfo
		{pages} {508} (\bibinfo {year} {1924})}\BibitemShut {NoStop}%
	\bibitem [{\citenamefont {Shen}\ \emph {et~al.}(2022)\citenamefont {Shen},
		\citenamefont {Zhang}, \citenamefont {Hu}, \citenamefont {Zhang},
		\citenamefont {Zhou}, \citenamefont {Qian}, \citenamefont {Lu},\ and\
		\citenamefont {Ji}}]{shenEffectSurfaceRoughness2022}%
	\BibitemOpen
	\bibfield  {author} {\bibinfo {author} {\bibfnamefont {G.}~\bibnamefont
			{Shen}}, \bibinfo {author} {\bibfnamefont {D.}~\bibnamefont {Zhang}},
		\bibinfo {author} {\bibfnamefont {Y.}~\bibnamefont {Hu}}, \bibinfo {author}
		{\bibfnamefont {X.}~\bibnamefont {Zhang}}, \bibinfo {author} {\bibfnamefont
			{F.}~\bibnamefont {Zhou}}, \bibinfo {author} {\bibfnamefont {Y.}~\bibnamefont
			{Qian}}, \bibinfo {author} {\bibfnamefont {X.}~\bibnamefont {Lu}},\ and\
		\bibinfo {author} {\bibfnamefont {X.}~\bibnamefont {Ji}},\ }\bibfield
	{title} {\bibinfo {title} {Effect of surface roughness on partition of ionic
			liquids in nanopores by a perturbed-chain {{SAFT}} density functional
			theory},\ }\href {https://doi.org/10.1063/5.0098924} {\bibfield  {journal}
		{\bibinfo  {journal} {J. Chem. Phys.}\ }\textbf {\bibinfo {volume} {157}},\
		\bibinfo {pages} {014701} (\bibinfo {year} {2022})}\BibitemShut {NoStop}%
	\bibitem [{\citenamefont {Ojha}\ \emph {et~al.}(2020)\citenamefont {Ojha},
		\citenamefont {Arulmozhi}, \citenamefont {Aranzales},\ and\ \citenamefont
		{Koper}}]{ojhaDoubleLayerPt111Aqueous2020}%
	\BibitemOpen
	\bibfield  {author} {\bibinfo {author} {\bibfnamefont {K.}~\bibnamefont
			{Ojha}}, \bibinfo {author} {\bibfnamefont {N.}~\bibnamefont {Arulmozhi}},
		\bibinfo {author} {\bibfnamefont {D.}~\bibnamefont {Aranzales}},\ and\
		\bibinfo {author} {\bibfnamefont {M.~T.~M.}\ \bibnamefont {Koper}},\
	}\bibfield  {title} {\bibinfo {title} {Double {{Layer}} at the
			{{Pt}}(111)--{{Aqueous Electrolyte Interface}}: {{Potential}} of {{Zero
					Charge}} and {{Anomalous Gouy}}--{{Chapman Screening}}},\ }\href
	{https://doi.org/10.1002/anie.201911929} {\bibfield  {journal} {\bibinfo
			{journal} {Angew. Chem., Int. Ed.}\ }\textbf {\bibinfo {volume} {59}},\
		\bibinfo {pages} {711} (\bibinfo {year} {2020})}\BibitemShut {NoStop}%
	\bibitem [{\citenamefont {Mostany}\ \emph {et~al.}(2003)\citenamefont
		{Mostany}, \citenamefont {Herrero}, \citenamefont {Feliu},\ and\
		\citenamefont {Lipkowski}}]{mostanyDeterminationGibbsExcess2003}%
	\BibitemOpen
	\bibfield  {author} {\bibinfo {author} {\bibfnamefont {J.}~\bibnamefont
			{Mostany}}, \bibinfo {author} {\bibfnamefont {E.}~\bibnamefont {Herrero}},
		\bibinfo {author} {\bibfnamefont {J.~M.}\ \bibnamefont {Feliu}},\ and\
		\bibinfo {author} {\bibfnamefont {J.}~\bibnamefont {Lipkowski}},\ }\bibfield
	{title} {\bibinfo {title} {Determination of the {{Gibbs}} excess of {{H}} and
			{{OH}} adsorbed at a {{Pt}}(1 ~ 1 ~ 1) electrode surface using a
			thermodynamic method},\ }\href
	{https://doi.org/10.1016/S0022-0728(03)00372-3} {\bibfield  {journal}
		{\bibinfo  {journal} {J. Electroanal. Chem.}\ }\textbf {\bibinfo {volume}
			{558}},\ \bibinfo {pages} {19} (\bibinfo {year} {2003})}\BibitemShut
	{NoStop}%
	\bibitem [{\citenamefont
		{Schmickler}(2021)}]{schmicklerEffectWeakAdsorption2021}%
	\BibitemOpen
	\bibfield  {author} {\bibinfo {author} {\bibfnamefont {W.}~\bibnamefont
			{Schmickler}},\ }\bibfield  {title} {\bibinfo {title} {The {{Effect}} of
			{{Weak Adsorption}} on the {{Double Layer Capacitance}}},\ }\href
	{https://doi.org/10.1002/celc.202100842} {\bibfield  {journal} {\bibinfo
			{journal} {ChemElectroChem}\ }\textbf {\bibinfo {volume} {8}},\ \bibinfo
		{pages} {4218} (\bibinfo {year} {2021})}\BibitemShut {NoStop}%
	\bibitem [{\citenamefont {{Doblhoff-Dier}}\ and\ \citenamefont
		{Koper}(2021)}]{doblhoff-dierModelingGouyChapman2021}%
	\BibitemOpen
	\bibfield  {author} {\bibinfo {author} {\bibfnamefont {K.}~\bibnamefont
			{{Doblhoff-Dier}}}\ and\ \bibinfo {author} {\bibfnamefont {M.~T.~M.}\
			\bibnamefont {Koper}},\ }\bibfield  {title} {\bibinfo {title} {Modeling the
			{{Gouy}}--{{Chapman Diffuse Capacitance}} with {{Attractive Ion}}--{{Surface
					Interaction}}},\ }\href {https://doi.org/10.1021/acs.jpcc.1c02381} {\bibfield
		{journal} {\bibinfo  {journal} {J. Phys. Chem. C}\ }\textbf {\bibinfo
			{volume} {125}},\ \bibinfo {pages} {16664} (\bibinfo {year}
		{2021})}\BibitemShut {NoStop}%
	\bibitem [{\citenamefont {Huang}(2023)}]{huangZoomingInnerHelmholtz2023}%
	\BibitemOpen
	\bibfield  {author} {\bibinfo {author} {\bibfnamefont {J.}~\bibnamefont
			{Huang}},\ }\bibfield  {title} {\bibinfo {title} {Zooming into the {{Inner
					Helmholtz Plane}} of {{Pt}}(111)--{{Aqueous Solution Interfaces}}:
			{{Chemisorbed Water}} and {{Partially Charged Ions}}},\ }\href
	{https://doi.org/10.1021/jacsau.2c00650} {\bibfield  {journal} {\bibinfo
			{journal} {JACS Au}\ }\textbf {\bibinfo {volume} {3}},\ \bibinfo {pages}
		{550} (\bibinfo {year} {2023})}\BibitemShut {NoStop}%
	\bibitem [{\citenamefont {{Doblhoff-Dier}}\ and\ \citenamefont
		{Koper}(2023)}]{doblhoff-dierElectricDoubleLayer2023}%
	\BibitemOpen
	\bibfield  {author} {\bibinfo {author} {\bibfnamefont {K.}~\bibnamefont
			{{Doblhoff-Dier}}}\ and\ \bibinfo {author} {\bibfnamefont {M.~T.~M.}\
			\bibnamefont {Koper}},\ }\bibfield  {title} {\bibinfo {title} {Electric
			double layer of {{Pt}}(111): {{Known}} unknowns and unknown knowns},\ }\href
	{https://doi.org/10.1016/j.coelec.2023.101258} {\bibfield  {journal}
		{\bibinfo  {journal} {Curr. Opin. Electrochem.}\ }\textbf {\bibinfo {volume}
			{39}},\ \bibinfo {pages} {101258} (\bibinfo {year} {2023})}\BibitemShut
	{NoStop}%
	\bibitem [{\citenamefont {Bruch}\ \emph
		{et~al.}(2024{\natexlab{a}})\citenamefont {Bruch}, \citenamefont {Binninger},
		\citenamefont {Huang},\ and\ \citenamefont
		{Eikerling}}]{bruchIncorporatingElectrolyteCorrelation2024}%
	\BibitemOpen
	\bibfield  {author} {\bibinfo {author} {\bibfnamefont {N.}~\bibnamefont
			{Bruch}}, \bibinfo {author} {\bibfnamefont {T.}~\bibnamefont {Binninger}},
		\bibinfo {author} {\bibfnamefont {J.}~\bibnamefont {Huang}},\ and\ \bibinfo
		{author} {\bibfnamefont {M.}~\bibnamefont {Eikerling}},\ }\bibfield  {title}
	{\bibinfo {title} {Incorporating {{Electrolyte Correlation Effects}} into
			{{Variational Models}} of {{Electrochemical Interfaces}}},\ }\href
	{https://doi.org/10.1021/acs.jpclett.3c03295} {\bibfield  {journal} {\bibinfo
			{journal} {J. Phys. Chem. Lett.}\ ,\ \bibinfo {pages} {2015}} (\bibinfo
		{year} {2024}{\natexlab{a}})}\BibitemShut {NoStop}%
	\bibitem [{\citenamefont
		{Gidopoulos}(1998)}]{gidopoulosKohnShamEquationsMulticomponent1998}%
	\BibitemOpen
	\bibfield  {author} {\bibinfo {author} {\bibfnamefont {N.}~\bibnamefont
			{Gidopoulos}},\ }\bibfield  {title} {\bibinfo {title} {Kohn-{{Sham}}
			equations for multicomponent systems: {{The}} exchange and correlation energy
			functional},\ }\href {https://doi.org/10.1103/PhysRevB.57.2146} {\bibfield
		{journal} {\bibinfo  {journal} {Phys. Rev. B}\ }\textbf {\bibinfo {volume}
			{57}},\ \bibinfo {pages} {2146} (\bibinfo {year} {1998})}\BibitemShut
	{NoStop}%
	\bibitem [{\citenamefont
		{Binninger}(2023)}]{binningerFirstprinciplesTheoryElectrochemical2023}%
	\BibitemOpen
	\bibfield  {author} {\bibinfo {author} {\bibfnamefont {T.}~\bibnamefont
			{Binninger}},\ }\bibfield  {title} {\bibinfo {title} {First-principles theory
			of electrochemical capacitance},\ }\href
	{https://doi.org/10.1016/j.electacta.2023.142016} {\bibfield  {journal}
		{\bibinfo  {journal} {Electrochim. Acta}\ }\textbf {\bibinfo {volume}
			{444}},\ \bibinfo {pages} {142016} (\bibinfo {year} {2023})}\BibitemShut
	{NoStop}%
	\bibitem [{\citenamefont
		{Jackson}(1999)}]{jacksonClassicalElectrodynamics1999}%
	\BibitemOpen
	\bibfield  {author} {\bibinfo {author} {\bibfnamefont {J.~D.}\ \bibnamefont
			{Jackson}},\ }\href@noop {} {\emph {\bibinfo {title} {Classical
				Electrodynamics}}},\ \bibinfo {edition} {3rd}\ ed.\ (\bibinfo  {publisher}
	{Wiley},\ \bibinfo {address} {New York, NY},\ \bibinfo {year}
	{1999})\BibitemShut {NoStop}%
	\bibitem [{\citenamefont {Kornyshev}\ \emph {et~al.}(1977)\citenamefont
		{Kornyshev}, \citenamefont {Rubinshtein},\ and\ \citenamefont
		{Vorotyntsev}}]{kornyshevImagePotentialDielectric1977}%
	\BibitemOpen
	\bibfield  {author} {\bibinfo {author} {\bibfnamefont {A.~A.}\ \bibnamefont
			{Kornyshev}}, \bibinfo {author} {\bibfnamefont {A.~I.}\ \bibnamefont
			{Rubinshtein}},\ and\ \bibinfo {author} {\bibfnamefont {M.~A.}\ \bibnamefont
			{Vorotyntsev}},\ }\bibfield  {title} {\bibinfo {title} {Image potential near
			a dielectric--plasma-like medium interface},\ }\href
	{https://doi.org/10.1002/pssb.2220840114} {\bibfield  {journal} {\bibinfo
			{journal} {physica status solidi (b)}\ }\textbf {\bibinfo {volume} {84}},\
		\bibinfo {pages} {125} (\bibinfo {year} {1977})}\BibitemShut {NoStop}%
	\bibitem [{\citenamefont {Hedley}\ \emph {et~al.}(2025)\citenamefont {Hedley},
		\citenamefont {Bhatt}, \citenamefont {Berthoumieux},\ and\ \citenamefont
		{Kornyshev}}]{hedleyWhatDoesIon2025}%
	\BibitemOpen
	\bibfield  {author} {\bibinfo {author} {\bibfnamefont {J.~G.}\ \bibnamefont
			{Hedley}}, \bibinfo {author} {\bibfnamefont {K.~K.}\ \bibnamefont {Bhatt}},
		\bibinfo {author} {\bibfnamefont {H.}~\bibnamefont {Berthoumieux}},\ and\
		\bibinfo {author} {\bibfnamefont {A.~A.}\ \bibnamefont {Kornyshev}},\
	}\bibfield  {title} {\bibinfo {title} {What does an ion feel at the
			electrochemical interface? {{Revisiting}} electrosorption through nonlocal
			electrostatics},\ }\href {https://doi.org/10.1063/5.0254033} {\bibfield
		{journal} {\bibinfo  {journal} {J. Chem. Phys.}\ }\textbf {\bibinfo {volume}
			{162}},\ \bibinfo {pages} {114703} (\bibinfo {year} {2025})}\BibitemShut
	{NoStop}%
	\bibitem [{\citenamefont
		{Attard}(1996)}]{attardElectrolytesElectricDouble1996}%
	\BibitemOpen
	\bibfield  {author} {\bibinfo {author} {\bibfnamefont {P.}~\bibnamefont
			{Attard}},\ }\bibfield  {title} {\bibinfo {title} {Electrolytes and the
			{{Electric Double Layer}}},\ }in\ \href
	{https://doi.org/10.1002/9780470141519.ch1} {\emph {\bibinfo {booktitle}
			{Advances in {{Chemical Physics}}}}}\ (\bibinfo  {publisher} {John Wiley \&
		Sons, Ltd},\ \bibinfo {year} {1996})\ pp.\ \bibinfo {pages}
	{1--159}\BibitemShut {NoStop}%
	\bibitem [{\citenamefont {Torrie}\ \emph {et~al.}(1982)\citenamefont {Torrie},
		\citenamefont {Valleau},\ and\ \citenamefont
		{Patey}}]{torrieElectricalDoubleLayers1982}%
	\BibitemOpen
	\bibfield  {author} {\bibinfo {author} {\bibfnamefont {G.~M.}\ \bibnamefont
			{Torrie}}, \bibinfo {author} {\bibfnamefont {J.~P.}\ \bibnamefont
			{Valleau}},\ and\ \bibinfo {author} {\bibfnamefont {G.~N.}\ \bibnamefont
			{Patey}},\ }\bibfield  {title} {\bibinfo {title} {Electrical double layers.
			{{II}}. {{Monte Carlo}} and {{HNC}} studies of image effects},\ }\href
	{https://doi.org/10.1063/1.443541} {\bibfield  {journal} {\bibinfo  {journal}
			{J. Chem. Phys.}\ }\textbf {\bibinfo {volume} {76}},\ \bibinfo {pages} {4615}
		(\bibinfo {year} {1982})}\BibitemShut {NoStop}%
	\bibitem [{\citenamefont {Henderson~*}\ \emph {et~al.}(2005)\citenamefont
		{Henderson~*}, \citenamefont {Gillespie}, \citenamefont {Nagy},\ and\
		\citenamefont {Boda}}]{henderson*MonteCarloSimulation2005}%
	\BibitemOpen
	\bibfield  {author} {\bibinfo {author} {\bibfnamefont {D.}~\bibnamefont
			{Henderson~*}}, \bibinfo {author} {\bibfnamefont {D.}~\bibnamefont
			{Gillespie}}, \bibinfo {author} {\bibfnamefont {{\relax
					T{\'i}}.}~\bibnamefont {Nagy}},\ and\ \bibinfo {author} {\bibfnamefont
			{D.}~\bibnamefont {Boda}},\ }\bibfield  {title} {\bibinfo {title} {Monte
			{{Carlo}} simulation of the electric double layer: Dielectric boundaries and
			the effects of induced charge},\ }\href
	{https://doi.org/10.1080/00268970500108668} {\bibfield  {journal} {\bibinfo
			{journal} {Mol. Phys.}\ }\textbf {\bibinfo {volume} {103}},\ \bibinfo {pages}
		{2851} (\bibinfo {year} {2005})}\BibitemShut {NoStop}%
	\bibitem [{\citenamefont {Geada}\ \emph {et~al.}(2018)\citenamefont {Geada},
		\citenamefont {{Ramezani-Dakhel}}, \citenamefont {Jamil}, \citenamefont
		{Sulpizi},\ and\ \citenamefont {Heinz}}]{geadaInsightInducedCharges2018}%
	\BibitemOpen
	\bibfield  {author} {\bibinfo {author} {\bibfnamefont {I.~L.}\ \bibnamefont
			{Geada}}, \bibinfo {author} {\bibfnamefont {H.}~\bibnamefont
			{{Ramezani-Dakhel}}}, \bibinfo {author} {\bibfnamefont {T.}~\bibnamefont
			{Jamil}}, \bibinfo {author} {\bibfnamefont {M.}~\bibnamefont {Sulpizi}},\
		and\ \bibinfo {author} {\bibfnamefont {H.}~\bibnamefont {Heinz}},\ }\bibfield
	{title} {\bibinfo {title} {Insight into induced charges at metal surfaces
			and biointerfaces using a polarizable {{Lennard}}--{{Jones}} potential},\
	}\href {https://doi.org/10.1038/s41467-018-03137-8} {\bibfield  {journal}
		{\bibinfo  {journal} {Nat Commun}\ }\textbf {\bibinfo {volume} {9}},\
		\bibinfo {pages} {716} (\bibinfo {year} {2018})}\BibitemShut {NoStop}%
	\bibitem [{\citenamefont {Ntim}\ and\ \citenamefont
		{Sulpizi}(2023)}]{ntimMolecularDynamicsSimulations2023}%
	\BibitemOpen
	\bibfield  {author} {\bibinfo {author} {\bibfnamefont {S.}~\bibnamefont
			{Ntim}}\ and\ \bibinfo {author} {\bibfnamefont {M.}~\bibnamefont {Sulpizi}},\
	}\bibfield  {title} {\bibinfo {title} {Molecular dynamics simulations of
			electrified interfaces including the metal polarisation},\ }\href
	{https://doi.org/10.1039/D3CP01472J} {\bibfield  {journal} {\bibinfo
			{journal} {Phys. Chem. Chem. Phys.}\ }\textbf {\bibinfo {volume} {25}},\
		\bibinfo {pages} {22619} (\bibinfo {year} {2023})}\BibitemShut {NoStop}%
	\bibitem [{\citenamefont {Son}\ and\ \citenamefont
		{Wang}(2021)}]{sonImagechargeEffectsIon2021}%
	\BibitemOpen
	\bibfield  {author} {\bibinfo {author} {\bibfnamefont {C.~Y.}\ \bibnamefont
			{Son}}\ and\ \bibinfo {author} {\bibfnamefont {Z.-G.}\ \bibnamefont {Wang}},\
	}\bibfield  {title} {\bibinfo {title} {Image-charge effects on ion adsorption
			near aqueous interfaces},\ }\href {https://doi.org/10.1073/pnas.2020615118}
	{\bibfield  {journal} {\bibinfo  {journal} {Proc. Natl. Acad. Sci. U.S.A.}\
		}\textbf {\bibinfo {volume} {118}},\ \bibinfo {pages} {e2020615118} (\bibinfo
		{year} {2021})}\BibitemShut {NoStop}%
	\bibitem [{\citenamefont {Levine}\ and\ \citenamefont
		{Bell}(1960)}]{levineTheoryModifiedPoissonBoltzmann1960}%
	\BibitemOpen
	\bibfield  {author} {\bibinfo {author} {\bibfnamefont {S.}~\bibnamefont
			{Levine}}\ and\ \bibinfo {author} {\bibfnamefont {{\relax GM}.}~\bibnamefont
			{Bell}},\ }\bibfield  {title} {\bibinfo {title} {Theory of a modified
			{{Poisson-Boltzmann}} equation. {{I}}. {{The}} volume effect of hydrated
			ions},\ }\href@noop {} {\bibfield  {journal} {\bibinfo  {journal} {J. Phys.
				Chem.}\ }\textbf {\bibinfo {volume} {64}},\ \bibinfo {pages} {1188} (\bibinfo
		{year} {1960})}\BibitemShut {NoStop}%
	\bibitem [{\citenamefont {Carnie}\ and\ \citenamefont
		{Torrie}(1984)}]{carnieStatisticalMechanicsElectrical1984}%
	\BibitemOpen
	\bibfield  {author} {\bibinfo {author} {\bibfnamefont {S.~L.}\ \bibnamefont
			{Carnie}}\ and\ \bibinfo {author} {\bibfnamefont {G.~M.}\ \bibnamefont
			{Torrie}},\ }\bibfield  {title} {\bibinfo {title} {The {{Statistical
					Mechanics}} of the {{Electrical Double Layer}}},\ }in\ \href
	{https://doi.org/10.1002/9780470142806.ch2} {\emph {\bibinfo {booktitle}
			{Advances in {{Chemical Physics}}}}}\ (\bibinfo  {publisher} {John Wiley \&
		Sons, Ltd},\ \bibinfo {year} {1984})\ pp.\ \bibinfo {pages}
	{141--253}\BibitemShut {NoStop}%
	\bibitem [{\citenamefont
		{Outhwaite}(1970)}]{outhwaiteModifiedPoissonboltzmannEquation1970}%
	\BibitemOpen
	\bibfield  {author} {\bibinfo {author} {\bibfnamefont {C.~W.}\ \bibnamefont
			{Outhwaite}},\ }\bibfield  {title} {\bibinfo {title} {A modified
			poisson-boltzmann equation in the double layer},\ }\href
	{https://doi.org/10.1016/0009-2614(70)87027-0} {\bibfield  {journal}
		{\bibinfo  {journal} {Chem. Phys. Lett.}\ }\textbf {\bibinfo {volume} {7}},\
		\bibinfo {pages} {636} (\bibinfo {year} {1970})}\BibitemShut {NoStop}%
	\bibitem [{\citenamefont {Bell}\ and\ \citenamefont
		{Rangecroft}(1972)}]{bellLinearizedPotentialEquation1972}%
	\BibitemOpen
	\bibfield  {author} {\bibinfo {author} {\bibfnamefont {G.}~\bibnamefont
			{Bell}}\ and\ \bibinfo {author} {\bibfnamefont {P.}~\bibnamefont
			{Rangecroft}},\ }\bibfield  {title} {\bibinfo {title} {A linearized potential
			equation for the interfacial region in an unsymmetrical electrolyte},\ }\href
	{https://doi.org/10.1080/00268977200101421} {\bibfield  {journal} {\bibinfo
			{journal} {J. Phys. Chem.}\ }\textbf {\bibinfo {volume} {24}},\ \bibinfo
		{pages} {255} (\bibinfo {year} {1972})}\BibitemShut {NoStop}%
	\bibitem [{\citenamefont {Henderson}\ \emph {et~al.}(1979)\citenamefont
		{Henderson}, \citenamefont {Blum},\ and\ \citenamefont
		{Smith}}]{hendersonApplicationHypernettedChain1979}%
	\BibitemOpen
	\bibfield  {author} {\bibinfo {author} {\bibfnamefont {D.}~\bibnamefont
			{Henderson}}, \bibinfo {author} {\bibfnamefont {L.}~\bibnamefont {Blum}},\
		and\ \bibinfo {author} {\bibfnamefont {W.~R.}\ \bibnamefont {Smith}},\
	}\bibfield  {title} {\bibinfo {title} {Application of the hypernetted chain
			approximation to the electric double layer at a charged planar interface},\
	}\href {https://doi.org/10.1016/0009-2614(79)87041-4} {\bibfield  {journal}
		{\bibinfo  {journal} {Chem. Phys. Lett.}\ }\textbf {\bibinfo {volume} {63}},\
		\bibinfo {pages} {381} (\bibinfo {year} {1979})}\BibitemShut {NoStop}%
	\bibitem [{\citenamefont {Kjellander}\ and\ \citenamefont {Marc{\v
				e}lja}(1984)}]{kjellanderCorrelationImageCharge1984}%
	\BibitemOpen
	\bibfield  {author} {\bibinfo {author} {\bibfnamefont {R.}~\bibnamefont
			{Kjellander}}\ and\ \bibinfo {author} {\bibfnamefont {S.}~\bibnamefont
			{Marc{\v e}lja}},\ }\bibfield  {title} {\bibinfo {title} {Correlation and
			image charge effects in electric double layers},\ }\href
	{https://doi.org/10.1016/0009-2614(84)87039-6} {\bibfield  {journal}
		{\bibinfo  {journal} {Chem. Phys. Lett.}\ }\textbf {\bibinfo {volume}
			{112}},\ \bibinfo {pages} {49} (\bibinfo {year} {1984})}\BibitemShut
	{NoStop}%
	\bibitem [{\citenamefont {Attard}\ \emph {et~al.}(1988)\citenamefont {Attard},
		\citenamefont {Mitchell},\ and\ \citenamefont
		{Ninham}}]{attardPoissonBoltzmannImages1988}%
	\BibitemOpen
	\bibfield  {author} {\bibinfo {author} {\bibfnamefont {P.}~\bibnamefont
			{Attard}}, \bibinfo {author} {\bibfnamefont {D.~J.}\ \bibnamefont
			{Mitchell}},\ and\ \bibinfo {author} {\bibfnamefont {B.~W.}\ \bibnamefont
			{Ninham}},\ }\bibfield  {title} {\bibinfo {title} {Beyond
			{{Poisson}}--{{Boltzmann}}: {{Images}} and correlations in the electric
			double layer. {{I}}. {{Counterions}} only},\ }\href
	{https://doi.org/10.1063/1.454678} {\bibfield  {journal} {\bibinfo  {journal}
			{J. Chem. Phys.}\ }\textbf {\bibinfo {volume} {88}},\ \bibinfo {pages} {4987}
		(\bibinfo {year} {1988})}\BibitemShut {NoStop}%
	\bibitem [{\citenamefont {Netz}\ and\ \citenamefont
		{Orland}(2000)}]{netzPoissonBoltzmannFluctuationEffects2000}%
	\BibitemOpen
	\bibfield  {author} {\bibinfo {author} {\bibfnamefont {R.}~\bibnamefont
			{Netz}}\ and\ \bibinfo {author} {\bibfnamefont {H.}~\bibnamefont {Orland}},\
	}\bibfield  {title} {\bibinfo {title} {Beyond {{Poisson-Boltzmann}}:
			{{Fluctuation}} effects and correlation functions},\ }\href
	{https://doi.org/10.1007/s101890050023} {\bibfield  {journal} {\bibinfo
			{journal} {Eur. Phys. J. E}\ }\textbf {\bibinfo {volume} {1}},\ \bibinfo
		{pages} {203} (\bibinfo {year} {2000})}\BibitemShut {NoStop}%
	\bibitem [{\citenamefont {Lau}(2008)}]{lauFluctuationCorrelationEffects2008}%
	\BibitemOpen
	\bibfield  {author} {\bibinfo {author} {\bibfnamefont {A.~W.~C.}\
			\bibnamefont {Lau}},\ }\bibfield  {title} {\bibinfo {title} {Fluctuation and
			correlation effects in a charged surface immersed in an electrolyte
			solution},\ }\href {https://doi.org/10.1103/PhysRevE.77.011502} {\bibfield
		{journal} {\bibinfo  {journal} {Phys. Rev. E}\ }\textbf {\bibinfo {volume}
			{77}},\ \bibinfo {pages} {011502} (\bibinfo {year} {2008})}\BibitemShut
	{NoStop}%
	\bibitem [{\citenamefont
		{Hubbard}(1959)}]{hubbardCalculationPartitionFunctions1959}%
	\BibitemOpen
	\bibfield  {author} {\bibinfo {author} {\bibfnamefont {J.}~\bibnamefont
			{Hubbard}},\ }\bibfield  {title} {\bibinfo {title} {Calculation of
			{{Partition Functions}}},\ }\href {https://doi.org/10.1103/PhysRevLett.3.77}
	{\bibfield  {journal} {\bibinfo  {journal} {Phys. Rev. Lett.}\ }\textbf
		{\bibinfo {volume} {3}},\ \bibinfo {pages} {77} (\bibinfo {year}
		{1959})}\BibitemShut {NoStop}%
	\bibitem [{\citenamefont
		{Stratonovich}(1957)}]{stratonovichMethodCalculatingQuantum1957}%
	\BibitemOpen
	\bibfield  {author} {\bibinfo {author} {\bibfnamefont {R.~L.}\ \bibnamefont
			{Stratonovich}},\ }\bibfield  {title} {\bibinfo {title} {On a {{Method}} of
			{{Calculating Quantum Distribution Functions}}},\ }\href@noop {} {\bibfield
		{journal} {\bibinfo  {journal} {Dokl. Phys.}\ }\textbf {\bibinfo {volume}
			{2}},\ \bibinfo {pages} {416} (\bibinfo {year} {1957})}\BibitemShut {NoStop}%
	\bibitem [{\citenamefont {Markovich}\ \emph {et~al.}(2016)\citenamefont
		{Markovich}, \citenamefont {Andelman},\ and\ \citenamefont
		{Orland}}]{markovichIonicProfilesClose2016}%
	\BibitemOpen
	\bibfield  {author} {\bibinfo {author} {\bibfnamefont {T.}~\bibnamefont
			{Markovich}}, \bibinfo {author} {\bibfnamefont {D.}~\bibnamefont
			{Andelman}},\ and\ \bibinfo {author} {\bibfnamefont {H.}~\bibnamefont
			{Orland}},\ }\bibfield  {title} {\bibinfo {title} {Ionic profiles close to
			dielectric discontinuities: {{Specific}} ion-surface interactions},\ }\href
	{https://doi.org/10.1063/1.4963083} {\bibfield  {journal} {\bibinfo
			{journal} {J. Chem. Phys.}\ }\textbf {\bibinfo {volume} {145}},\ \bibinfo
		{pages} {134704} (\bibinfo {year} {2016})}\BibitemShut {NoStop}%
	\bibitem [{\citenamefont {Buyukdagli}\ \emph {et~al.}(2010)\citenamefont
		{Buyukdagli}, \citenamefont {Manghi},\ and\ \citenamefont
		{Palmeri}}]{buyukdagliVariationalApproachElectrolyte2010}%
	\BibitemOpen
	\bibfield  {author} {\bibinfo {author} {\bibfnamefont {S.}~\bibnamefont
			{Buyukdagli}}, \bibinfo {author} {\bibfnamefont {M.}~\bibnamefont {Manghi}},\
		and\ \bibinfo {author} {\bibfnamefont {J.}~\bibnamefont {Palmeri}},\
	}\bibfield  {title} {\bibinfo {title} {Variational approach for electrolyte
			solutions: {{From}} dielectric interfaces to charged nanopores},\ }\href
	{https://doi.org/10.1103/PhysRevE.81.041601} {\bibfield  {journal} {\bibinfo
			{journal} {Phys. Rev. E}\ }\textbf {\bibinfo {volume} {81}},\ \bibinfo
		{pages} {041601} (\bibinfo {year} {2010})}\BibitemShut {NoStop}%
	\bibitem [{\citenamefont {Lue}(2017)}]{lueVariationalPerturbationTheory2017}%
	\BibitemOpen
	\bibfield  {author} {\bibinfo {author} {\bibfnamefont {L.}~\bibnamefont
			{Lue}},\ }\bibfield  {title} {\bibinfo {title} {Variational {{Perturbation
					Theory}} for~{{Electrolyte Solutions}}},\ }in\ \href
	{https://doi.org/10.1007/978-981-10-2502-0_5} {\emph {\bibinfo {booktitle}
			{Variational {{Methods}} in {{Molecular Modeling}}}}},\ \bibinfo {series and
		number} {Molecular {{Modeling}} and {{Simulation}}},\ \bibinfo {editor}
	{edited by\ \bibinfo {editor} {\bibfnamefont {J.}~\bibnamefont {Wu}}}\
	(\bibinfo  {publisher} {Springer},\ \bibinfo {address} {Singapore},\ \bibinfo
	{year} {2017})\ pp.\ \bibinfo {pages} {137--154}\BibitemShut {NoStop}%
	\bibitem [{\citenamefont {Wang}\ and\ \citenamefont
		{Wang}(2013)}]{wangEffectsImageCharges2013}%
	\BibitemOpen
	\bibfield  {author} {\bibinfo {author} {\bibfnamefont {R.}~\bibnamefont
			{Wang}}\ and\ \bibinfo {author} {\bibfnamefont {Z.-G.}\ \bibnamefont
			{Wang}},\ }\bibfield  {title} {\bibinfo {title} {Effects of image charges on
			double layer structure and forces},\ }\href
	{https://doi.org/10.1063/1.4821636} {\bibfield  {journal} {\bibinfo
			{journal} {J. Chem. Phys.}\ }\textbf {\bibinfo {volume} {139}},\ \bibinfo
		{pages} {124702} (\bibinfo {year} {2013})}\BibitemShut {NoStop}%
	\bibitem [{\citenamefont {Wang}\ and\ \citenamefont
		{Wang}(2015)}]{wangTheoreticalDescriptionWeakly2015}%
	\BibitemOpen
	\bibfield  {author} {\bibinfo {author} {\bibfnamefont {R.}~\bibnamefont
			{Wang}}\ and\ \bibinfo {author} {\bibfnamefont {Z.-G.}\ \bibnamefont
			{Wang}},\ }\bibfield  {title} {\bibinfo {title} {On the theoretical
			description of weakly charged surfaces},\ }\href
	{https://doi.org/10.1063/1.4914170} {\bibfield  {journal} {\bibinfo
			{journal} {J. Chem. Phys.}\ }\textbf {\bibinfo {volume} {142}},\ \bibinfo
		{pages} {104705} (\bibinfo {year} {2015})}\BibitemShut {NoStop}%
	\bibitem [{\citenamefont {Zhou}\ \emph {et~al.}(2024)\citenamefont {Zhou},
		\citenamefont {Bruch},\ and\ \citenamefont
		{Wang}}]{zhouImageChargeEffects2024}%
	\BibitemOpen
	\bibfield  {author} {\bibinfo {author} {\bibfnamefont {T.}~\bibnamefont
			{Zhou}}, \bibinfo {author} {\bibfnamefont {D.}~\bibnamefont {Bruch}},\ and\
		\bibinfo {author} {\bibfnamefont {Z.-G.}\ \bibnamefont {Wang}},\ }\bibfield
	{title} {\bibinfo {title} {Image charge effects under metal and dielectric
			boundary conditions},\ }\href {https://doi.org/10.1103/PhysRevE.110.044129}
	{\bibfield  {journal} {\bibinfo  {journal} {Phys. Rev. E}\ }\textbf {\bibinfo
			{volume} {110}},\ \bibinfo {pages} {044129} (\bibinfo {year}
		{2024})}\BibitemShut {NoStop}%
	\bibitem [{\citenamefont
		{Schmickler}(1996)}]{schmicklerElectronicEffectsElectric1996}%
	\BibitemOpen
	\bibfield  {author} {\bibinfo {author} {\bibfnamefont {W.}~\bibnamefont
			{Schmickler}},\ }\bibfield  {title} {\bibinfo {title} {Electronic {{Effects}}
			in the {{Electric Double Layer}}},\ }\href
	{https://doi.org/10.1021/cr940408c} {\bibfield  {journal} {\bibinfo
			{journal} {Chem. Rev.}\ }\textbf {\bibinfo {volume} {96}},\ \bibinfo {pages}
		{3177} (\bibinfo {year} {1996})}\BibitemShut {NoStop}%
	\bibitem [{\citenamefont {Bruch}\ \emph
		{et~al.}(2024{\natexlab{b}})\citenamefont {Bruch}, \citenamefont {Binninger},
		\citenamefont {Huang},\ and\ \citenamefont
		{Eikerling}}]{bruchVariationalFunctionalTheory2024}%
	\BibitemOpen
	\bibfield  {author} {\bibinfo {author} {\bibfnamefont {N.}~\bibnamefont
			{Bruch}}, \bibinfo {author} {\bibfnamefont {T.}~\bibnamefont {Binninger}},
		\bibinfo {author} {\bibfnamefont {J.}~\bibnamefont {Huang}},\ and\ \bibinfo
		{author} {\bibfnamefont {M.}~\bibnamefont {Eikerling}},\ }\href
	{https://doi.org/10.48550/arXiv.2411.16521} {\bibinfo {title} {Variational
			functional theory for coulombic correlations in the electric double layer}}
	(\bibinfo {year} {2024}{\natexlab{b}}),\ \Eprint
	{https://arxiv.org/abs/2411.16521} {arXiv:2411.16521 [physics]} \BibitemShut
	{NoStop}%
	\bibitem [{\citenamefont {Taddei}\ \emph {et~al.}(2009)\citenamefont {Taddei},
		\citenamefont {Mendes},\ and\ \citenamefont
		{Farina}}]{taddeiSubtletiesEnergyCalculations2009}%
	\BibitemOpen
	\bibfield  {author} {\bibinfo {author} {\bibfnamefont {M.~M.}\ \bibnamefont
			{Taddei}}, \bibinfo {author} {\bibfnamefont {T.~N.~C.}\ \bibnamefont
			{Mendes}},\ and\ \bibinfo {author} {\bibfnamefont {C.}~\bibnamefont
			{Farina}},\ }\bibfield  {title} {\bibinfo {title} {Subtleties in energy
			calculations in the image method},\ }\href
	{https://doi.org/10.1088/0143-0807/30/5/005} {\bibfield  {journal} {\bibinfo
			{journal} {Eur. J. Phys.}\ }\textbf {\bibinfo {volume} {30}},\ \bibinfo
		{pages} {965} (\bibinfo {year} {2009})}\BibitemShut {NoStop}%
	\bibitem [{Note1()}]{Note1}%
	\BibitemOpen
	\bibinfo {note} {Multiplicative constants do not play a role for
		thermodynamic averages and are therefore neglected. This is due to the fact
		that averages are normalized with respect to the partition function.
		Therefore, multiplicative constants simply cancel out and can therefore be
		omitted.}\BibitemShut {Stop}%
	\bibitem [{\citenamefont {Boettcher}\ \emph {et~al.}(2021)\citenamefont
		{Boettcher}, \citenamefont {Oener}, \citenamefont {Lonergan}, \citenamefont
		{Surendranath}, \citenamefont {Ardo}, \citenamefont {Brozek},\ and\
		\citenamefont {Kempler}}]{boettcherPotentiallyConfusingPotentials2021}%
	\BibitemOpen
	\bibfield  {author} {\bibinfo {author} {\bibfnamefont {S.~W.}\ \bibnamefont
			{Boettcher}}, \bibinfo {author} {\bibfnamefont {S.~Z.}\ \bibnamefont
			{Oener}}, \bibinfo {author} {\bibfnamefont {M.~C.}\ \bibnamefont {Lonergan}},
		\bibinfo {author} {\bibfnamefont {Y.}~\bibnamefont {Surendranath}}, \bibinfo
		{author} {\bibfnamefont {S.}~\bibnamefont {Ardo}}, \bibinfo {author}
		{\bibfnamefont {C.}~\bibnamefont {Brozek}},\ and\ \bibinfo {author}
		{\bibfnamefont {P.~A.}\ \bibnamefont {Kempler}},\ }\bibfield  {title}
	{\bibinfo {title} {Potentially {{Confusing}}: {{Potentials}} in
			{{Electrochemistry}}},\ }\href
	{https://doi.org/10.1021/acsenergylett.0c02443} {\bibfield  {journal}
		{\bibinfo  {journal} {ACS Energy Lett.}\ }\textbf {\bibinfo {volume} {6}},\
		\bibinfo {pages} {261} (\bibinfo {year} {2021})}\BibitemShut {NoStop}%
	\bibitem [{\citenamefont
		{{Zinn-Justin}}(2021)}]{zinn-justinQuantumFieldTheory2021}%
	\BibitemOpen
	\bibfield  {author} {\bibinfo {author} {\bibfnamefont {J.}~\bibnamefont
			{{Zinn-Justin}}},\ }\href@noop {} {\emph {\bibinfo {title} {Quantum {{Field
						Theory}} and {{Critical Phenomena}}}}},\ Oxford {{Graduate Texts}}\ (\bibinfo
	{publisher} {Oxford University Press, USA},\ \bibinfo {year}
	{2021})\BibitemShut {NoStop}%
	\bibitem [{\citenamefont {Podgornik}\ and\ \citenamefont {{\v Z}ek{\v
				s}}(1988)}]{podgornikInhomogeneousCoulombFluid1988}%
	\BibitemOpen
	\bibfield  {author} {\bibinfo {author} {\bibfnamefont {R.}~\bibnamefont
			{Podgornik}}\ and\ \bibinfo {author} {\bibfnamefont {B.}~\bibnamefont {{\v
					Z}ek{\v s}}},\ }\bibfield  {title} {\bibinfo {title} {Inhomogeneous coulomb
			fluid. {{A}} functional integral approach},\ }\href
	{https://doi.org/10.1039/F29888400611} {\bibfield  {journal} {\bibinfo
			{journal} {J. Chem. Soc., Faraday Trans. 2}\ }\textbf {\bibinfo {volume}
			{84}},\ \bibinfo {pages} {611} (\bibinfo {year} {1988})}\BibitemShut
	{NoStop}%
	\bibitem [{\citenamefont {Brown}\ and\ \citenamefont
		{Yaffe}(2001)}]{brownEffectiveFieldTheory2001}%
	\BibitemOpen
	\bibfield  {author} {\bibinfo {author} {\bibfnamefont {L.~S.}\ \bibnamefont
			{Brown}}\ and\ \bibinfo {author} {\bibfnamefont {L.~G.}\ \bibnamefont
			{Yaffe}},\ }\bibfield  {title} {\bibinfo {title} {Effective field theory for
			highly ionized plasmas},\ }\href
	{https://doi.org/10.1016/S0370-1573(00)00068-5} {\bibfield  {journal}
		{\bibinfo  {journal} {Physics Reports}\ }\textbf {\bibinfo {volume} {340}},\
		\bibinfo {pages} {1} (\bibinfo {year} {2001})}\BibitemShut {NoStop}%
	\bibitem [{\citenamefont {Schmickler}\ and\ \citenamefont
		{Santos}(2010)}]{schmicklerInterfacialElectrochemistry2010}%
	\BibitemOpen
	\bibfield  {author} {\bibinfo {author} {\bibfnamefont {W.}~\bibnamefont
			{Schmickler}}\ and\ \bibinfo {author} {\bibfnamefont {E.}~\bibnamefont
			{Santos}},\ }\href {https://doi.org/10.1007/978-3-642-04937-8} {\emph
		{\bibinfo {title} {Interfacial {{Electrochemistry}}}}}\ (\bibinfo
	{publisher} {Springer Berlin Heidelberg},\ \bibinfo {address} {Berlin,
		Heidelberg},\ \bibinfo {year} {2010})\BibitemShut {NoStop}%
	\bibitem [{\citenamefont
		{Bikerman}(1942)}]{bikermanStructureCapacityElectrical1942}%
	\BibitemOpen
	\bibfield  {author} {\bibinfo {author} {\bibfnamefont {J.~J.}\ \bibnamefont
			{Bikerman}},\ }\bibfield  {title} {\bibinfo {title} {Structure and capacity
			of electrical double layer},\ }\href
	{https://doi.org/10.1080/14786444208520813} {\bibfield  {journal} {\bibinfo
			{journal} {Philos. Mag.}\ }\textbf {\bibinfo {volume} {33}},\ \bibinfo
		{pages} {384} (\bibinfo {year} {1942})}\BibitemShut {NoStop}%
	\bibitem [{\citenamefont {Bazant}\ \emph {et~al.}(2011)\citenamefont {Bazant},
		\citenamefont {Storey},\ and\ \citenamefont
		{Kornyshev}}]{bazantDoubleLayerIonic2011}%
	\BibitemOpen
	\bibfield  {author} {\bibinfo {author} {\bibfnamefont {M.~Z.}\ \bibnamefont
			{Bazant}}, \bibinfo {author} {\bibfnamefont {B.~D.}\ \bibnamefont {Storey}},\
		and\ \bibinfo {author} {\bibfnamefont {A.~A.}\ \bibnamefont {Kornyshev}},\
	}\bibfield  {title} {\bibinfo {title} {Double {{Layer}} in {{Ionic Liquids}}:
			{{Overscreening}} versus {{Crowding}}},\ }\href
	{https://doi.org/10.1103/PhysRevLett.106.046102} {\bibfield  {journal}
		{\bibinfo  {journal} {Phys. Rev. Lett.}\ }\textbf {\bibinfo {volume} {106}},\
		\bibinfo {pages} {046102} (\bibinfo {year} {2011})}\BibitemShut {NoStop}%
	\bibitem [{\citenamefont {Netz}(1999)}]{netzDebyeHUckelTheory1999}%
	\BibitemOpen
	\bibfield  {author} {\bibinfo {author} {\bibfnamefont {R.~R.}\ \bibnamefont
			{Netz}},\ }\bibfield  {title} {\bibinfo {title}
		{Debye-{{H}}{\textbackslash}"uckel theory for interfacial geometries},\
	}\href {https://doi.org/10.1103/PhysRevE.60.3174} {\bibfield  {journal}
		{\bibinfo  {journal} {Phys. Rev. E}\ }\textbf {\bibinfo {volume} {60}},\
		\bibinfo {pages} {3174} (\bibinfo {year} {1999})}\BibitemShut {NoStop}%
	\bibitem [{Note2()}]{Note2}%
	\BibitemOpen
	\bibinfo {note} {If the gauge is chosen to be $\mu _\protect \text
		{g}=q_i^2V_0$, then the the 1L expansion has an activity coefficient of DH
		form $\protect \qopname \relax o{log}\gamma _i = -\protect \frac {\beta
			q_{i}^{2}}{8\pi \epsilon _{B}\lambda _{D}}$. This shows that the 1L level
		that is being discussed here is, in fact, equivalent to the DH theory \cite
		{Debye_1923}. In the case of symmetric electrolytes, both anion and cation
		densities are equivalently modified, so electroneutrality is still satisfied
		in the bulk. However, this means that the 1L ion densities in the electrolyte
		bulk are different from the MF densities.}\BibitemShut {Stop}%
	\bibitem [{\citenamefont {Foresti}\ \emph {et~al.}(1993)\citenamefont
		{Foresti}, \citenamefont {Guidelli},\ and\ \citenamefont
		{Hamelin}}]{forestiModelEffectRoughness1993}%
	\BibitemOpen
	\bibfield  {author} {\bibinfo {author} {\bibfnamefont {M.~L.}\ \bibnamefont
			{Foresti}}, \bibinfo {author} {\bibfnamefont {R.}~\bibnamefont {Guidelli}},\
		and\ \bibinfo {author} {\bibfnamefont {A.}~\bibnamefont {Hamelin}},\
	}\bibfield  {title} {\bibinfo {title} {A model for the effect of roughness of
			single-crystal electrodes on {{Parsons-Zobel}} plots},\ }\href
	{https://doi.org/10.1016/0022-0728(93)85004-Z} {\bibfield  {journal}
		{\bibinfo  {journal} {J. Electroanal. Chem.}\ }\bibinfo {series} {An
			{{International Journal Devoted}} to All {{Aspects}} of {{Electrode
					Kinetics}}, {{Interfacial Structure}}, {{Properties}} of {{Electrolytes}},
			{{Colloid}} and {{Biological Electrochemistry}}},\ \textbf {\bibinfo {volume}
			{346}},\ \bibinfo {pages} {73} (\bibinfo {year} {1993})}\BibitemShut
	{NoStop}%
	\bibitem [{\citenamefont {Parsons}\ and\ \citenamefont
		{Zobel}(1965)}]{parsonsInterphaseMercuryAqueous1965}%
	\BibitemOpen
	\bibfield  {author} {\bibinfo {author} {\bibfnamefont {R.}~\bibnamefont
			{Parsons}}\ and\ \bibinfo {author} {\bibfnamefont {F.~G.~R.}\ \bibnamefont
			{Zobel}},\ }\bibfield  {title} {\bibinfo {title} {The interphase between
			mercury and aqueous sodium dihydrogen phosphate},\ }\href
	{https://doi.org/10.1016/0022-0728(65)85029-X} {\bibfield  {journal}
		{\bibinfo  {journal} {J. Electroanal. Chem.}\ }\textbf {\bibinfo {volume}
			{9}},\ \bibinfo {pages} {333} (\bibinfo {year} {1965})}\BibitemShut {NoStop}%
	\bibitem [{\citenamefont {Daikhin}\ \emph {et~al.}(1996)\citenamefont
		{Daikhin}, \citenamefont {Kornyshev},\ and\ \citenamefont
		{Urbakh}}]{daikhinDoublelayerCapacitanceRough1996}%
	\BibitemOpen
	\bibfield  {author} {\bibinfo {author} {\bibfnamefont {L.~I.}\ \bibnamefont
			{Daikhin}}, \bibinfo {author} {\bibfnamefont {A.~A.}\ \bibnamefont
			{Kornyshev}},\ and\ \bibinfo {author} {\bibfnamefont {M.}~\bibnamefont
			{Urbakh}},\ }\bibfield  {title} {\bibinfo {title} {Double-layer capacitance
			on a rough metal surface},\ }\href {https://doi.org/10.1103/PhysRevE.53.6192}
	{\bibfield  {journal} {\bibinfo  {journal} {Phys. Rev. E}\ }\textbf {\bibinfo
			{volume} {53}},\ \bibinfo {pages} {6192} (\bibinfo {year}
		{1996})}\BibitemShut {NoStop}%
	\bibitem [{\citenamefont {Gabovich}\ \emph {et~al.}(2006)\citenamefont
		{Gabovich}, \citenamefont {Reznikov},\ and\ \citenamefont
		{Voitenko}}]{gabovichExcessNonspecificCoulomb2006}%
	\BibitemOpen
	\bibfield  {author} {\bibinfo {author} {\bibfnamefont {A.~M.}\ \bibnamefont
			{Gabovich}}, \bibinfo {author} {\bibfnamefont {{\relax Yu}.~A.}\ \bibnamefont
			{Reznikov}},\ and\ \bibinfo {author} {\bibfnamefont {A.~I.}\ \bibnamefont
			{Voitenko}},\ }\bibfield  {title} {\bibinfo {title} {Excess nonspecific
			{{Coulomb}} ion adsorption at the metal electrode/electrolyte solution
			interface: {{Role}} of the surface layer},\ }\href
	{https://doi.org/10.1103/PhysRevE.73.021606} {\bibfield  {journal} {\bibinfo
			{journal} {Phys. Rev. E}\ }\textbf {\bibinfo {volume} {73}},\ \bibinfo
		{pages} {021606} (\bibinfo {year} {2006})}\BibitemShut {NoStop}%
	\bibitem [{\citenamefont {Gonz{\'a}lez}\ \emph {et~al.}(2016)\citenamefont
		{Gonz{\'a}lez}, \citenamefont {Goikolea}, \citenamefont {Barrena},\ and\
		\citenamefont {Mysyk}}]{gonzalezReviewSupercapacitorsTechnologies2016}%
	\BibitemOpen
	\bibfield  {author} {\bibinfo {author} {\bibfnamefont {A.}~\bibnamefont
			{Gonz{\'a}lez}}, \bibinfo {author} {\bibfnamefont {E.}~\bibnamefont
			{Goikolea}}, \bibinfo {author} {\bibfnamefont {J.~A.}\ \bibnamefont
			{Barrena}},\ and\ \bibinfo {author} {\bibfnamefont {R.}~\bibnamefont
			{Mysyk}},\ }\bibfield  {title} {\bibinfo {title} {Review on supercapacitors:
			{{Technologies}} and materials},\ }\href
	{https://doi.org/10.1016/j.rser.2015.12.249} {\bibfield  {journal} {\bibinfo
			{journal} {Renewable and Sustainable Energy Reviews}\ }\textbf {\bibinfo
			{volume} {58}},\ \bibinfo {pages} {1189} (\bibinfo {year}
		{2016})}\BibitemShut {NoStop}%
	\bibitem [{\citenamefont
		{Grahame}(1954)}]{grahameDifferentialCapacityMercury1954}%
	\BibitemOpen
	\bibfield  {author} {\bibinfo {author} {\bibfnamefont {D.~C.}\ \bibnamefont
			{Grahame}},\ }\bibfield  {title} {\bibinfo {title} {Differential {{Capacity}}
			of {{Mercury}} in {{Aqueous Sodium Fluoride Solutions}}. {{I}}. {{Effect}} of
			{{Concentration}} at 25{$^\circ$}},\ }\href
	{https://doi.org/10.1021/ja01648a014} {\bibfield  {journal} {\bibinfo
			{journal} {J. Am. Chem. Soc.}\ }\textbf {\bibinfo {volume} {76}},\ \bibinfo
		{pages} {4819} (\bibinfo {year} {1954})}\BibitemShut {NoStop}%
	\bibitem [{\citenamefont {Valette}(1989)}]{valetteDoubleLayerSilver1989}%
	\BibitemOpen
	\bibfield  {author} {\bibinfo {author} {\bibfnamefont {G.}~\bibnamefont
			{Valette}},\ }\bibfield  {title} {\bibinfo {title} {Double layer on silver
			single crystal electrodes in contact with electrolytes having anions which
			are slightly specifically adsorbed: {{Part III}}. {{The}} (111) face},\
	}\href {https://doi.org/10.1016/0022-0728(89)80112-3} {\bibfield  {journal}
		{\bibinfo  {journal} {J. Electroanal. Chem.}\ }\textbf {\bibinfo {volume}
			{269}},\ \bibinfo {pages} {191} (\bibinfo {year} {1989})}\BibitemShut
	{NoStop}%
	\bibitem [{\citenamefont {Hamelin}\ and\ \citenamefont
		{Stoicoviciu}(1987)}]{hamelinStudyGoldLow1987}%
	\BibitemOpen
	\bibfield  {author} {\bibinfo {author} {\bibfnamefont {A.}~\bibnamefont
			{Hamelin}}\ and\ \bibinfo {author} {\bibfnamefont {L.}~\bibnamefont
			{Stoicoviciu}},\ }\bibfield  {title} {\bibinfo {title} {Study of gold low
			index faces in {{KPF6}} solutions},\ }\href
	{https://doi.org/10.1016/0022-0728(87)80164-X} {\bibfield  {journal}
		{\bibinfo  {journal} {J. Electroanal. Chem.}\ }\textbf {\bibinfo {volume}
			{234}},\ \bibinfo {pages} {93} (\bibinfo {year} {1987})}\BibitemShut
	{NoStop}%
	\bibitem [{\citenamefont {Shatla}\ \emph {et~al.}(2021)\citenamefont {Shatla},
		\citenamefont {Landstorfer},\ and\ \citenamefont
		{Baltruschat}}]{shatlaDifferentialCapacitancePotential2021}%
	\BibitemOpen
	\bibfield  {author} {\bibinfo {author} {\bibfnamefont {A.~S.}\ \bibnamefont
			{Shatla}}, \bibinfo {author} {\bibfnamefont {M.}~\bibnamefont
			{Landstorfer}},\ and\ \bibinfo {author} {\bibfnamefont {H.}~\bibnamefont
			{Baltruschat}},\ }\bibfield  {title} {\bibinfo {title} {On the {{Differential
					Capacitance}} and {{Potential}} of {{Zero Charge}} of {{Au}}(111) in {{Some
					Aprotic Solvents}}},\ }\href {https://doi.org/10.1002/celc.202100316}
	{\bibfield  {journal} {\bibinfo  {journal} {ChemElectroChem}\ }\textbf
		{\bibinfo {volume} {8}},\ \bibinfo {pages} {1817} (\bibinfo {year}
		{2021})}\BibitemShut {NoStop}%
	\bibitem [{\citenamefont {Eberhardt}\ \emph {et~al.}(1996)\citenamefont
		{Eberhardt}, \citenamefont {Santos},\ and\ \citenamefont
		{Schmickler}}]{eberhardtImpedanceStudiesReconstructed1996}%
	\BibitemOpen
	\bibfield  {author} {\bibinfo {author} {\bibfnamefont {D.}~\bibnamefont
			{Eberhardt}}, \bibinfo {author} {\bibfnamefont {E.}~\bibnamefont {Santos}},\
		and\ \bibinfo {author} {\bibfnamefont {W.}~\bibnamefont {Schmickler}},\
	}\bibfield  {title} {\bibinfo {title} {Impedance studies of reconstructed and
			non-reconstructed gold single crystal surfaces},\ }\href
	{https://doi.org/10.1016/S0022-0728(96)04872-3} {\bibfield  {journal}
		{\bibinfo  {journal} {J. Electroanal. Chem.}\ }\textbf {\bibinfo {volume}
			{419}},\ \bibinfo {pages} {23} (\bibinfo {year} {1996})}\BibitemShut
	{NoStop}%
	\bibitem [{\citenamefont {Li}\ \emph {et~al.}(2025)\citenamefont {Li},
		\citenamefont {Eggert}, \citenamefont {Reuter},\ and\ \citenamefont
		{H{\"o}rmann}}]{liElectronSpilloverWater2025}%
	\BibitemOpen
	\bibfield  {author} {\bibinfo {author} {\bibfnamefont {L.}~\bibnamefont
			{Li}}, \bibinfo {author} {\bibfnamefont {T.}~\bibnamefont {Eggert}}, \bibinfo
		{author} {\bibfnamefont {K.}~\bibnamefont {Reuter}},\ and\ \bibinfo {author}
		{\bibfnamefont {N.~G.}\ \bibnamefont {H{\"o}rmann}},\ }\bibfield  {title}
	{\bibinfo {title} {Electron {{Spillover}} into {{Water Layers}}: {{A Quantum
					Leap}} in {{Understanding Capacitance Behavior}}},\ }\bibfield  {journal}
	{\bibinfo  {journal} {Electron spillover into water layers: A quantum leap in
			understanding capacitance behavior}\ }\href
	{https://doi.org/10.26434/chemrxiv-2025-hx7t5} {10.26434/chemrxiv-2025-hx7t5}
	(\bibinfo {year} {2025})\BibitemShut {NoStop}%
	\bibitem [{\citenamefont {Gim}\ \emph {et~al.}(2019)\citenamefont {Gim},
		\citenamefont {Cho}, \citenamefont {Lim},\ and\ \citenamefont
		{Kim}}]{gimStructureDynamicsWettability2019}%
	\BibitemOpen
	\bibfield  {author} {\bibinfo {author} {\bibfnamefont {S.}~\bibnamefont
			{Gim}}, \bibinfo {author} {\bibfnamefont {K.~J.}\ \bibnamefont {Cho}},
		\bibinfo {author} {\bibfnamefont {H.-K.}\ \bibnamefont {Lim}},\ and\ \bibinfo
		{author} {\bibfnamefont {H.}~\bibnamefont {Kim}},\ }\bibfield  {title}
	{\bibinfo {title} {Structure, {{Dynamics}}, and {{Wettability}} of {{Water}}
			at {{Metal Interfaces}}},\ }\href
	{https://doi.org/10.1038/s41598-019-51323-5} {\bibfield  {journal} {\bibinfo
			{journal} {Sci Rep}\ }\textbf {\bibinfo {volume} {9}},\ \bibinfo {pages}
		{14805} (\bibinfo {year} {2019})}\BibitemShut {NoStop}%
	\bibitem [{\citenamefont {Valette}(1982)}]{valetteDoubleLayerSilver1982}%
	\BibitemOpen
	\bibfield  {author} {\bibinfo {author} {\bibfnamefont {G.}~\bibnamefont
			{Valette}},\ }\bibfield  {title} {\bibinfo {title} {Double layer on silver
			single crystal electrodes in contact with electrolytes having anions which
			are slightly specifically adsorbed: {{Part II}}. {{The}} (100) face},\ }\href
	{https://doi.org/10.1016/0022-0728(82)87126-X} {\bibfield  {journal}
		{\bibinfo  {journal} {J. Electroanal. Chem.}\ }\textbf {\bibinfo {volume}
			{138}},\ \bibinfo {pages} {37} (\bibinfo {year} {1982})}\BibitemShut
	{NoStop}%
	\bibitem [{\citenamefont {Huang}\ \emph {et~al.}(2024)\citenamefont {Huang},
		\citenamefont {{Dom{\'i}nguez-Flores}},\ and\ \citenamefont
		{Melander}}]{huangVariantsSurfaceCharges2024}%
	\BibitemOpen
	\bibfield  {author} {\bibinfo {author} {\bibfnamefont {J.}~\bibnamefont
			{Huang}}, \bibinfo {author} {\bibfnamefont {F.}~\bibnamefont
			{{Dom{\'i}nguez-Flores}}},\ and\ \bibinfo {author} {\bibfnamefont
			{M.}~\bibnamefont {Melander}},\ }\bibfield  {title} {\bibinfo {title}
		{Variants of {{Surface Charges}} and {{Capacitances}} in
			{{Electrocatalysis}}: {{Insights}} from {{Density-Potential Functional Theory
					Embedded}} with an {{Implicit Chemisorption Model}}},\ }\href
	{https://doi.org/10.1103/PRXEnergy.3.043008} {\bibfield  {journal} {\bibinfo
			{journal} {PRX Energy}\ }\textbf {\bibinfo {volume} {3}},\ \bibinfo {pages}
		{043008} (\bibinfo {year} {2024})}\BibitemShut {NoStop}%
	\bibitem [{\citenamefont {Altland}\ and\ \citenamefont
		{Simons}(2010)}]{altlandCondensedMatterField2010}%
	\BibitemOpen
	\bibfield  {author} {\bibinfo {author} {\bibfnamefont {A.}~\bibnamefont
			{Altland}}\ and\ \bibinfo {author} {\bibfnamefont {B.~D.}\ \bibnamefont
			{Simons}},\ }\href {https://doi.org/10.1017/CBO9780511789984} {\emph
		{\bibinfo {title} {Condensed {{Matter Field Theory}}}}}\ (\bibinfo
	{publisher} {Cambridge University Press},\ \bibinfo {address} {Cambridge,
		England},\ \bibinfo {year} {2010})\BibitemShut {NoStop}%
	\bibitem [{\citenamefont {Debye}\ and\ \citenamefont
		{H{\"u}ckel}(1923)}]{Debye_1923}%
	\BibitemOpen
	\bibfield  {author} {\bibinfo {author} {\bibfnamefont {P.}~\bibnamefont
			{Debye}}\ and\ \bibinfo {author} {\bibfnamefont {E.}~\bibnamefont
			{H{\"u}ckel}},\ }\bibfield  {title} {\bibinfo {title} {Zur theorie der
			elektrolyte. {{I}}. {{Gefrierpunktserniedrigung}} und verwandte
			erscheinungen},\ }\href@noop {} {\bibfield  {journal} {\bibinfo  {journal}
			{Phys. Z.}\ }\textbf {\bibinfo {volume} {24}},\ \bibinfo {pages} {305}
		(\bibinfo {year} {1923})}\BibitemShut {NoStop}%
\end{thebibliography}
\end{document}